\documentclass[a4paper,12pt]{article}
\usepackage{graphicx,graphics,xcolor}
\usepackage{epsfig}
\usepackage{amsmath}
\usepackage{amssymb}
\usepackage{mathtools}
\usepackage[top=1cm, bottom=1.5cm, left=1.5cm, right=1.5cm]{geometry}
\usepackage{microtype}
\usepackage{lmodern}
\usepackage[utf8]{inputenc}
\usepackage{float}
\usepackage{hyperref}
\usepackage{cite}
\usepackage{orcidlink}
\newcommand{\ds}{\displaystyle }

\newcommand{\beq}{\begin{equation} }
\newcommand{\eeq}{\end{equation}}
\begin{document}
\author{Thomas M. Michelitsch  
\orcidlink{0000-0001-7955-6666} 
$^a$, 
Alejandro P. Riascos 
\orcidlink{0000-0002-9243-3246} 
$^b$
\\[1ex]
\footnotesize{$^a$ Sorbonne Université, CNRS, Institut Jean Le Rond d’Alembert, F-75005 Paris, France } \\
\footnotesize{E-mail: thomas.michelitsch@sorbonne-universite.fr} \,\, (corresponding) \\[1ex]
\footnotesize{$^b$ Departamento de Física, Universidad Nacional de Colombia, Bogot\'a, Colombia}  \\
\footnotesize{E-mail: alperezri@unal.edu.co} 
}
\title{Evanescent random walker on networks: Hitting times, budget renewal, and survival dynamics}
\maketitle
\begin{abstract}
We consider a mortal random walker evolving with discrete time on a network, where transitions follow a degree-biased Markovian navigation strategy. The walker starts with a random initial budget $T_1 \in \mathbb{N}$ and must maintain a strictly positive budget to remain alive. Each step incurs a unit cost, decrementing the budget by one; the walker perishes (is ruined) upon depletion of the budget. However, when the walker reaches designated target nodes, the budget is renewed by an independent and identically distributed (IID) copy of its initial value. The degree bias is tuned to either favor or disfavor visits to these target nodes. Our model exhibits connections with stochastic resetting. The evolution of the budget can be interpreted as a deterministic drift on the integer line toward negative values, where the walker is intermittently reset to positive IID random positions  and dies at the first hit of the origin. The first part of the paper focuses on the target-hitting statistics of an immortal Markovian walker. We analyze the \textit{target hitting counting process} (THCP) for an arbitrary set of target nodes. In the special case where a single target node coincides with the starting node, the THCP reduces to a renewal counting process. We establish connections with classical results from the literature. Within this framework, the second part of the paper addresses the dynamics of the evanescent walker. We derive analytical results for arbitrary configurations of target nodes, including the evanescent propagator matrix, the survival probability, the mean residence time on a set of nodes during the walker's lifetime, and the expected lifetime itself. Additionally, we compute the expected number of target hits (i.e., budget renewals) in a lifetime of the walker and related distributions. We explore both analytically and numerically a set of characteristic scenarios, including a \textit{forager scenario}, in which frequent encounters with target nodes extend the walker's lifetime, and a \textit{detrimental scenario}, where such encounters instead reduce it. Finally, we identify a \textit{neutral scenario} in which frequent visits to target nodes have no effect on the walker's lifetime. Our analytical results are validated through random walk simulations.
\end{abstract}

\clearpage
%
%
%
%
%
%
\section*{}
%
%
%
%
{\bf
Dynamical phenomena in which agents have a limited lifespan are ubiquitous in nature and occur in virtually all disciplines. In random walk dynamics, the mortality of a walker may drastically alter dynamic characteristics, such as the first-passage statistics of a target. 
In many cases, the survival of an agent is governed by the limitation of available resources (energy, food, money, etc.), which can be quantified by a `budget'. In this picture, survival depends on the balance between the steady cost of resources needed to maintain the functionality of `life' and their acquisition. Once the budget reaches a critically low value for the first time, the life of the agent ends. Applications of such models are cross-disciplinary. 
In animal foraging, this picture corresponds to the competition between ``metabolic resilience'' to resist starvation and the frequency of food consumption. The economic `survival' of a company depends on its limited monetary resources, which evolve through the struggle between steady costs (salaries of employees, investments in development, and others) and the generated income from operations. In a large class of such systems, the dynamics can be reduced to a mortal random walker whose survival depends on limited resources.
\\[2mm]
In this paper, we study a generic model of such an evanescent motion: a mortal random walker travelling in an ergodic network. The survival of the walker requires a positive budget, which is governed by the competition between the costs of the steps and budget renewals occurring upon visits of a designated target. 
We derive analytical solutions for the pertinent distributions characterizing the survival and target-hitting statistics of this walker. We explore several aspects of the resulting complex dynamics, including the expected lifetime of the walker, the counting process of target visits, and related quantities.
} 

\section{Introduction}
\label{Intro}
The interest in random walks on networks has continuously grown during recent decades due to the capacity of graphs to capture a wide range of dynamical phenomena observed in nature and human society \cite{Hughes1996}. 
Network science has become a vast interdisciplinary field with cross-disciplinary applications in physics, mathematical biology, epidemiology, population dynamics, social science, economy, computer science, and others. 
The upswing of network science was enhanced by several seminal works which include the discovery of the small world property by Watts and Strogatz \cite{WS1998}, and of the scale-free feature by Barabási and Albert \cite{BA1999,Albert-etal1999} with the development of the World Wide Web at about the same time \cite{BrinPage1998}.
Fundamental aspects of random walks in networks have been extensively studied 
\cite{MontrollWeiss1965,NohRieger2004,AlbertBarabasi2002,Newman2010,VanMiegem2011,fractionalbook-2019}. Applications of random walks are widespread and include anomalous diffusion and transport phenomena 
\cite{Scher_Montroll1975,Shlesinger2017,MetzlerKlafter2000,Blanchard-Volchenkov2011},
dynamical models to optimize traffic flows in communication networks \cite{Wang-etal2006}, spreading of epidemics \cite{Satoras-Vespignani-etal2015,Pastor-SatorrasVespignani2001,OkabeShudo2021,BasnakovSandev-etal2020,BesMi-etal2021,BesMiRias2022,Granger-et-al2023,SISI_Entropy_2024} with implications of passenger traveling \cite{Lawyer2016,Mao-etal2015},  
human mobility dynamics \cite{RiaMateo2020}, and random walks with optimal mixing features based on maximum entropy \cite{Sinatra_etal2011}. Further works consider biased walks in weighted networks \cite{Medina_et_al2022,Riascos_Mateos2021} with models of ageing of complex systems \cite{Riascos_WangMi2019}.
An interesting class of biased random walks in networks constitute the `degree biased' walks \cite{Calva-Riascos2022,Fronczak2009,Yang2005}. 
Degree biased walks are also considered in the present paper. 

A large field of research is dedicated to optimize random search strategies. Several studies of random walks with stochastic resetting \cite{Mujamda2011} 
in complex networks have been developed with emphasis to improve or optimize search strategies \cite{Huang_Chen2021,Chen_Ye2022,Singh_Metzler_Sandev_Chaos2025,Boyer_Ria_Chaos2025,Mi_Dono_Po_Ria_Chaos2025}. Indeed, SR may significantly reduce first passage quantities to a specific target, where in some cases an optimal resetting rate exists
\cite{Kusmierz-etal2014,Pal_Reuveni2017,Chechkin2018,Das2022,Jalakovsky-et-al2023,Herringer_Riascos2020,Boyer-et-al2023,Radice-etal2025}. Further works include the analysis of multiple competing walkers under resetting \cite{SinghMetzlerSandev2025},
ultraslow diffusion with resetting \cite{Cherstvy_etal2025}, emerging non-ergodicity in various diffusion processes 
\cite{WeiWang_etal2021,Vinod_etal2022},
and effects of mortality on random search under resetting \cite{Radice2023}.

Compared to persistent random motions of immortal random walkers, the problem class of mortal or evanescent random walks is relatively rarely addressed in the literature \cite{Yuste_etal2013}. Mortal walkers ``die'' (disappear) in the course of their motion.
Evanescent walks have interesting applications in foraging.
In this context, foragers with depletion of food resources and mortality from starvation have been 
thoroughly analyzed \cite{Pyke1984,Benichou-etal2011,Chupeau_Beni2016} including variants with regeneration of food resources 
\cite{Chupeau_eta2016} and effects of ``greed'' and ``food aversion'' \cite{Bhat_etal2017,Sanhedrai_etal2021}. 
The original version of the so called `starving random walk' \cite{Starving_walk_beni2014} is an idealized model for the foraging process of a mortal walker unaware of food resources, which travels in a Polya walk through the d-dimensional infinite lattice.
This walker survives ${\it S}$ steps without food encounter, where ${\it S}$ is constant.
In the beginning, each lattice point contains a food unit which is consumed and depleted upon visitations of the walker. 
In order to survive, the walker needs sufficiently often hop on new sites. For a given observation time,
the exploration of new sites in a Polya walk increases with the dimension $d$ of the lattice. The expected lifetime of this walker is therefore a monotonously increasing function of lattice dimension $d$ 
for walkers of the same ${\cal S}$ \cite{Chupeau_Beni2016}.

In the present paper, we analyze a mortal walker which travels with Markovian (degree-biased) steps through an ergodic network. In order to stay alive and continue its walk, our walker needs to maintain a positive (energy) budget.
Each step has a cost and reduces the budget by one unit, bringing the walker closer to the ruin condition (death) for which the budget reaches null and the walk ends. 
Our walk begins with a random budget $T_1 \in \mathbb{N}$, which is renewed at each visit of specific target nodes (t-nodes) by an independent identically distributed (IID) copy of $T_1$.
The survival of our walker depends on repeated visits of t-nodes.
The dynamics of the budget itself can be thought as a deterministic motion on the integer line with unit steps in the negative direction and stochastic relocations (renewals of the budget) to positive IID integer positions, where this motion ends (death of the walker) 
when the origin (zero budget) is reached for the first time.
%
%
The random motion of the budget evolution has a connection with resetting: Whenever the walker hits
any target node, the budget is reset.
%
%
%
%
The budget dynamics also allows for interpretations in gambling where
the capital (budget) is renewed under certain random events and the game ends at the ruin condition.
Several random walk models in which the motion is associated with a cost have been studied recently \cite{Mujandar_Mori_etal2023,Burunev_Mujamdar2025}.

Our paper has the following structure. In Sect. \ref{prelim} we recall some basic features of Markovian degree-biased random walks.
The parameter $\alpha$, which governs the bias of the steps serves us as control parameter to induce preferential or anti-preferential 
visits of a given set of t-nodes. Sect. \ref{general_THCP} is devoted to the analysis 
of the hitting process of an arbitrary set of t-nodes by an immortal Markov walker.
By involving specific `defective' transition matrices, which contain the information of the target, 
characterizing the first hitting statistics of the target, we derive the discrete first hitting PDF 
(probabilities of first passage to the target) and related pertinent distributions. 
In Sect. \ref{hitting_process_part} we focus on the `{\it target hitting counting process}' (THCP) for an arbitrary set of t-nodes.

With this framework, we analyze in
Sect. \ref{starved_random_walks} aforementioned mortal walker.
We refer the associated motion on the network to as `{\it mortal random walk}' (MRW).
In a MRW, the walker navigates during its lifetime with Markovian steps through a network,
steadily reducing its budget, which is reset when hopping on a t-node.
In this picture, the walker induces resets to the motion of the budget.
The MRW is therefore different from random walks under resetting in networks, where a walker is reset to specific or random nodes.
Examples for the latter include cases (among many others) 
in which the resets depend on the instantaneous position node of the walker. 
We refer to reference \cite{YeChen2022b} for such a model.

In Sect \ref{gen_mod} we give a general definition and outline of some basic features of the MRW
and also highlight the connection with stochastic resetting.
In Sect. \ref{arbitrary_target_MRW} we connect the evolution of the budget with the target hitting process.
We provide analytical solutions for the evanescent propagator matrix containing the probabilities that the walker reaches a specific node alive at a given time. 
From this result we derive the `mean residential time' the walker spends on a specific set of nodes during its life, the survival probability and the expected lifetime of the walker. In addition, by introducing 
a non-evanescent auxiliary propagator matrix, we deduce the expected number of visits 
of the target in a lifetime of the walker and related distributions such as the infinite time limit of the state probability distribution.
In the limit of immortality, the results of the immortal Markov walker of Sect. \ref{general_THCP} are recovered.
In Sect. \ref{generic_example} we perform a numerical study 
of the mentioned quantities, where we investigate scenarios in which frequent target hits (budget renewals) have different effects on the survival of the walker.
Our analytical results are corroborated by random walk simulations on a Barabási-Albert graph and we provide complementary derivations and technical details in the appendices.

\section{Background}
\label{prelim}

\subsection{Markovian degree-biased random walks}
\label{degree-based}

We consider a connected ergodic network of $N$ nodes, which we label with $i=1,\ldots, N$.
The propagator matrix of an immortal Markov walker, that is the transition matrix for $t$ steps, fulfills the master equation 
\beq
\label{master_eq}
P_{ij}(t+1) =  \sum_{k=1}^N P_{ik}(t)W_{kj} , \hspace{1cm}  P_{ij}(0) =\delta_{ij}, \hspace{1cm} t \in \mathbb{N}_0,
\eeq
where $P_{ij}(t)$ indicates the probability that the walker occupies node $j$ at time $t$, starting from departure node $i$ at time $0$ and $\pmb{W} = [W_{ij}]$ stands for the transition matrix for a single step specified hereafter,
which is by construction row-stochastic ($\sum_{j=1}^N W_{ij} =1$). We employ the notation $\pmb{1} = [\delta_{ij}]$ for the $N \times N$ identity matrix and $\delta_{ij}$ indicates the Kronecker symbol. For unbiased walks and the degree-biased walks considered in the present paper, the propagator can be diagonalized and has the canonical form \cite{NohRieger2004,AlbertBarabasi2002,Newman2010,VanMiegem2011,Mi_Ria_frac_walk2017,fractionalbook-2019} (see Appendix \ref{degree_biased} for details) 
\beq
\label{purely_markovian}
\mathbf{P}(t) = \pmb{W}^t = |\phi_1\rangle \langle {\overline \phi}_1| + \sum_{m=2}^N  \lambda_m^t | \phi_m\rangle \langle {\overline \phi}_m| , \hspace{1cm} \mathbf{P}(0) = \pmb{1},
\eeq
where we employ throughout the paper Dirac's bra-ket notation. $\mathbf{P}(t)$ in equation (\ref{purely_markovian}) is a Markov chain inheriting row-stochasticity 
from $\pmb{W}$. The part $|\phi_1\rangle \langle {\overline \phi}_1| = \lim_{t\to \infty} \pmb{W}^t = \pmb{W}^{(\infty)} =[W_j^{(\infty)}]$ is the stationary
transition matrix (large time limit of the Markov propagator).
For unbiased walks on undirected networks,
the steps from $i$ to $j$ are drawn from $\pmb{W} = [A_{ij}/K_i]$ where
$\pmb{A}=[A_{ij}]$ stands for the symmetric adjacency matrix $A_{ij}=A_{ji}$ with
$A_{ij}=1$ if nodes $i$ and $j$ are connected with an edge i.e. are neighbor nodes, and $A_{ij}=0$ else. Further, we have $A_{ii}=0$ to avoid self-loops. $K_i=\sum_{r=1}^NA_{ir}$ is the degree of node $i$ (number of its neighbor nodes).
The transition matrix $\pmb{W}$ is ergodic and has uniquely real eigenvalues 
with $|\lambda_m|< 1$ ($m=2,\ldots, N$) and one largest unique Perron–Frobenius eigenvalue $\lambda_1=1$ reflecting row-normalization of $\pmb{W}$. This implies that at least one return path of odd length exists to avoid that the graph is bipartite \cite{NohRieger2004,Lovasz1993}.
The quantities $|\phi_k\rangle$, $\langle {\overline \phi}_l|$ indicate the right- and left  
eigenvectors of $\pmb{W}$, respectively with  $\langle {\overline \phi}_r|\phi_s\rangle =\delta_{rs}$ and $\pmb{1} = 
\sum_{r=1}^N |\phi_r\rangle \langle {\overline \phi}_r|$. Note that $\pmb{W}$ is not symmetric unless the degrees are constant $K_i=K$ for all nodes. 
In this paper, we will consider degree-biased walks, since the bias offers us an elegant tool to
encourage or discourage the walker to visit certain (target) nodes. Among the biased walks, we consider degree-biased walks since they have spectral features which are as nice as in unbiased walks (see Appendix \ref{degree_biased}).
In our degree biased walks the steps are drawn from a transition matrix of the general form
\beq
\label{preferential_steps}
W_{ij} = \frac{A_{ij} f_i f_j}{\sum_{r=1}^N A_{ir} f_i f_r} =\frac{{\cal A}_{ij}}{{\cal A}_i}  , \hspace{0.5cm} {\cal A}_i = \sum_{r=1}^N {\cal A}_{ir}, \hspace{1cm} f_r=f(K_r;\alpha) .
\eeq
Here we have introduced the symmetric matrix $\pmb{{\cal A}} =[ {\cal A}_{ij}]$ with the elements ${\cal A}_{ij} = [A_{ij}f_if_j]$, in which $\pmb{A} =[A_{ij}]$ refers to the above introduced symmetric adjacency matrix. 
The matrix $\pmb{{\cal A}}$ can be seen as the weighted adjacency matrix of a specific class of weighted networks \cite{Riascos_Mateos2021}. The definition in equation 
(\ref{preferential_steps}) contains
the strictly positive function $f_j=f(K_j;\alpha)$, which is for $\alpha >0$ ($\alpha<0$)
monotonously increasing (decreasing) with the degree and fulfils $f(K;0) =1$ recovering the unbiased standard walk $W_{ij}= [A_{ij}/K_i]$, which is also retrieved for (regular) networks with constant degrees $K_j=K$ ($j=1,\ldots,N$). We further assume that the function $f$ is such that 
$f(K_j;\infty) =\infty$ and $f(K_j;-\infty) =0$.
Although the steps drawn from (\ref{preferential_steps}) are biased, the network is undirected. 
We focus in the present paper on degree-biased steps following a power-law $f_j =K_j^{\alpha}$, a class which was extensively studied in the literature
\cite{Wang-etal2006,Calva-Riascos2022,Fronczak2009,Yang2005}. 
To amplify the effects of degree bias, we consider networks with broad degree distributions, specifically, Barabási–Albert (BA) graphs exhibiting scale-free structure and power-law degree distribution \cite{AlbertBarabasi2002,Newman2010}.

Let us briefly introduce the notations we use throughout our paper. We indicate the sum of entries in row $i$ of an $N \times N$ matrix $\pmb{M}$ by the equivalent notations
\beq
\label{notation}
M_i = [\pmb{M}]_i =: \sum_{j=1}^NM_{ij} = \langle i|\pmb{M} |\phi_1\rangle  ,\hspace{1cm} \pmb{M}  = [M_{ij}] ,
\eeq
where $|\phi_1 \rangle$ stands for the column vector which has constant components equal to one $\langle i|\phi_1 \rangle = 1$.
We make extensive use of generating functions. Let $g(t)$ be a function supported on $t \in \mathbb{N}_0$. 
We define its generating function (GF) by
\beq
\label{GF_def}
{\overline g}(u) = \sum_{t=0}^{\infty} g(t) u^t
\eeq
for suitably chosen $u$ within the radius of convergence.
One establishes the connection with discrete Laplace transforms by $u=e^{-s}$.
We employ the symbol ``$\star$" to indicate discrete convolutions of two functions $g_i(t)$ supported on $\mathbb{N}_0$, namely
\beq
\label{convols}
(g_1 \star g_2)(t) =: \sum_{r_1=0}^{\infty}\sum_{r_2=0}^{\infty} g_1(r_1)g_2(r_2) \delta_{t,r_1+r_2} = \sum_{r=0}^t g_1(r)g_2(t-r) , \hspace{1cm} t = \{0,1,2,\ldots\} \in \mathbb{N}_0 .
\eeq
We use the synonymous notations $\delta_{ab} =\delta_{a,b}$ for the Kronecker symbol.
The convolution (\ref{convols}) has the GF ${\overline g}_1(u){\overline g}_2(u)$ 
and we use notation $g^{(n)}(t) = (g \star g^{(n-1)})(t)$ for the $n$th convolution power of $g$. We indicate with 
$\big\langle \ldots \big\rangle$ averaging with respect to all contained random variables and we use notation $\mathbb{P}[..]$ for the probability of an event $[..]$.
\section{Target hitting process in Markov walks}
\label{general_THCP}

In this section, we elaborate some technical details regarding the {\it target hitting counting process} (THCP) of an arbitrary stationary target, which is hit by an immortal Markov walker. Although there are many works considering random walks in networks, to the best of our knowledge, some aspects of the THCP have not been addressed in the literature so far.
We will use the framework of this section, in order to investigate the dynamics of our mortal walker in Sect. \ref{starved_random_walks}. 

We assume that the transition matrix $\pmb{W}$ of the walker's steps is primitive, that is $\pmb{W}^t$ realizes an ergodic, irreducible and aperiodic Markov chain. 
Recall that a non-negative $N \times N$ matrix $\pmb{M}$ is referred to as {\it primitive} (synonymously ``irreducible aperiodic" or ``ergodic"), if there is a positive integer $k_0$ such that 
all entries $[\pmb{M}^k]_{ij} >0$ for $k \geq k_0$ \cite{Newman2010,VanMiegem2011,fractionalbook-2019,Birkhoff1931,Lefevre2007}. 

\subsection{First passage statistics of an arbitrary set of target nodes}
\label{first_passage_stat}

Let ${\cal B}$ be an arbitrary and stationary set of t-nodes, and introduce the $N \times N$ diagonal matrix
\beq
\label{char_mat}
\pmb{\Theta}({\cal B}) = [\Theta(j;{\cal B}) \delta_{ij}] 
\eeq
containing ``$1$'' at diagonal positions labelling t-nodes and ``$0$'' elsewhere. $\Theta(r;{\cal B})$ stands for the target indicator function telling us whether a node $r$ is in ${\cal B}$ or not, defined as
\beq
\label{t_indicator}
\Theta(r;{\cal B}) = \sum_{b \in {\cal B}} \delta_{rb} =\left\{\begin{array}{clr} 1 ,&  r\in {\cal B} & \\[2ex] 
                                    0 ,&  r\notin {\cal B} .&
                                   \end{array}\right. 
\eeq
Moreover, we will employ a further diagonal matrix
\beq
\label{diag_xis}
\pmb{J}(\xi;{\cal B}) = [ \delta_{ij} \xi^{\Theta(j;{\cal B})} ] , \hspace{1cm} \xi \in [0,1] ,
\eeq
which has the entries $J_{jj}= \xi$ at diagonal positions labelling t-nodes $j \in {\cal B}$ 
and $J_{jj}= 1$ else. Then we define the generating matrix
\beq
\label{Omega}
\pmb{\Omega}(\xi;{\cal B}) = \pmb{W} \cdot \pmb{J}(\xi;{\cal B}) = 
{\widetilde {\pmb W}} + \xi \pmb{W}^{({\cal B})} =
[W_{ij} \xi^{\Theta(j;{\cal B})} ]   , \hspace{1cm}    \Omega_{ij}(\xi;{\cal B}) = \left\{\begin{array}{clr} \xi \, W_{ij} ,&  j\in {\cal B} & \\[2ex] 
                                    W_{ij} ,&  j\notin {\cal B} &
                                   \end{array}\right. 
\eeq
with $\pmb{\Omega}(1;{\cal B}) = \pmb{W}$. The matrices
 $\pmb{\Omega}(0;{\cal B}) = {\widetilde {\pmb W}}$ and $ \pmb{W}^{({\cal B})}$ are `defective' transition matrices containing the complete stochastic first passage information of t-node set ${\cal B}$ \cite{Mi_Dono_Po_Ria_Chaos2025}. They can be represented as\footnote{We skip in this notation the obvious dependence on ${\cal B}$ in ${\widetilde {\pmb W}}$.}
\beq
\label{transition_mat_mod} 
 {\widetilde {\pmb W}} =   \pmb{W} \cdot [\pmb{1} - \pmb{\Theta}({\cal B})]  = [W_{ij}(1-\Theta(j;{\cal B}))] ; \hspace{0.5cm} 
 \pmb{W}^{({\cal B})} = \pmb{W} \cdot \pmb{\Theta}({\cal B}) = [W_{ij} \Theta(j;{\cal B})] .
\eeq
${\widetilde {\pmb W}}$ has the entries ${\widetilde W}_{ij} =0$ for $j \in {\cal B}$, i.e. the columns labelling
t-nodes are set to zero, and ${\widetilde W}_{ij} = W_{ij}$ otherwise; $\pmb{W}^{({\cal B})}$ has the entries
$W^{({\cal B})}_{ij}=W_{ij}$ if $j \in {\cal B}$ and $W^{({\cal B})}_{ij}=0$ otherwise.
The matrix element $[{\widetilde {\pmb W}}^t]_{ij}$ indicates the probability that the walker has reached node $j$ at time $t$ without having hit a t-node (not considering as a hit if the departure node $i \in {\cal B}$). Note that integer powers $[{\widetilde {\pmb W}}^t]_{ij}=0$ for $j \in {\cal B}$ ($t\geq 1$). Therefore, contrarily to $\pmb{W}$, the defective matrix ${\widetilde {\pmb W}}$ is not primitive.
Let ${\cal T}_{i \to {\cal B}} =\{1,2,\ldots\} \in \mathbb{N}$ be the first hitting time (first passage time) of target ${\cal B}$ for departure node $i$.
The probability that the walker has not hit a t-node up to and including time instant $t$ then is 
\beq
\label{survival_proba}
\mathbb{P}[{\cal T}_{i \to {\cal B}} > t] = \Lambda_i(t) = [{\widetilde {\pmb W}}^t]_i = \sum_{j=1}^N [{\widetilde {\pmb W}}^t]_{ij} , \hspace{1cm} t \in \mathbb{N}_0
\eeq
with $\Lambda_i(0)=1$ for all $i=1,\ldots,N$, since we define ${\widetilde {\pmb W}}^0 = \pmb{1}$ yielding the N$\times N$ unity matrix, reflecting that per construction departures from $i \in {\cal B}$ are not counted as t-node hit.
By referring ${\widetilde {\pmb W}}$ to as {\it defective}, we mean that this matrix is not properly row-normalized $[{\widetilde {\pmb W}}]_i \leq 1$. Especially $[{\widetilde {\pmb W}}]_i < 1$ if node $i$ is a neighbor node of a t-node. Therefore, $\sum_{i=1}^N \Lambda_i(t) < N$
for any non-empty t-node set ${\cal B}$. We point out that ${\widetilde {\pmb W}}$ has spectral radius 
$\rho({\widetilde {\pmb W}}) < 1$ for any non-empty target ${\cal B}$. This feature is reflected by 
\beq
\label{zero_limit}
\lim_{t\to \infty} {\widetilde {\pmb W}}^t = \pmb{0} ,
\eeq
and consequently $\Lambda_i(\infty)=0$ ($i=1,\ldots, N$).
This asymptotic relation tells us that the probability of never hitting ${\cal B}$ in an infinitely long Markov walk is null, being equivalent to recurrence and ergodicity of the Markovian walk, see \cite{Mi_Dono_Po_Ria_Chaos2025} or Appendix \ref{rec_mar_cha}).
We show a little later that our approach involving above introduced defective transition matrices contains the classical `first passage' approach of Noh and Rieger \cite{NohRieger2004} when considering a single t-node. 
It is instructive to construct the discrete `first hitting PDF' 
(first passage probability) of an arbitrary t-node set ${\cal B}$, i.e. the probability that the walker hits 
a t-node (no matter which one) for the first time at time instant $t \in \mathbb{N}$. 
The first hitting PDF of target ${\cal B}$ (the probability to hit ${\cal B}$ at time $t$ for the first time) can be retrieved from the `{\it first hitting matrix}' (FHM) given by
\beq
\label{first_hitting_mat}
\pmb{\chi}(t) = {\widetilde {\pmb W}}^{t-1} \cdot \pmb{W}^{({\cal B})} =  
\left[ ({\widetilde {\pmb W}}^{t-1} \cdot \pmb{W})_{ij}\Theta(j;{\cal B}) \right] , \hspace{1cm} t=\{1,2,\ldots\} \in \mathbb{N} ,
\eeq
which has non-zero entries only in columns labelling t-nodes. We define $\pmb{\chi}(0)  =\pmb{0}$ to avoid t-node hits at $t=0$.
The first hitting PDF of target ${\cal B}$ for departure node $i$ is given by
\beq
\label{first_hitting_PDF}
\chi_{i \to {\cal B}}(t)= \Lambda_i(t-1) - \Lambda_i(t) = 
[ {\widetilde {\pmb W}}^{t-1} \cdot \pmb{W}^{({\cal B})} ]_i = \sum_{k=1}^N[ {\widetilde {\pmb W}}^{t-1}]_{ik} 
\sum_{b \in {\cal B}} W_{kb}  , \hspace{1cm} t \in \mathbb{N}
\eeq
with $\chi_{i \to {\cal B}}(0) =0$. 
This expression has an evocative interpretation:
$[{\widetilde {\pmb W}}^{t-1}]_{ik}$ is the probability that the walker reaches the node $k$ in $t-1$ steps without having
visited ${\cal B}$, and $\sum_{b \in {\cal B}} W_{kb} = [\pmb{W}^{({\cal B})}]_k = \chi_{k \to {\cal B}}(1)$ is the probability that one of the t-nodes is hit in one step.
Instructive is to consider the GF of (\ref{survival_proba}) given by
\beq
\label{survival_GF}
{\overline \Lambda}_i(u) =\sum_{t=0}^{\infty}u^t \Lambda_i(t)  = [(\pmb {1}-u{\widetilde {\pmb W}})^{-1}]_i , \hspace{1cm} |u| \leq 1
\eeq
together with the GF of the first hitting PDF
\beq
\label{first_hitting_GF}
{\overline \chi}_{i\to {\cal B}}(u) = \sum_{t=1}^{\infty} u^t  [ {\widetilde {\pmb W}}^{t-1} \cdot \pmb{W}^{({\cal B})} ]_i
= u [(\pmb {1}-u{\widetilde {\pmb W}})^{-1}\cdot \pmb{W}^{({\cal B})}]_i
\eeq
to observe that
\beq
\label{basic}
\frac{1- {\overline \chi}_{i\to {\cal B}}(u)}{1-u} =  \frac{1}{1-u} [ (\pmb {1}-u{\widetilde {\pmb W}})^{-1} \cdot \left(\pmb {1}-u{\widetilde {\pmb W}} - u \pmb{W}^{({\cal B})}\right)]_i = [ (\pmb {1}-u{\widetilde {\pmb W}})^{-1}]_i = {\overline \Lambda}_i(u) ,
\eeq
which recovers (\ref{survival_GF}) and confirming $\Lambda_i(t) = 1-\sum_{r=1}^t \chi_{i\to {\cal B}}(r)$. 
Recall that we denote the first hitting time (FHT) or first passage time of ${\cal B}$ for departure node $i$ with ${\cal T}_{i \to {\cal B}} =\{1,2,\ldots\} \in \mathbb{N}$.
We can then retrieve the mean first passage time (MFPT) of ${\cal B}$ from 
\beq
\label{MFPT_to_B}
\big\langle {\cal T}_{i \to {\cal B}} \big\rangle = \sum_{t=1}^{\infty} t \chi_{i \to {\cal B}}(t) =  {\overline \Lambda}_i(1) = [(\pmb {1}-{\widetilde {\pmb W}})^{-1}]_i ,
\eeq
generalizing the well-known classical result \cite{NohRieger2004} to an arbitrary t-node set ${\cal B}$.
Since $\rho({\widetilde {\pmb W}}) < 1$ the matrix $\pmb {1}-{\widetilde {\pmb W}}$ is invertible, and therefore the MFPT 
and all moments of ${\cal T}_{i \to {\cal B}}$ are finite for any departure node and t-node set. Consider for a moment the special case that the target comprises the complete network; thus, ${\widetilde {\pmb W}}=\pmb{0}$, and hence $\big\langle {\cal T}_{i \to {\cal B}} \big\rangle =1$, i.e. the walker is hitting the target at the first step.
On the other hand, for ${\cal B} = \emptyset$ we have ${\widetilde {\pmb W}}= \pmb{W}$ (where $\pmb{1}-\pmb{W}$ is not invertible) thus the MFPT is infinite.

\subsection{Target hitting counting process of an arbitrary target}
\label{hitting_process_part}

Here we investigate the THCP of an arbitrary t-node set ${\cal B}$.
Let ${\cal N}_i(t; path, {\cal B}) \in \mathbb{N}_0$ be the non-decreasing THCP counting variable, which is increased by one at each hit of a t-node (no matter which one). We define ${\cal N}_i(0; path, {\cal B})=0$ ($i=1,\ldots,N$) 
even if the departure node $i$ is a t-node. This quantity counts the number of t-nodes contained in a specific sample path of $t$ steps ($path = i \to j_1\to \ldots j_{t-1} \to j_t$), not counting the departure node if it is a t-node. The THCP counting variable can be represented as
\beq
\label{arrivals_on_B}
{\cal N}(i \to j_1 \to j_{t-1} \to j_t; {\cal B}) =:  {\cal N}_{i}(t; path, {\cal B}) = \sum_{r=1}^t \Theta(j_r;{\cal B}) 
\eeq
in which $\Theta(r;{\cal B})$ is the target indicator function given in (\ref{t_indicator}), and $0 \leq {\cal N}_{ij}(t; path, {\cal B}) \leq t$. 
For a specific sample path ${\cal N}_{i}(t; path, {\cal B})$ is a deterministic quantity, 
however, appears as a random variable in a random walk. From now on, we use the simpler notation,
${\cal N}_{i}(t; path, {\cal B}) = {\cal N}_i(t; {\cal B})$ omitting the dependence of the path for brevity.
To capture the statistics of the THCP, we define a moment GF
by averaging the random function $
\xi^{{\cal N}_i(t; {\cal B})}$ over all paths starting from $i$ of $t$ steps length, namely
\beq
\label{moment_gen}
\begin{array}{clr} 
\ds \Big\langle \xi^{{\cal N}_i(t; {\cal B})} \Big\rangle & = \ds
\sum_{j_1=1}^N \sum_{j_2=1}^N \ldots 
\sum_{j_{t-1}=1}^N \sum_{j_t=1}^N  W_{i,j_1}W_{j_1,j_2} \ldots W_{j_{t-1},j_t} 
\xi^{\Theta(j_1;{\cal B})}\xi^{\Theta(j_2;{\cal B})}\ldots \xi^{\Theta(j_{t-1};{\cal B})}\xi^{\Theta(j_t;{\cal B})}  & \\ \\
 & = \ds \sum_{j_1=1}^N \sum_{j_2=1}^N \ldots 
\sum_{j_{t-1}=1}^N  \sum_{j_t=1}^N  \Omega_{i,j_1}\Omega_{j_1,j_2} \ldots \Omega_{j_{t-1},j_t} 
&\\ \\
&  = \ds  \sum_{j=1}^N [(\pmb{\Omega}(\xi; {\cal B}))^t]_{ij} =   [(\pmb{\Omega}(\xi; {\cal B}))^t]_{i}  = \ds \sum_{n=0}^t Q^{(n)}_{i}(t) \xi^n , \hspace{1cm} t \in \mathbb{N}_0 . &
\end{array} 
\eeq
This GF involves the $t$th matrix power of generating matrix (\ref{Omega}) and
is a polynomial of degree $t$ since ${\cal N}_{i}(t; {\cal B}) \in \{0,1,\ldots, t\}$.
Observe that $(\pmb{\Omega}(\xi; {\cal B}))^t = \sum_{n=0}^t \xi^n \mathbf{Q}^{(n)}(t)$ recovering for $\xi=1$ the Markov chain. 
We refer to the matrices $\mathbf{Q}^{(n)}(t)= [ Q^{(n)}_{ij}(t) ]$ to as `state probability matrices'. They are given by
\beq
\label{state_hitting_ptoba}
Q^{(n)}_{ij}(t) = 
\frac{1}{n!}\frac{d^n}{d\xi^n}\left[\pmb{\Omega}(\xi;{\cal B})]^t\right]_{ij}\bigg|_{\xi=0}  , \hspace{0.5cm} t, n \in \mathbb{N}_0,
\hspace{1cm} Q^{(n)}_{ij}(0)  = \delta_{n0} \delta_{ij} ,
\eeq
and for $n=0$ one recovers $Q^{(0)}_{ij}(t) = [{\widetilde {\pmb W}}^t]_{ij}$.
The state probability matrices are even
for unbiased walks non-symmetric (except for $t=0$).
The entries $Q^{(n)}_{ij}(t)$ indicate the probabilities that the walker reaches the node $j$ at time instant $t$ by choosing paths that contain
$n$ t-nodes. They lead to the state probabilities
\beq
\label{stqte_relation}
\mathbb{P}[{\cal N}_i(t; {\cal B}) =n] = Q^{(n)}_i(t) = [\pmb{Q}^{(n)}(t)]_i = \sum_{j=1}^N Q^{(n)}_{ij}(t) , \hspace{1cm} n, t \in \mathbb{N}_0
\eeq
with $Q^{(0)}_i(t) = \Lambda_i(t) = [{\widetilde {\pmb W}}^t]_{i}$ of (\ref{survival_proba}).
The state probabilities are the probabilities that the walker has hit target ${\cal B}$ $n$ times in a walk of $t$ steps
starting from departure node $i$. They are properly normalized 
\beq
\label{norm_state}
\sum_{n=0}^t Q^{(n)}_i(t) = [\pmb{W}^t]_i=1 .
\eeq
We will show hereafter that the expected number of t-node hits in a walk of $t$ steps starting on $i$ is given by
\beq
\label{total_hits}
 \big\langle {\cal N}_i(t; {\cal B}) \big\rangle = 
 \frac{d}{d\xi} [(\pmb{\Omega}(\xi; {\cal B}))^t]_i\Big|_{\xi=1} = \sum_{n=0}^t n Q_i^{(n)}(t) = \sum_{r=1}^t [\pmb{W}^{r-1}\cdot \pmb{W}^{({\cal B})}]_i .
\eeq
This quantity is obtained by evaluating ($' = \frac{d}{d\xi} $)
\beq
\label{mean_hits_eval}
\begin{array}{clr}
\Big\langle \pmb{{\cal N}}(t; {\cal B})\Big\rangle & =\ds  \frac{d}{d\xi}
\left(\widetilde{\pmb{W}}+\xi \pmb{W}^{({\cal B})}\right)^t \Big|_{\xi=1} & \\ \\ 
& =\left\{\left(\widetilde{\pmb{W}}+\xi \pmb{W}^{({\cal B})}\right)'\cdot \left(\widetilde{\pmb{W}}+\xi \pmb{W}^{({\cal B})}\right)^{t-1}+\ldots
\left(\widetilde{\pmb{W}}+\xi \pmb{W}^{({\cal B})}\right) \cdot \left(\widetilde{\pmb{W}}+\xi \pmb{W}^{({\cal B})}\right)'\cdot 
\left(\widetilde{\pmb{W}}+\xi \pmb{W}^{({\cal B})}\right)^{t-2}  \right.& \\ \\ & \left. + \ldots + \left(\widetilde{\pmb{W}}+\xi \pmb{W}^{({\cal B})}\right)^{t-1}\cdot \left(\widetilde{\pmb{W}}+\xi \pmb{W}^{({\cal B})}\right)' \right\}\Big|_{\xi=1}& \\ \\
& =\ds \sum_{r=1}^t [\pmb{W}^{r-1}\cdot \pmb{W}^{({\cal B})}] \cdot \pmb{W}^{t-r} . &
\end{array}
\eeq   
We call the matrix in the brackets `{\it mean hitting rate matrix}' (MHR matrix) of target ${\cal B}$ 
\beq
\label{brackets}
\pmb{H}(t;{\cal B}) =  \pmb{W}^{t-1}\cdot \pmb{W}^{({\cal B})} = \pmb{W}^t \cdot \pmb{\Theta}({\cal B}) , \hspace{1cm} t =\{1,2,\ldots \} \in \mathbb{N} .
\eeq
It makes sense to define $\pmb{H}(0;{\cal B}) = \pmb{0}$ as per construction, no target hit may happen at $t=0$. The MHR matrix must not be confused with the `first hitting matrix' - (FHM) defined in (\ref{first_hitting_mat}).
Relation (\ref{total_hits}) is summing up the `{\it mean hitting rate}' (MHR) of ${\cal B}$ 
\beq
\label{hitting_rate_B}
H_{i \to {\cal B}}(t) = \Big\langle  {\cal N}_i(t;{\cal B})\Big\rangle - \Big\langle  {\cal N}_i(t-1;{\cal B})\Big\rangle =
[\pmb{H}(t;{\cal B})]_i = [\pmb{W}^{t-1}\cdot \pmb{W}^{({\cal B})}]_i
\hspace{1cm} t = \{1,2,\ldots \} \in \mathbb{N}
\eeq
with $H_{i \to {\cal B}}(0) =0$ and the obvious feature $H_{i \to {\cal B}}(t) \in [0,1]$. 
For standard Markov walks the inequality holds (see (\ref{mean_hits_eval}))
\beq
\label{inequality_holds}
\frac{d}{d\xi}[(\pmb{\Omega}(\xi; {\cal B}))^t]_i \neq t [(\pmb{\Omega}(\xi; {\cal B}))^{t-1} \cdot \pmb{W}^{{\cal B}}]_i
\eeq
reflecting the non-commuting feature $[{\widetilde {\pmb W}}\cdot \pmb{W}^{({\cal B})}]_i \neq [ \pmb{W}^{({\cal B})} \cdot {\widetilde {\pmb W}} ]_i$ introducing a memory on when a t-node hitting event happens.
The THCP ${\cal N}_i(t;{\cal B})$ is non-Markovian, apart of the distinguished
class of stationary Markov chains (see Appendix \ref{stat_Markov_chains} where (\ref{inequality_holds}) turns into an equality). 
Instructive is to consider the large time asymptotics of the MHR matrix (\ref{brackets})
\beq
\label{MHR_large_time}
\pmb{H}(\infty;{\cal B}) = \pmb{W}^{(\infty)} \cdot \pmb{\Theta}({\cal B})
\eeq
where $ \pmb{W}^{(\infty)} =|\phi_1\rangle \langle \overline{\phi}_1|=[W^{(\infty)}_j] $ is the stationary transition matrix. The large time limit of the MHR of target ${\cal B}$ is therefore
\beq
\label{large_time_hitting_rate_B}
H_{i\to {\cal B}}(\infty)= [\pmb{H}(\infty;{\cal B})]_i = \sum_{j \in {\cal B}} W^{(\infty)}_j = p_{\cal B}
\eeq
having lost the memory of the departure node.
The THCP converges for large time to a (Markovian) Bernoulli counting process, in which the target ${\cal B}$ is hit at each time step with probability $p_{\cal B}$, thus 
\beq
\label{large_time_THCP_asymptotics}
\Big\langle  {\cal N}_i(t;{\cal B})\Big\rangle  \to   p_{\cal B} t , \hspace{1cm} (t \to \infty), 
\eeq
and consult Appendix \ref{stat_Markov_chains} for complementary details.

It is useful to connect MHR (\ref{hitting_rate_B}) and the THCP with first hitting distributions.
To this end, we will show hereafter by considering GFs that the following matrix identities ((\ref{useful_hitting_rel}), (\ref{thefollowing_important})) are holding true:
\beq
\label{useful_hitting_rel}
\begin{array}{clr}
\ds [\pmb{\Omega}(\xi;{\cal B})]^t & = \ds {\widetilde {\pmb W}}^t + \sum_{r=1}^t \pmb{H}(r,\xi; {\cal B})\cdot  
{\widetilde {\pmb W}}^{t-r} , \hspace{1cm} t \in \mathbb{N}_0 . &
\end{array}
\eeq
For $t=0$ the sum $\sum_{r=1}^t(..)$ has to be read as null (and so in all convolution relations hereafter).
This expression contains
\beq
\label{H_matrix}
\begin{array}{clr}
\ds \pmb{H}(t,\xi;{\cal B}) & = \ds  \sum_{n=1}^{\infty} \xi^n\pmb{\chi}^{(n)}(t) & \\ \\ 
& = \ds  [\pmb{\Omega}(\xi;{\cal B})]^{t-1}\cdot \pmb{W}^{({\cal B})} \xi &
\end{array} \hspace{0.5cm} t\in \mathbb{N}, \hspace{0.5cm}  \pmb{H}(0,\xi;{\cal B}) = \pmb{0} ,
\eeq
retrieving the MHR matrix $\pmb{H}(t,1;{\cal B})= \pmb{H}(t;{\cal B})$ (\ref{brackets}). 
%
%
The equivalence of the two equalities of (\ref{H_matrix}) can be seen
by taking the GF of the second one, yielding
\beq
\label{important_GF}
\begin{array}{clr}
\overline{\pmb{H}}(u,\xi;{\cal B}) & = 
\left[\pmb{1} - u\pmb{\Omega}\right]^{-1} \cdot \pmb{W}^{({\cal B})} u \xi = \left\{ [\pmb{1} - u {\widetilde {\pmb W}} ] \cdot \left[1- \overline{\pmb{\chi}}(u)\xi \right] \right\}^{-1}\cdot \pmb{W}^{({\cal B})} u \xi & \\[1ex]
& =
\left[1- \overline{\pmb{\chi}}(u) \xi \right]^{-1} \cdot \pmb{\chi}(u) \xi &
\end{array}
\eeq
in which we used the GF $\overline{\pmb{\chi}}(u) = (\pmb{1} - u{\widetilde {\pmb W}})^{-1} \cdot \pmb{W}^{({\cal B})} u $ of the FHM (\ref{first_hitting_mat}).
The second line in (\ref{important_GF}) indeed is the GF of the first equality of (\ref{H_matrix}) and 
in which $\overline{\pmb{H}}(u, 1; {\cal B})= \overline{\pmb{H}}(u; {\cal B})$ is the GF of the 
MHR matrix (\ref{brackets}).
%
%
%
%

The second identity reads
\beq
\label{thefollowing_important}
\begin{array}{clr}
\ds [\pmb{\Omega}(\xi;{\cal B})]^t &  = \ds {\widetilde {\pmb W}}^t + \xi \sum_{r=1}^t \pmb{\chi}(r) 
\cdot [\pmb{\Omega}(\xi;{\cal B})]^{t-r} ,\hspace{1cm} t \in \mathbb{N}_0  &
\end{array}
\eeq
and is solved by (\ref{useful_hitting_rel}).
Keep in mind that in all these relations, the sequence of matrix multiplications matters.
Setting $\xi=1$ retrieves the corresponding equations for the Markov propagator.
(\ref{H_matrix}) contains the n$th$ (matrix) convolution 
powers of the FHM (\ref{first_hitting_mat}) $\pmb{\chi}^{(n)}(t)  = (\pmb{\chi}^{(n-1)} \star \pmb{\chi})(t) $ (fulfilling the initial condition $\pmb{\chi}^{(n)}(t) = \pmb{1} \delta_{t,0} \delta_{n,0}$ with 
$\pmb{\chi}^{(0)}(t)=\pmb{1} \delta_{t,0}$). 
Relation (\ref{thefollowing_important}) has a `matrix type' renewal structure involving the FHM. We will show in the next section that
in the case of a single target node and departures from this node, this equation indeed boils down to a proper renewal equation (Eq. (\ref{renewal_propagator_i_eq_b})).
We will show hereafter that the state probability matrices (\ref{state_hitting_ptoba}) are given by
\beq
\label{state_proba_matrix}
\pmb{Q}^{(n)}(t) = \sum_{r=0}^t \pmb{\chi}^{(n)}(r) \cdot {\widetilde {\pmb W}}^{t-r} , \hspace{1cm} n, t \in \mathbb{N}_0 ,
\eeq
with $\pmb{Q}^{(0)}(t) =  {\widetilde {\pmb W}}^t$. To this end,
let us prove validity of (\ref{useful_hitting_rel}) - (\ref{thefollowing_important}) by considering GFs and using the following identity
\beq
\label{useful_identity}
\begin{array}{clr}
\ds  \left(\pmb{1}-u \pmb{\Omega}(\xi;{\cal B})\right)^{-1} & = \ds \left[\pmb{1}-\xi \overline{\pmb{\chi}}(u)\right]^{-1} \cdot \left[\pmb{1}-\overline{\pmb{\chi}}(u)\right]  \cdot \left[\pmb{1}-u\pmb{W}\right]^{-1} & \\[2ex]
 & =  \ds \left[\pmb{1}-\xi \overline{\pmb{\chi}}(u)\right]^{-1} \cdot   [\pmb{1} - u \widetilde{\pmb{W}}]^{-1} .     & 
 \end{array}
\eeq
Expanding the second line with respect to $\xi$ yields the GFs of the state probability matrices 
as
\beq
\label{extraction_state_proba_mats}
\overline{\pmb{Q}}^{(n)}(u)  =
 [\overline{\pmb{\chi}}(u)]^n \cdot  [\pmb{1} - u \widetilde{\pmb{W}}]^{-1} , \hspace{1cm} n \in \mathbb{N}_0 
\eeq
being consistent with (\ref{state_proba_matrix}). On the other hand, plugging $\left[\pmb{1}-\xi \overline{\pmb{\chi}}(u)\right]^{-1} =
\pmb{1} + \xi \overline{\pmb{\chi}}(u) \cdot \left[\pmb{1}-\xi \overline{\pmb{\chi}}(u)\right]^{-1}$ into the second line of (\ref{useful_identity}) 
and inverting the resulting expression yields (\ref{useful_hitting_rel}).
In addition, we prove Eq. (\ref{thefollowing_important}) by considering its GF
$$\left(\pmb{1}-u\pmb{\Omega}(\xi;{\cal B})\right)^{-1} =   [\pmb{1} - u \widetilde{\pmb{W}}]^{-1} + \xi  \overline{\pmb{\chi}}(u) \cdot \left(\pmb{1}-u\pmb{\Omega}(\xi;{\cal B})\right)^{-1} ,$$ where some simple manipulations take us to the second line of (\ref{useful_identity}).
Finally, representation (\ref{useful_identity}) also allows us to retrieve the GF of the matrices (\ref{mean_hits_eval}) yielding
\beq
\label{mean_counting}
\begin{array}{clr}
\ds \big\langle \overline{\pmb{\mathcal{N}}}(u;{\cal B}) \big\rangle & = 
\ds   \big\langle \overline{\mathcal{N}}_{ij}(u;{\cal B}) \big\rangle   = \frac{d}{d\xi} \left[\pmb{1}-u\pmb{\Omega}(\xi;{\cal B})\right]^{-1}\Big|_{\xi=1} & \\[2ex]
&= \ds \frac{d}{d\xi}  [\pmb{1} - \xi \overline{\pmb{\chi}}(u)]^{-1} \Big|_{\xi=1}\cdot 
[\pmb{1} -u {\widetilde {\pmb W}}]^{-1} & \\[2ex]
 & = \ds [\pmb{1} -  \overline{\pmb{\chi}}(u)]^{-1} \cdot  \overline{\pmb{\chi}}(u) \cdot 
  [1-u\pmb{W}]^{-1} & \\[2ex]
 & = \ds  \overline{\pmb{H}}(u; {\cal B}) \cdot [1-u\pmb{W}]^{-1} .       & 
\end{array}
\eeq
The third line can be rewritten as 
\beq
\label{aux_rel_THCP_GF}
\big\langle \overline{\pmb{\mathcal{N}}}(u;{\cal B}) \big\rangle  = 
\overline{\pmb{\chi}}(u) \cdot [1-u\pmb{W}]^{-1} +  \overline{\pmb{\chi}}(u) \cdot \big\langle \overline{\pmb{\mathcal{N}}}(u;{\cal B}) \big\rangle .
\eeq
Inverting this relation leads to a matrix type renewal equation for the THCP counting variable
\beq
\label{further_mean}
\Big\langle  {\cal N}_i(t;{\cal B})\Big\rangle = \sum_{r=1}^t \chi_{i\to{\cal B}}(t) + 
\sum_{r=1}^t \left[\pmb{\chi}(r)  \cdot \big\langle \pmb{\mathcal{N}}(t-r;{\cal B}) \big\rangle\right]_i .
\eeq
Moreover, inverting (\ref{mean_counting}) to time brings us back to relation (\ref{mean_hits_eval}).
In addition, from (\ref{mean_counting}) one has
\beq
\label{row-xcount_sum}
\big\langle \overline{\mathcal{N}}_i(u;{\cal B}) \big\rangle =
\frac{[\overline{\pmb{H}}(u; {\cal B})]_i}{1-u} 
\eeq
being consistent with (\ref{total_hits}). The general framework of this section, which holds for an immortal Markov walker and arbitrary t-node sets, has a wide range of potential applications in random walk dynamics and will be crucial for the analysis of our mortal random walker later on.
\hfill
\begin{figure}[t!]
\centerline{
\includegraphics[width=1.0\textwidth]{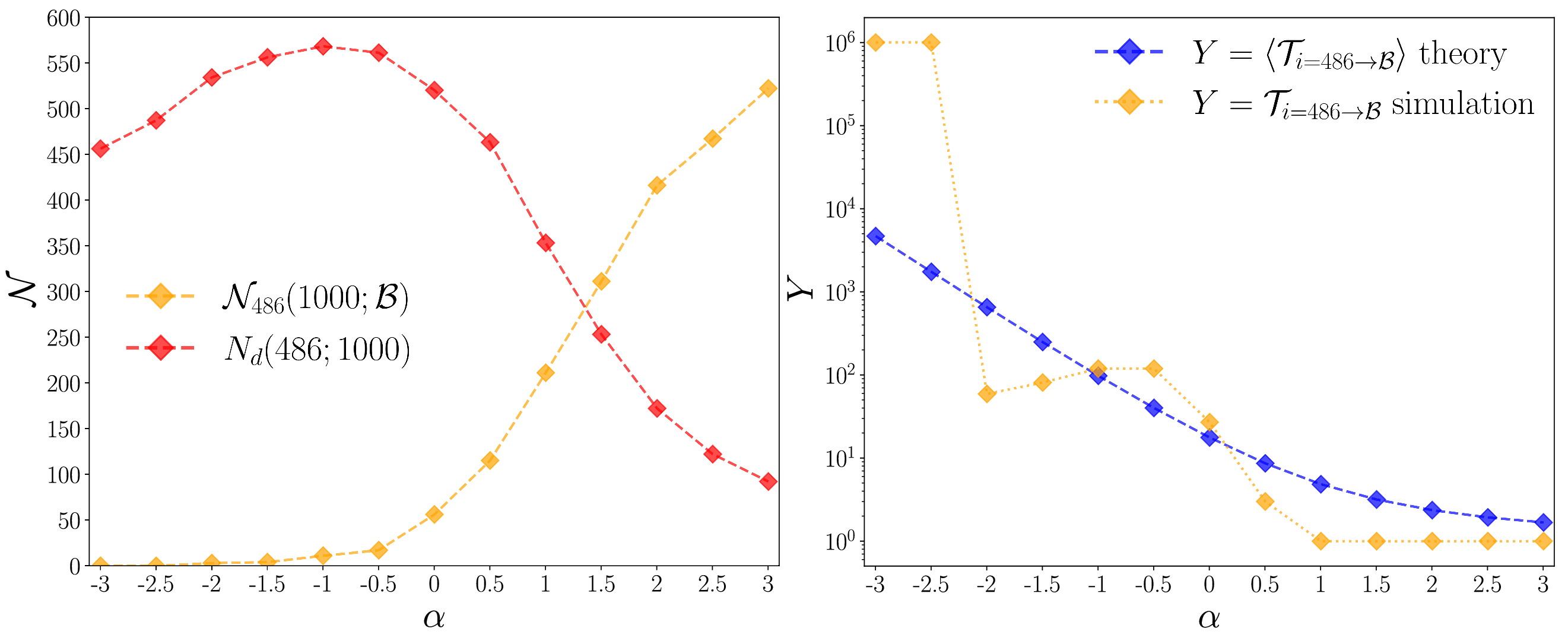}
}
\caption{
Numerical results for the target hitting counting process. We consider the same BA network and setting as in Fig. \ref{fig2}. (Left frame) Numbers of distinct nodes visited $N_d(i=486;t=1000)$ and number of target hits ${\cal N}_{i=486}(t=1000;{\cal B})$, respectively,  in a power-law degree-biased walk [Eq.~(\ref{preferential_steps})], plotted as functions of the tuning parameter \( \alpha \), as obtained from random walk simulations.  (Right frame) MFPT to the target set  ${\cal B}$  as a function of $\alpha$, from the analytical expression in Eq. (\ref{MFPT_to_B}) (blue curve), compared with first hitting times (FHTs) obtained using Monte Carlo simulations (orange curve). For $\alpha <-2$, the target is not reached within the simulation time  (FHTs $\gg$ runtime).
}
\label{fig3}
\end{figure}
\\[2mm]
To give a visual impression of what happens in a THCP, we explore in Figs. \ref{fig3} and \ref{fig2} the THCP and the MFPT in power-law degree-biased walks (\ref{preferential_steps}) ($f_j= K_j^{\alpha}$) for runtime $1000$.
We focus on the effect of the bias and the connectivity of a target ${\cal B}$.
All plots of Figs.\ref{fig3} and \ref{fig2} refer to the same Barabási-Albert (BA) graph of $1000$ nodes (generated with Python NetworkX library $nx.barabasi\_albert\_graph(1000, m=7, seed=0)$\,) considering departure node $i=486$ and a
target ${\cal B}$ consisting of $10$ highly connected nodes (average degree of the t-nodes $\langle K\rangle_{\cal B} = 82.4$). 
The t-nodes are marked in cyan color in the left frames of Fig. \ref{fig2}.
To generate such a highly connected target, the t-nodes are selected among all 
nodes with probabilities proportional to $K^{\beta}$ ($\beta=30$).
The target also contains the node ${\hat j}=8$ with the largest degree $K_8=141$.
The higher the degree of a node, the more central its location is in the graphical representation (see Appendix \ref{degree_biased} for complementary details).

The MFPT $\big\langle {\cal T}_{486 \to {\cal B}} \big\rangle$ of Eq. (\ref{MFPT_to_B}) is depicted in Fig.
\ref{fig3} (blue curve, right frame) and compared with the first hitting time (FHT) to ${\cal B}$ obtained from random walk simulations (orange curve).
We observe that the MFPT decreases and the number of target hits $ {\cal N}_{486}(t;{\cal B})$ increases monotonously with $\alpha$ which reflects the effect of bias. For large $\alpha$ the walker travels preferentially on highly connected nodes which include the t-nodes. The larger $\alpha$ the faster the walker reaches one of the highly connected t-nodes. Consequently the MFPT is decreasing and $ {\cal N}_{486}(1000;{\cal B})$ is increasing with $\alpha$. For very large $\alpha$ the walker navigates almost exclusively on t-nodes, with small MFPT close to one and large hitting numbers.
For negative $\alpha$ the walker travels preferentially on nodes with small degrees (peripherical nodes), with large MFPT and small hitting numbers 
of ${\cal B}$.
For $\alpha < -2$ the target is not any more hit within the runtime $1000$. This corresponds to extremely large values of the MFPT 
(largely exceeding the runtime), and $ {\cal N}_{486}(1000;{\cal B}) \to 0 $.
\begin{figure}[t!]
\centerline{
\includegraphics[width=1.0\textwidth]{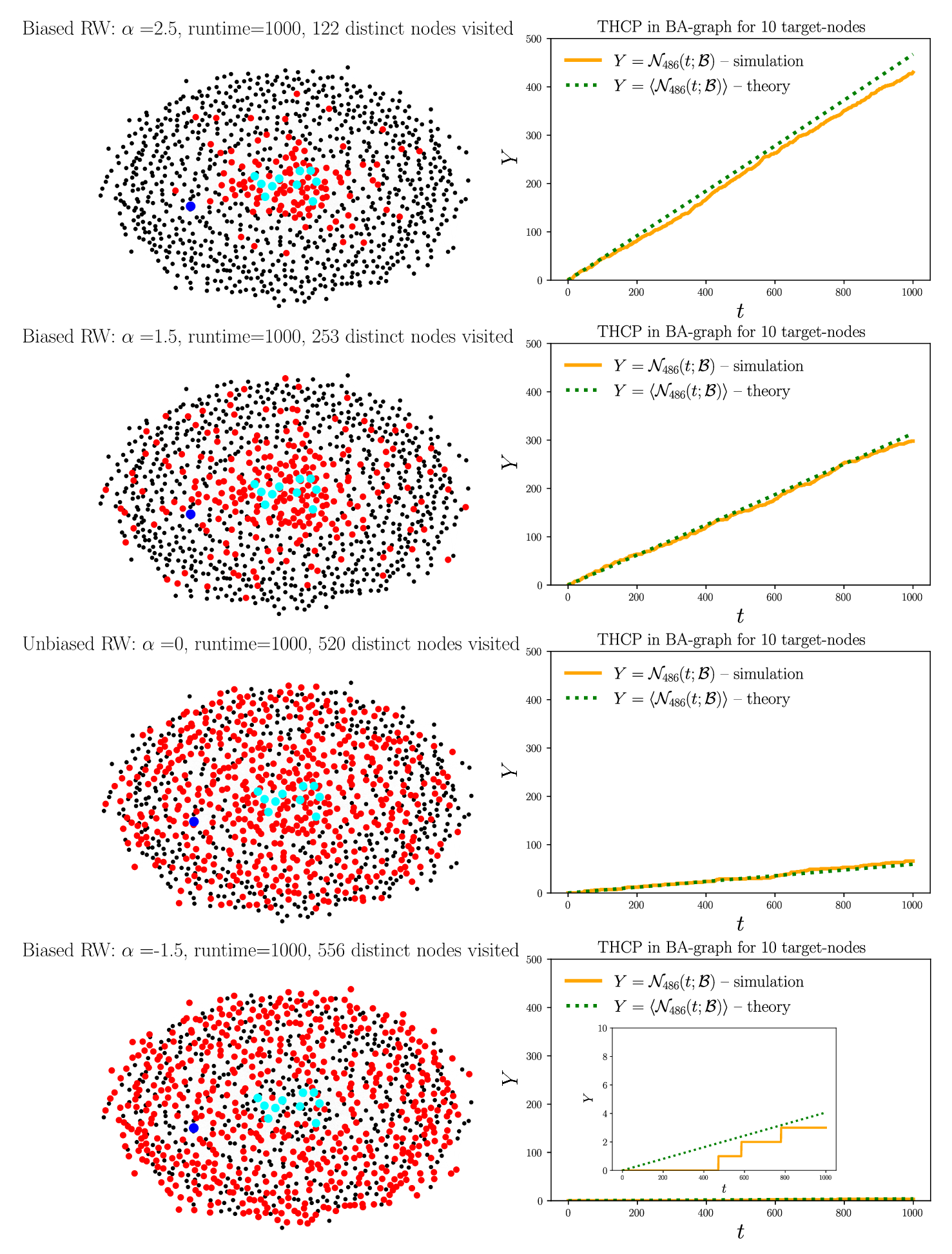}}
\vspace{-5mm}
\caption{THCP of $10$ t-nodes for the biased walk (\ref{preferential_steps}) in a BA network 
(Python NetworkX library $G=nx.barabasi\_albert\_graph(1000, m=7, seed=0)$] of $1000$ nodes. Left frames: Highly connected target ${\cal B}$ in cyan color. Departure node ($i=486$) in blue, distinct nodes visited during the runtime $1000$ are drawn in red. Right frames: 
Orange curves: ${\cal N}_{i=486}(t;{\cal B})$ (\ref{arrivals_on_B}) recorded in random walk simulations; Green curves: expected hitting number $\langle {\cal N}_{i=486}(t;{\cal B})\rangle $  of Eq. (\ref{total_hits}).
}
\label{fig2}
\end{figure}

Fig. \ref{fig2} depicts $\langle  {\cal N}_{486}(t;{\cal B}) \rangle$ of Eq. (\ref{total_hits}) 
(green dotted curves), and hitting numbers ${\cal N}_{486}(t;{\cal B})$ recorded in random walk simulations (orange curves).
Both quantities behave similarly, exhibiting an approximate linear increase. 
This property corresponds to fast relaxation to the stationary
distribution, where the THCP converges rapidly to a Bernoulli counting process 
with linear large time asymptotics of $\langle  {\cal N}_i(t;{\cal B}) \rangle$ (independent of the departure node), see
(\ref{large_time_THCP_asymptotics}).
We prove in Appendix \ref{stat_Markov_chains} for stationary Markov chains
that the slope of  $\langle  {\cal N}_i(t;{\cal B}) \rangle$
is given by the inverse of the mean recurrence time to ${\cal B}$, generalizing the Kac theorem to arbitrary t-node sets ${\cal B}$.
This explains also why the slope of $\langle  {\cal N}_i(t;{\cal B}) \rangle$ increases with $\alpha$, as the mean recurrence time
to the highly connected target ${\cal B}$ then decreases. 

The red curve of Fig. \ref{fig3} depicts the number of distinct nodes visited within the runtime $1000$ recorded in random walk simulations, considering the same BA network as in Fig. \ref{fig2}.
The distinct nodes visited within the runtime are marked in red in the left frames of Figs. \ref{fig2}.

The $\alpha$-dependence of the number of the distinct nodes visited (red curve of Fig. \ref{fig3}) reflects the interplay of the degree-bias and the topology of the considered BA network.
We observe that the number of distinct nodes visited 
(which we denote with $N_d(i;t)$ and consider departure node $i=486$ and observation time $t=1000$) initially increases with $\alpha$, reaching a maximum at an intermediate value not far from zero, and then decreases as $\alpha$ becomes larger. To explain this behavior, we notice first that in  BA networks the degree is power-law distributed, there are many (peripheral) nodes with low degrees and a few (central) nodes (`hub nodes') with high degrees.
For $\alpha< 0$ far from zero, the walker navigates preferentially on the numerous nodes with small degrees. As $\alpha$ approaches zero, the preference is gradually removed. The walker then moves without preference over the entire network, performing a nearby unbiased walk that increases the number of distinct nodes visited.
Finally, increasing $\alpha$ further toward positive values, introduces newly a
preferential node selection toward the fewer number of nodes with small degrees, explaining the fall-off to very small values of the number of distinct nodes visited. We corroborate this behavior a little later by considering the
mean of the number of distinct nodes visited for a given observation time (see Fig. \ref{newfig}).
%
%
%
%

\subsection{Hitting process of a single target node}
\label{renewal}
A particular instructive case with connections to classical well-known results emerges when ${\cal B}$ comprises a single t-node, say $b$. 
Recall the defective transition matrices (\ref{transition_mat_mod}) taking here the form
\beq
\label{defectives}
\widetilde{\pmb{W}}=[W_{ij}(1-\delta_{jb})] , \hspace{1cm} \pmb{W}^{({\cal B})} = \pmb{W}^{(b)} = [W_{ij} \delta_{jb}] ,
\eeq
thus FHM (\ref{first_hitting_mat}) reads 
\beq
\label{FHT_PDF_to_b}
 \pmb{\chi}(t) = \widetilde{\pmb{W}}^{t-1} \cdot \pmb{W}^{(b)} = [\chi_{ib}(t)\delta_{jb}],  
 \hspace{1cm} \chi_{ib}(t) = [\widetilde{\pmb{W}}^{t-1} \cdot \pmb{W}]_{ib}  , \hspace{1cm} t \in \mathbb{N} ,
\eeq
with $\chi_{ij}(0) =0$. The FHM has only non-zero entries in the column $b$ containing the first hitting PDFs (first passage probabilities to reach $b$)
$[\pmb{\chi}(t)]_i =  \chi_{ib}(t)$ with $\Lambda_i(t) = 1-\sum_{r=1}^t \chi_{ib}(t)$ (see (\ref{first_hitting_PDF})).
%
%

\begin{figure}[t!]
\centerline{
\includegraphics[width=0.7\textwidth]{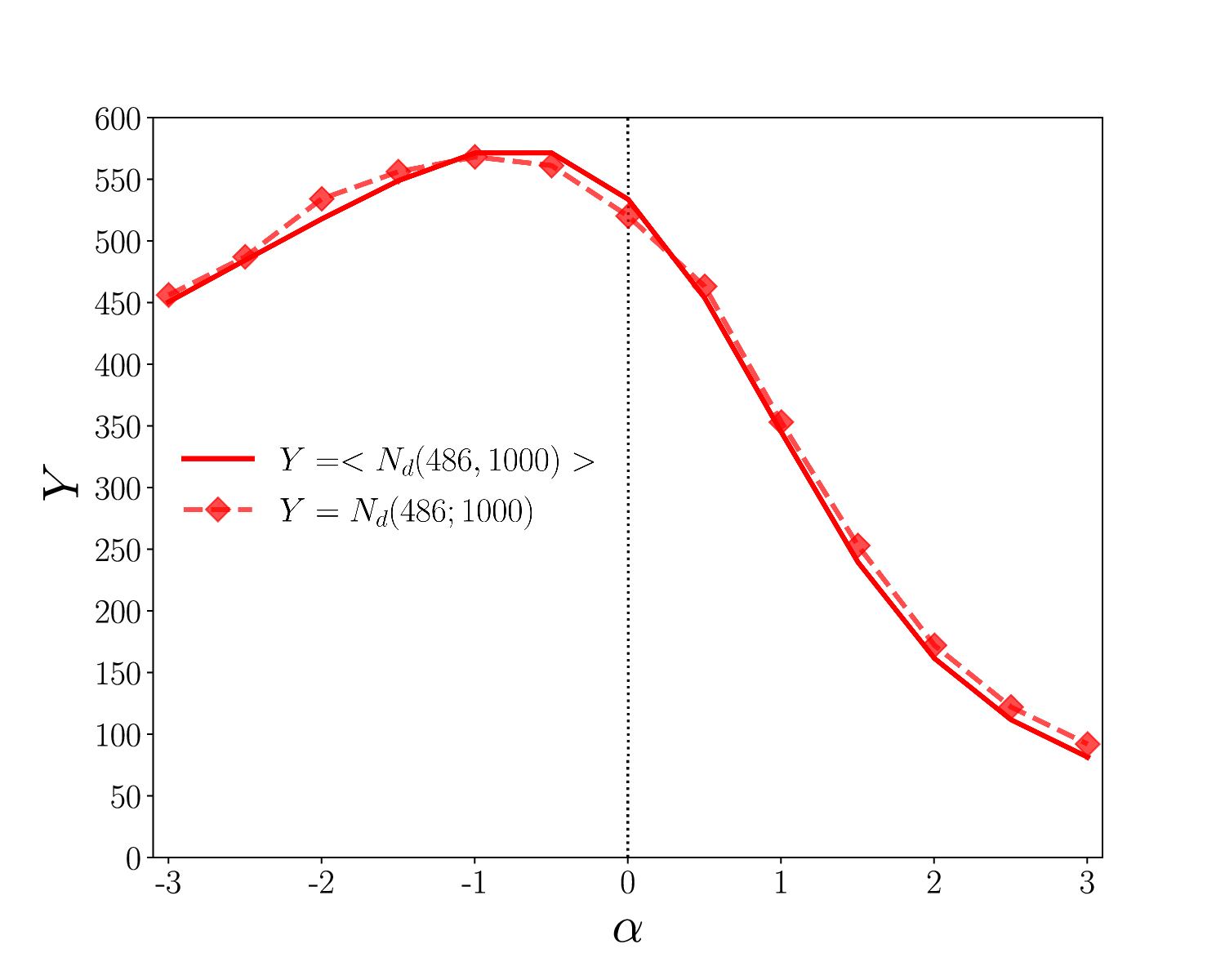}
}
%
%
\caption{
We depict the average number of distinct nodes visited 
$\big\langle N_d(486;1000) \big\rangle$ from Eq. (\ref{dis_nodes_vis}), and the recorded values of 
$N_d(486;1000)$ from Fig. \ref{fig3} versus $\alpha$. For this setting and observation time, the exploration of the network is maximal at values slightly shifted to negative values from $\alpha=0$ (unbiased walk).}
\label{newfig}
\end{figure}

We point out that these ingredients enable us to investigate the statistics of the
number of distinct nodes visited $N_d(i;t)$ up to and including observation time $t$ for a given departure node $i$. In particular, its mean is obtained as \cite{Biroli-etal2022}
\beq
\label{dis_nodes_vis}
\big\langle N_d(i;t) \big\rangle  = \sum_{b=1}^N \sum_{r=1}^t\chi_{ib}(r) = N - \sum_{b=1}^N \Lambda^{(b)}_i(t) .
\eeq
We use here the notation $\Lambda^{(b)}_i(t) =[\widetilde{\pmb{W}}]_i$ highlighting the t-node dependence of the involved defective transition matrix (\ref{defectives}). 
Considering that the Markov walk is ergodic, we have 
$\sum_{r=1}^{\infty}\chi_{ib}(r) =1$ (see Appendix \ref{rec_mar_cha}),
thus in the infinite time limit $\big\langle N_d(i;t) \big\rangle \to N $ i.e., is approaching the total number $N$ of nodes of the network. All nodes of the network are eventually visited by the (immortal) Markov walker. This asymptotic feature remains valid for degree-biased walks, independent of the bias parameter $\alpha$.
In Fig. \ref{newfig} we plot the average number of distinct nodes visited $\big\langle N_d(i=486;t=1000) \big\rangle$ from Eq. (\ref{dis_nodes_vis}) for the same setting as in Figs. \ref{fig3} and \ref{fig2}, and the record of $N_d(i=486;t=1000)$ from Fig. \ref{fig3}
as a function of $\alpha$. 
Both curves agree impressively, corroborating the analysis.
%
%
%
%

Subsequently, we will use the GFs 
\beq
\label{GF_IB}
{\overline \chi}_{ib}(u) = \sum_{t=1}^{\infty}  u^t\chi_{ib}(t) = u [(\pmb{1}-u {\widetilde {\pmb W}} )^{-1}\cdot \pmb{W}]_{ib} , \hspace{1cm} |u| \leq 1
\eeq
and
\beq
\label{proba_not_GF}
{\overline \Lambda}_i(u) = \sum_{t=0}^{\infty} u^t \Lambda_i(t) =[(\pmb{1}-u {\widetilde {\pmb W}} )^{-1}]_i .
\hspace{1cm} |u| \leq 1 .
\eeq
We explore now the connection of our framework with the classical first passage approach 
introduced by Noh and Rieger \cite{NohRieger2004}.
Their approach holds for Markovian walks and considers a single arrival node, which we identify with the t-node.
They established, invoking conditional arguments and using the Markov property of the walk the following relation 
\beq
\label{NohRieger_relation}
P_{ij}(t) = \delta_{ij}\delta_{t,0} +\sum_{r=0}^t F_{ij}(r)P_{jj}(t-r) ,
\eeq
where departure node is $i$ and $\pmb{P}(t)= \pmb{W}^t$ is the Markov propagator, $F_{ij}(t)$ 
the first passage probability (probability to reach node $j$ at time $t$ for the first time [with $F_{ij}(0) =0$]). 
We identify the arrival node with the t-node $j=b$, and the first passage probabilities
with the first hitting PDFs $F_{ib}(t) = \chi_{ib}(t)$ of (\ref{FHT_PDF_to_b}). Taking GF of (\ref{NohRieger_relation}) takes us to
\beq
\label{first_passage_GF}
{\overline P}_{ib}(u) = \delta_{ib} + {\overline \chi}_{ib}(u){\overline P}_{bb}(u) , \hspace{1cm} |u|< 1 .
\eeq
Plugging in the GF (\ref{GF_IB}) and the GF of the Markov propagator yields 
\beq
\label{first_passage_GF_b}
\left[(1-u\pmb{W})^{-1}\right]_{ib} = \delta_{ib} +  u \left[ (\pmb{1}-u\widetilde{\pmb{W}})^{-1}\cdot \pmb{W} \right]_{ib} \left[(\pmb{1}-u\pmb{W})^{-1}\right]_{bb} .
\eeq
Multiplying this relation with the matrix $(\pmb{1}-u\widetilde{\pmb{W}})$ from the left 
(where $[\pmb{1} -u\widetilde{\pmb{W}}]_{ib} = \delta_{ib}$ as ${\widetilde W}_{ib}=0$) leads to
\beq
\label{first_passage_GF_c}
 [\left(\pmb{1}-u\widetilde{\pmb{W}}\right) \cdot \left(\pmb{1}-u\pmb{W}\right)^{-1}]_{ib}  = \delta_{ib} +u W_{ib} \left[(1-u\pmb{W})^{-1}\right]_{bb} ,
 \eeq
where correctness of this relation is easily confirmed by rewriting the LHS as
\beq
 \label{first_passage_GF_d}
[\left(\pmb{1}-u\widetilde{\pmb{W}}\right) \cdot \left(\pmb{1}-u\pmb{W}\right)^{-1}]_{ib}   = [\left(\pmb{1}-u \pmb{W} + u \pmb{W}^{(b)} \right) \cdot \left(\pmb{1}-u\pmb{W}\right)^{-1}]_{ib} ,
\eeq
proving that the first hitting PDF (\ref{FHT_PDF_to_b}) indeed solves Eq. (\ref{NohRieger_relation}).
%
%
%
%

It will be useful a little later to introduce the time of the $n$th hit of $b$ for departure node $i$ as
\beq
\label{indepndent_rand_interims}
\tau_n(i,b)=\Delta t_{ib}+\sum_{r=2}^n\Delta t^{(r)}_{bb}  , \hspace{0.5cm} n \in \mathbb{N} ,  \hspace{0.5cm} \tau_0(i,b) = 0  .
\eeq
We define $\tau_0(i,b) =0$ for all $i$ even if $i=b$, meaning that a departure from $b$ is not considered as a hitting event.
$\Delta t^{(r)}_{bb}$ are IID copies of the recurrence time 
$\Delta t_{bb}$ and $\Delta t_{ib}$ is the FHT of $b$ for departure node $i$.
One observes that the probability that the $n$th hit of $b$ happens at time instant $t \in \mathbb{N}$ 
is contained in the sequence of FHM convolutions 
\beq
\label{chi_n}
\begin{array}{clr}
\ds  \pmb{\chi}^{(n)}(t) & = \ds (\pmb{\chi} \star \pmb{\chi}^{(n-1)})(t) = \sum_{r=1}^t \pmb{\chi}(r) \cdot \pmb{\chi}^{(n-1)}(t-r)    , \hspace{1cm} n, t = \{1,2,\ldots\} \in \mathbb{N} & \\ \\
\ds   \pmb{\chi}^{(n)}(t)   & = \ds [\chi^{(n)}_{ib}(t) \delta_{bj}] = [\delta_{jb} \left(\chi_{ib} \star \chi^{(n-1)}_{bb}\right)(t) ], &
 \end{array}
\eeq
in which $\pmb{\chi}(t) = \pmb{\chi}^{(1)}(t)$ is the FHM (\ref{FHT_PDF_to_b}). 
This matrix convolution power boils down to a scalar one, involving only the first hitting PDF $\chi_{ib}(t)$ and the PDF of first return $\chi_{bb}(t)$, namely
\beq
\label{chi_n_b}
 \mathbb{P}[\tau_n(i,b) =t] = \chi^{(n)}_{ib}(t) =\big\langle \delta_{t,\tau_n(i,b)} \big\rangle = \sum_{r=0}^t \chi_{ib}(r) \cdot \chi^{(n-1)}_{bb}(t-r)    , \hspace{1cm} n, t \in \mathbb{N}_0 .
\eeq
Clearly, $\chi^{(n)}_{ib}(t) \neq \chi^{(n)}_{bi}(t)$ is non-symmetric, even for unbiased walks.
The state probability matrices (\ref{state_proba_matrix}) take here the particular simple form
\beq
\label{probabilitiespath_n_hits}
 Q_{ij}^{(n)}(t) = \sum_{k=0}^t  [\pmb{\chi}^{(n)}(r) \cdot 
 {\widetilde {\pmb W}}^{t-r}]_{ij} = \left\{\begin{array}{l} \ds [{\widetilde {\pmb W}}^{t}]_{ij}   , \hspace{1cm} n=0 \\ \\
\ds  \sum_{r=0}^t \chi^{(n)}_{ib}(r)[{\widetilde {\pmb W}}^{t-r}]_{bj} , \hspace{1cm}  n \in \mathbb{N} = \{1,2,\ldots\}
 \end{array}\right.
\eeq
with $Q_{ij}^{(0)}(t) = [{\widetilde {\pmb W}}^t]_{ij}$ as $[\pmb{\chi}^{(0)}(t)]_{ij} = \delta_{ij} \delta_{t,0}$. 
They lead us to the state probabilities 
\beq
\label{state_probas_N}
\mathbb{P}[{\cal N}_i(t;b)=n] = Q^{(n)}_i(t) = \left\{\begin{array}{l} \Lambda_i(t) , \hspace{1cm} n=0 \\ \\
\ds \sum_{k=0}^t \chi^{(n)}_{ib}(r)\Lambda_b(t-r) , \hspace{1cm} n \in \mathbb{N} = \{1,2,\ldots\} .
\end{array}\right.
\eeq
Moreover, from (\ref{thefollowing_important}) we have the relation
\beq
\label{Omega_evol}
[(\pmb{\Omega}(\xi;b))^t]_{ij} = [{\widetilde {\pmb W}}^t]_{ij} + \xi \sum_{r=1}^t \chi_{ib}(r) [(\pmb{\Omega}(\xi;b))^{t-r}]_{bj} .
\eeq
Noteworthy is the case when $j=b$ (arrival node coincides with the t-node) which leads to
\beq
\label{consistent_with_classic}
[\pmb{W}^t]_{ib} = \delta_{ib} \delta_{t,0} + \sum_{r=1}^t \chi_{ib}(r)[\pmb{W}^{t-r}]_{bb} ,
\eeq
where we have used for the first term that $[{\widetilde {\pmb W}}^t]_{ib} =\delta_{ib} \delta_{t,0}$. 
We identify this relation indeed with the classical equation (\ref{NohRieger_relation}) of Noh and Rieger.
Considering $i=b$ in (\ref{Omega_evol}) gives the renewal equation
\beq
\label{renewal_propagator_i_eq_b}
[\pmb{W}^t]_{bj} = [{\widetilde {\pmb W}}^t]_{bj} +\sum_{r=1}^t \chi_{bb}(r) [\pmb{W}^{t-r}]_{bj} ,
\eeq
highlighting that the hitting process of $b$ is a renewal process when the walk starts from this node. 
This renewal process has the IID inter-arrival time $\Delta t_{bb}$, which is the recurrence time to node $b$.
From (\ref{Omega_evol}) one can also retrieve
\beq
\label{special_case}
1- \sum_{r=1}^t \chi_{ib}(r) = \Lambda_i(t)
\eeq
being consistent with (\ref{first_hitting_PDF}).
In addition, from (\ref{useful_hitting_rel}) we get
\beq
\label{Markov_propagator}
[\pmb{W}^t]_{ij} = [{\widetilde {\pmb W}}^t]_{ij} +
\sum_{r=1}^t H_{ib}(r) [{\widetilde {\pmb W}}^{t-r}]_{bj} .
\eeq
Setting $j=b$ retrieves the MHR (see (\ref{hitting_rate_B}))
\beq
\label{MHR_single_t_node_b}
H_{ib}(t) = [\pmb{W}^{t}]_{ib}  ,\hspace{1cm} t \in \mathbb{N}
\eeq
and $H_{ib}(0) = 0$.
Finally, relation (\ref{Markov_propagator}) takes us to  
\beq
\label{survival_renewal}
1 = \Lambda_i(t) + \sum_{r=0}^t H_{ib}(t-r) \Lambda_b(r) .
\eeq
As a consequence of recurrence of the walk, one retrieves from (\ref{row-xcount_sum}) the infinite time asymptotics
\beq
\label{MHT_infinite_time}
\big\langle  {\cal N}_i(\infty;b) \big\rangle =\lim_{u\to 1} (1-u) \big\langle {\overline {\cal N}}_i(u;b) \big\rangle =
\lim_{u\to 1} \frac{{\overline \chi}_{ib}(u)}{1-{\overline \chi}_{bb}(u)} = \infty .
\eeq
Any t-node $b$ is infinitely often hit, as ${\overline \chi}_{bb}(1) = \sum_{t=1}^{\infty} \chi_{bb}(t)= 1$ (see Appendix \ref{rec_mar_cha}).
Interesting is the large-time limit of the MHR
\beq
\label{asymptotics_oh_H}
H_{ib}(\infty) = \lim_{u\to 1} (1-u) {\overline H}_{ib}(u) = \lim_{u\to 1}\frac{1-u}{1-{\overline \chi}_{bb}(u)} {\overline \chi}_{ib}(u) 
= \frac{1}{\overline{\Lambda}_b(1)} =
\frac{1}{\big\langle \Delta t_{bb}\big\rangle}  = W^{(\infty)}_b ,
\eeq
where $\Delta t_{bb}$ is the recurrence time to $b$. (\ref{asymptotics_oh_H}) is independent of the departure node and consistent with (\ref{large_time_hitting_rate_B}).
The stationary probability $W^{(\infty)}_b$ can be identified with the
inverse of the mean recurrence time of the node $b$ reflecting the Kac theorem.
As shown previously, the THCP converges in the large time limit to a Bernoulli counting process. Here the t-node $b$ is hit in each time increment with Bernoulli success probability (\ref{asymptotics_oh_H}) leading to the linear large time asymptotics 
$\big\langle {\cal N}_i(t;b) \big\rangle \to t W^{(\infty)}_b$ (Eq. (\ref{large_time_THCP_asymptotics})).

\section{Evanescent random walker}
\label{starved_random_walks}
\subsection{General model}
\label{gen_mod}
In the remainder of this paper, we consider a mortal random walker, which navigates with Markovian (degree-biased) steps through an ergodic network. We refer to this walk as `{\it mortal random walk}' (MRW). 
In order to survive and continue the walk, the walker needs to maintain a strictly positive budget ${\cal C}(t) \in \mathbb{N}$. The walker starts at $t=0$ with a random positive initial budget 
${\cal C}(0)=T_1 \in \mathbb{N}$. Each step has a cost reducing the budget
by one unit, except upon hopping on a t-node $b \in {\cal B}$ (no matter which one), where 
the budget is renewed with an IID copy of the initial value $T_1$. We identify the hitting times of the target with the renewal times of the budget. 
The steps of the walker in the network and the renewed budget values $T_n =\{1,2,\ldots \} \in \mathbb{N}$ are considered to be independent.
The walker dies (is dead) at the moment $t_{*} \in \mathbb{N}$ when the budget attains zero
(${\cal C}(t_{*}) = 0$).  
The time $t_{*}$ is the lifetime of the walker. Conceiving the budget ${\cal C}(t)$ as an auxiliary random walk on the integer line, the lifetime $t_{*}$ is the first hitting time of the origin in this auxiliary random walk.
In what follows we will employ the discrete Heaviside function defined by
\beq
\label{discrete-Heaviside}
\Theta(r) =  \left\{\begin{array}{cl} 1 ,\hspace{1cm} r \geq 0  \\[2mm]
    0, \hspace{1cm} r<0 \end{array}\right.    \hspace{1cm} r \in \mathbb{Z},  
\eeq
where $\Theta(0) =1$, together with the following indicator function
which is defined for $a,b,t \in \mathbb{N}_0$ and $a < b$ as
\begin{align}
\label{Theta-a-b}
\Theta[a,t,b] &=\Theta(t-a)-\Theta(t-b)
= \left\{\begin{array}{l} 1, \hspace{1cm} {\rm for} \hspace{0.5cm} a \leq t \leq b-1,  \\[2mm]
0, \hspace{1cm} {\rm otherwise}.\end{array}\right. 
\end{align}
Especially one has $\Theta[0,t,b] = \Theta(b-1-t)$. 
The time evolution of a sample budget is given by the following stochastic process
\beq
\label{budget}
{\cal C}(t) = \sum_{n=0}^{\infty} {\cal M}_n \Theta[\tau_n, t, \tau_{n+1}] \Theta\left(T_{n+1}-1-(t-\tau_n)\right) (T_{n+1}-(t-\tau_n)) , \hspace{1cm} \in \mathbb{N}_0 .
\eeq
In this relation $\tau_n = \Delta t_1+\ldots+\Delta t_n \in \mathbb{N}_0$ (with $\tau_0=0$) indicates the renewal time of the budget, where
${\cal C}(\tau_n)= T_{n+1}$, given the walker is alive $t=\tau_n$.
The walker is alive at time $t$ if ${\cal C}(t) \geq 1$ and dies at
$t=t_{*}$ when ${\cal C}(t_{*})=0$. 
We will identify later the budget renewal times with the
target hitting times of the Markov walker, however, at the moment it is sufficient to assume that $\tau_n$ is a strictly increasing integer 
function of $n \in \mathbb{N}_0$ with positive integer increments. The increments
$\Delta t_k \in \mathbb{N}$ indicate the time intervals between subsequent budget renewals. It has to be emphasized that the increments $\Delta t_k$ may be, but are in general, not IID. That is, the process of budget renewals is in general not necessarily a proper renewal process.
The indicator function $\Theta[\tau_n, t, \tau_{n+1}]$ in (\ref{budget}) is such that $\Theta[\tau_n, t, \tau_{n+1}]=1$ if $\tau_n \leq t \leq \tau_{n+1}-1$ and $\Theta[\tau_n, t, \tau_{n+1}]=0$ else, recall (\ref{Theta-a-b}). We point out that the $\Delta t_k$ and the renewed budget values $T_m$ are considered to be independent random variables.

\begin{figure}[t!]
\centerline{
\includegraphics[width=1.0\textwidth]{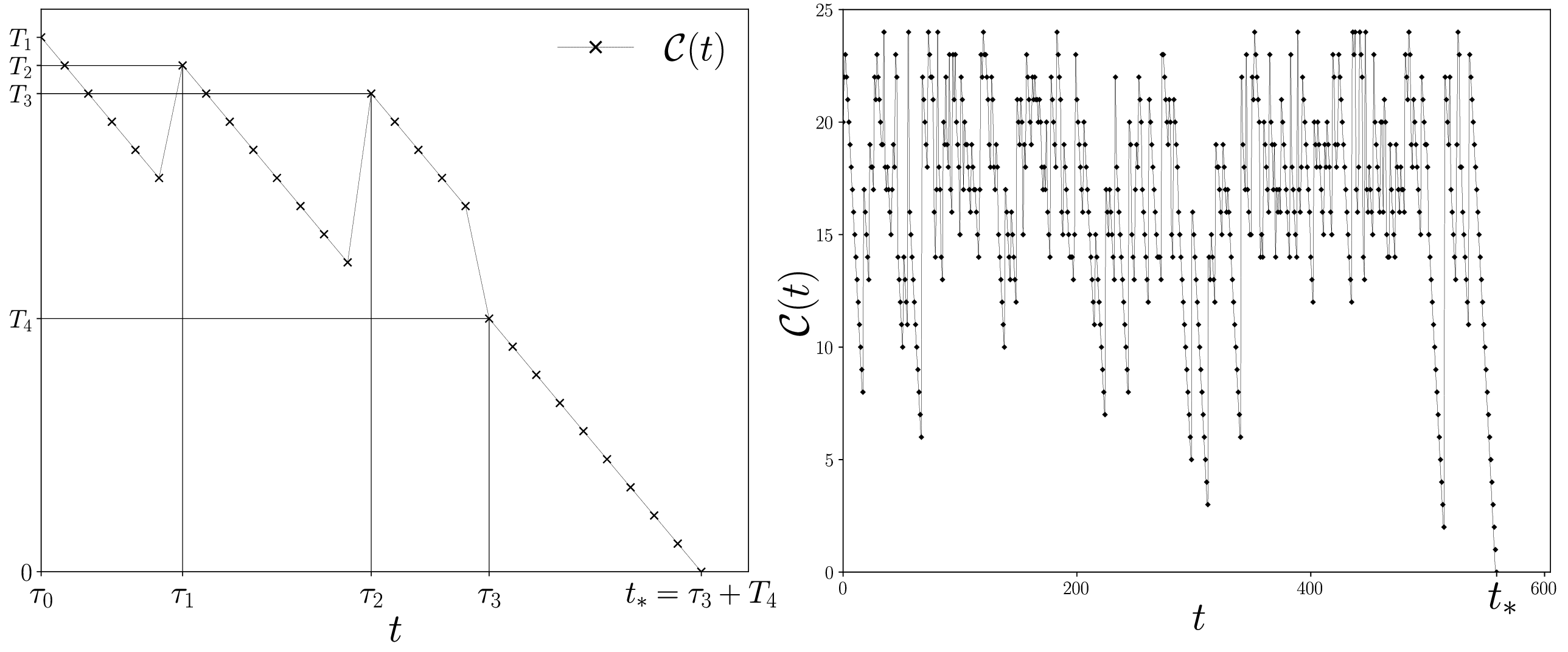}
}
\caption{ Left frame: Schematic: Sample budget evolution of (\ref{budget}) until walker's death at $t=t_{*}$ when ${\cal C}(t_{*})=0$.
Right frame: Budget evolution (\ref{budget}) until walker's death ($t_* = 559$) recorded in a random walk simulation for power-law biased steps (bias parameter $\alpha=1.6$) for the same BA network, target and 
departure node as in Figs. \ref{fig3} and \ref{fig2}.  In this simulation, the budget is renewed 
with uniformly distributed random integers ($T \in [14, 24]$). }
\label{budget_fig}
\end{figure}

%
%
In (\ref{budget}) we introduced the indicator function ${\cal M}_n$ containing a memory of the survival history of the walker up to and including the budget renewal time $\tau_n$, 
which fulfills ${\cal M}_n=1$ if the walker reaches the
$n$th budget renewal (alive), and ${\cal M}_n=0$ otherwise. In addition, we define ${\cal M}_0 = 1$.
It can be expressed by the `survival indicator functions'
$\Theta(T_k-1 -\Delta t_k)$ which equal one, if $T_k > \Delta t_k$, and zero else.
So ${\cal M}_n=1$ only if $T_k > \Delta t_k$ $\forall k=1,\ldots, n$, and therefore
\beq
\label{memory_budget_renewals}
{\cal M}_n =  \prod_{k=1}^n \Theta(T_k-1 -\Delta t_{k}) , \hspace{1cm} n \in \mathbb{N}.
\eeq
%
%
%
We define that the walker is alive at time instant $t$ if ${\cal C}(t) \geq 1$, which implies that 
\beq
\label{survival_cond}
T_{n+1} -(t-\tau_n)  \geq 1 , \hspace{1cm} \tau_n \leq t < \tau_{n+1} =\tau_n+\Delta t_{n+1} .
\eeq
An indicator function for this condition is $\Theta[\tau_n, t, \tau_{n+1}] \Theta(T_{n+1}-1 -(t-\tau_n) )$ (recall $\Theta(0)=1$), which equals one if (\ref{survival_cond}) is fulfilled, and null else. An indicator function 
telling us whether the walker is alive at time instant $t$ reads
\beq
\label{survival_indicator_function}
{\cal S}(t) = \sum_{n=0}^{\infty} {\cal M}_n \Theta[\tau_n, t, \tau_{n+1}]  \Theta\left(T_{n+1}-1-(t-\tau_n)\right) ,  \hspace{1cm} n \leq t \in \mathbb{N}_0
\eeq
with ${\cal S}(0)=1$.
%
%
The walker survives the time period from the $n$th to the $(n+1)$th budget renewal if 
\beq
\label{survival_condition}
{\cal C}(\tau_n) = T_{n+1} > \Delta t_{n+1}  , \hspace{1cm} n= \{0,1,2,\ldots \}  \in \mathbb{N}_0
\eeq
and dies if $T_{n+1}\leq \Delta t_{n+1}$ at instant $t_{*}=\tau_n+T_{n+1} \leq \tau_{n+1}$ (see left frame of Fig. \ref{budget_fig}). In the case where $T_{n+1}=\Delta t_{n+1}$, the budget is not any more renewed and the walker dies at 
time instant $\tau_{n+1}$. 
A schematic illustration of a sample budget (\ref{budget}) until the walker's death is given in Fig. \ref{budget_fig} (left frame).
The survival probability ${\cal P}_{\cal S}(t)$ of the walker (probability that the walker is alive at time $t$)
is obtained by averaging the survival indicator function (\ref{survival_indicator_function}) over all contained random variables
\beq
\label{survival_proba_MRW}
{\cal P}_{\cal S}(t) = \big\langle {\cal S}(t) \big\rangle ,
\eeq
where we define hereafter the pertinent distributions to do so.
We assume that the $T_n$ are IID and drawn from the discrete PDF
\beq
\label{life_time_PDF}
 \mathbb{P}[T=t] = \psi(t) = \big\langle \delta_{t,T} \big\rangle  , \hspace{1cm}  t = \{1,2,\ldots\} \in \mathbb{N} ,
\eeq
where $\psi(0)=0$ as $T\geq 1$ and we will use its GF ${\overline \psi}(u) = \big\langle u^T \big\rangle = \sum_{r=1}^{\infty} \psi(r) u^r$ ($|u| \leq 1$)
with ${\overline \psi}(1)=1$ ($\psi(t)$ is properly normalized).
The probability that the walker is alive at time instant $t$ counted from the time instant of the last budget renewal is
\beq
\label{walker_survival_T}
\mathbb{P}[T > t] = \Phi_{\cal S}(t) = \Big\langle \Theta(T-1-t) \Big\rangle = \Big\langle \Theta(0, t,T) \Big\rangle = \sum_{r=t+1}^{\infty} \psi(r) , \hspace{1cm} 
t \in \mathbb{N}_0 = \{0,1,2,\ldots\} ,
\eeq
where $\Phi_{\cal S}(0) =1$ and $\Phi_{\cal S}(\infty) =0$ with the GF
\beq
\label{GF_walker_Gf}
{\overline \Phi}_{\cal S}(u) =\big\langle \sum_{t=0}^{\infty}  \Theta(T-1-t) u^t \big\rangle =\frac{1- \big\langle u^T \big\rangle}{1-u} =\frac{1-{\overline \psi}(u)}{1-u} .
\eeq

As mentioned, we may think the process (\ref{budget}) as an auxiliary random walk on the integer line, in 
which a walker performs deterministic unit steps in the negative direction and is relocated
at the budget renewal times $\tau_n$ to the new positive positions ${\cal C}(\tau_n) = T_{n+1}$, see Fig. \ref{budget_fig}. 
The so defined auxiliary random walk has therefore a connection with stochastic resetting and is susceptible for a wide range of generalizations \cite{Singh_Metzler_Sandev_Chaos2025,Boyer_Ria_Chaos2025,Mi_Dono_Po_Ria_Chaos2025}.
Are frequent budget renewals advantageous for the survival of the walker? 
On the one hand, repeated renewals are vital for the survival of the walker, however frequent budget renewals may be detrimental in some cases.
This happens when ${\cal C}(\tau_n -1) > {\cal C}(\tau_n)$, where a budget renewal is not increasing the budget, but is coming along with a cost. 
Such a detrimental situation is illustrated at budget renewal time $\tau_3$ in Fig. \ref{budget_fig}. 

Consider two budget renewals, one at $t=0$ and the second at $t_1 >0$. The associated renewed budget values $T_1,T_2$ are IID, therefore
the survival probability at $t_1+t_2$ ($t_2>0$) is
$\Phi_{\cal S}(t_1)\Phi_{\cal S}(t_2)$. In contrast, when there is only a budget renewal at $t=0$, the survival probability at $t_1+t_2$
is $\Phi_{\cal S}(t_1+t_2)$. The two budget renewals are advantageous compared to the single one (at $t=0$) if
\beq
\label{condition_ineq}
\Phi_{\cal S}(t_1)\Phi_{\cal S}(t_2) \geq \Phi_{\cal S}(t_1+t_2) ,
\eeq
and $\Phi_{\cal S}(t_1)\Phi_{\cal S}(t_2) > \Phi_{\cal S}(t_1+t_2)$ for some $t_1,t_2$.
We observe that $\Phi_{\cal S}(t_1)\Phi_{\cal S}(t_2) = \Phi_{\cal S}(t_1+t_2)$ is fulfilled when $T$ 
is geometrically distributed, i.e. $\Phi_{\cal S}(t) = Q^t$ with $\psi(t)= PQ^{t-1}$ ($Q=1-P$), $P \in (0,1)$. 
In this case, (frequent) budget renewals have no effect on the survival probability of the walker 
compared to the situation of a single budget acquisition at $t=0$. 
We can write any $\Phi_{\cal S}(t)$ in the form \cite{Pachon_etal2025}
\beq
\label{represPhiS}
\Phi_{\cal S}(t)= \prod_{k=1}^t (1-\alpha_k) , \hspace{0.5cm} \alpha_k \in [0,1], \hspace{1cm} t \in \mathbb{N}
\eeq
with $\Phi_{\cal S}(0)= 1$ and $\psi(t)= \alpha_t \Phi_{\cal S}(t-1)$. 
Indeed, in the geometric case one has $\alpha_t=P$ being independent of $t$ reflecting the memoryless (Markov) 
feature of the corresponding Bernoulli (trial) process. 
The following observations are worthy of mention:
\\[2ex]
\noindent (i) We expect for geometrically distributed $T$ that
the survival statistics of the MRW does not depend on characteristics and topological details of the network and on the target.
Frequent budget renewals (visits of the target) do not affect the expected lifetime of this walker, thus the survival probability of the walker
is ${\cal P}_{\cal S}(t) = \Phi_{\cal S}(t)$ depending uniquely on the first budget acquisition at $t=0$.
\\[2ex]
\noindent (ii) In all other cases of non-geometrically distributed $T$, there is a memory on when budget renewals happen, since
$\alpha_t$ depends on $t$. Consequently the lifetime of the walker is sensitive on budget renewals and depends on the network and target
features.
\\[2ex]
\noindent (iii) Frequent budget renewals are advantageous 
when inequality (\ref{condition_ineq}) is holding true, which implies that
\beq
\label{advant}
\alpha_{t+1} \geq  \alpha_t , \hspace{1cm} t \in \mathbb{N} ,
\eeq
in which there is at least one integer $t_0 \in \mathbb{N}$ such that $\alpha_{t_0+1} >  \alpha_{t_0}$. Consequently, $ \Phi_{\cal S}(t)$ 
decays faster than geometrically: There is a $t_0 \in \mathbb{N}$ such that $\Phi_{\cal S}(t) < (1-\alpha_1)^t$ for
$t>t_0$.
We refer this scenario to as the `{\it forager's scenario}'. Budget renewals (t-node hits) are expected to
increase the expected lifetime, mimicking the effect of food consumption of a forager. We point out that the deterministic case $T=k_0$ (with $\alpha_t = \delta_{t,k_0}$, $\Phi_{\cal S}(t)= \Theta(k_0-1-t)$) refers to the
forager's scenario.
\\[2ex]
\noindent (iv) In the situation when $\alpha_{t+1} \leq \alpha_t $ (inequality for at least one time instant $t$), we expect that frequent budget renewals 
are detrimental and reduce the expected lifetime of the walker.
\\[2ex]
We will explore the complexity of the resulting MRW dynamics and corroborate the remarks (i)-(iv) in a numerical study at the end of the paper (Sect. \ref{generic_example}).
\subsection{MRW for an arbitrary target}
\label{arbitrary_target_MRW}
The above results enable us to derive heuristically the pertinent distributions of the MRW for an arbitrary t-node set ${\cal B}$.
We will do so by generalizing 
relations (\ref{useful_hitting_rel})-(\ref{thefollowing_important}).
The mortal counterpart of (\ref{H_matrix}) reads
\beq
\label{MRW_H_matrix}
\pmb{\mathcal{H}}(t,\xi;{\cal B}) = \sum_{n=1}^{\infty} \xi^n \pmb{\Upsilon}^{(n)}(t) ,
\hspace{0.5cm} t \in \mathbb{N} , \hspace{0.5cm} \pmb{\mathcal{H}}(0,\xi;{\cal B}) = 0 , \hspace{0.5cm} \xi \in [0,1],
\eeq
where $\pmb{\mathcal{H}}(t;{\cal B}) =\pmb{\mathcal{H}}(t,1;{\cal B})$ retrieves the MHR matrix of the mortal walker.
It contains matrix convolution powers $ \pmb{\Upsilon}^{(n)}(t) = \left(\pmb{\Upsilon}^{(n-1)} \star \pmb{\Upsilon}\right)(t)$, 
where $\pmb{\Upsilon}^{0}(t) = \delta_{t,0} \pmb{1}$ and $\pmb{\Upsilon}(t) = \pmb{\Upsilon}^{(1)}(t)$ is the FHM of the mortal
walker given by
\beq
\label{FHM_mortal_walker}
\pmb{\Upsilon}(t) = \pmb{\chi}(t) \Phi_{\cal S}(t) = 
\widetilde{\pmb{W}}^{t-1}\cdot \pmb{W}^{({\cal B})} \, \Phi_{\cal S}(t) , \hspace{0.5cm} t \in \mathbb{N} ,\hspace{0.5cm} \pmb{\Upsilon}(0)=0 .
\eeq
This matrix contains the (defective) first hitting PDF 
$[\pmb{\Upsilon}(t)]_i =\Phi_{\cal S}(t) \chi_{i \to {\cal B}}(t)$
of ${\cal B}$ indicating the probability that the walker hits the target ${\cal B}$ for the first time at time $t$ and is alive. 
We refer to \cite{dono_etal2024} for details concerning defective distributions. 
The probability that the mortal walker ever hits ${\cal B}$ in its lifetime is
\beq
\label{not_properly}
\sum_{t=1}^{\infty} \Phi_{\cal S}(t) \chi_{i \to {\cal B}}(t) < \sum_{t=1}^{\infty} \chi_{i \to {\cal B}}(t) = 1,
\eeq
and recall Appendix \ref{rec_mar_cha} for the RHS.
From this inequality follows that the MRW is not recurrent and lacks ergodicty: There is indeed a positive probability 
$1-\sum_{t=1}^{\infty} \Phi_{\cal S}(t) \chi_{i \to {\cal B}}(t)$ that the walker never hits ${\cal B}$ (in its lifetime).
We formulate the MRW counterpart of (\ref{useful_hitting_rel}) as
\beq
\label{hitting_rel_MRW}
\pmb{P}^{(*)}(t,\xi;{\cal B}) = \widetilde{\pmb{W}}^t \, \Phi_{\cal S}(t) + 
\sum_{r=1}^t \pmb{\mathcal{H}}(r,\xi;{\cal B}) \cdot \widetilde{\pmb{W}}^{t-r} \, \Phi_{\cal S}(t-r) , \hspace{1cm} t \in \mathbb{N}_0 ,
\eeq
in which $\pmb{P}^{(*)}(t,1;{\cal B})= \pmb{P}^{(*)}(t;{\cal B}) = [P_{ij}^{(*)}(t,{\cal B})]$ is the MRW propagator. 
The entry $P_{ij}^{(*)}(t,{\cal B})$ is the probability that the walker (starting from node $i$) reaches node $j$ at time $t$ (alive). 
In the immortal limit 
($T \to \infty, \Phi_{\cal S}(t) \to 1$), we recover $\pmb{P}^{(*)}(t,\xi;{\cal B}) \to [\pmb{\Omega}(\xi;{\cal B})]^t$
and all the below derived relations turn into those of the immortal Markov walker.
We establish the MRW counterpart of (\ref{thefollowing_important}) in the form
\beq
\label{thefollowing_important_MRW}
\pmb{P}^{(*)}(t,\xi;{\cal B}) = {\widetilde {\pmb W}}^t \, \Phi_{\cal S}(t) + 
\xi \sum_{r=1}^t \pmb{\Upsilon}(r) \cdot \pmb{P}^{(*)}(t-r,\xi;{\cal B}) ,  \hspace{1cm} t \in \mathbb{N}_0 ,
\eeq
which is solved by (\ref{hitting_rel_MRW}).
For the following derivations, we introduce
\beq
\label{Pi_dist}
\pmb{\Pi}(t) = {\widetilde {\pmb W}}^t \, \Phi_{\cal S}(t) , \hspace{1cm} t \in \mathbb{N}_0 .
\eeq
It contains the evanescent probability $[\pmb{\Pi}(t)]_i =  \Phi_{\cal S}(t) \Lambda_i(t)$ that 
the walker has not hit ${\cal B}$ {\it and} is alive at time instant $t$.
Be aware, that this quantity is different from the non-evanescent probability
$1 - \sum_{r=1}^t [\pmb{\Upsilon}(r)]_i$, which is the probability that 
the walker has not hit ${\cal B}$ up to time $t$, {\it no matter} whether
the walker is alive at time instant $t$. 
We also will need the GF
\beq
\label{Pi_GF_MRW}
\overline{\pmb{\Pi}}(u) = \sum_{t=0}^{\infty} [u {\widetilde {\pmb W}}]^t \, \Phi_{\cal S}(t) = 
\overline{ \Phi}_{\cal S}(u {\widetilde {\pmb W}}), 
\eeq
together with the GF of the FHM of the mortal walker
\beq
\label{GF_upsilon_arb_tar_MRW}
\overline{\pmb{\Upsilon}}(u) = \left\{\sum_{t=1}^{\infty} u^t \widetilde{\pmb{W}}^{t-1} \, \Phi_{\cal S}(t)  \right\} \cdot \pmb{W}^{({\cal B})}  = u \overline{G}_{\cal S}(u \widetilde{\pmb{W}}) \cdot \pmb{W}^{({\cal B})} 
\eeq 
containing the GF of the auxiliary function 
$G_{\cal S}(t) = \Phi_{\cal S}(t+1)$ ($t \in \mathbb{N}_0$) 
\beq
\label{hrel_auxiliary}
\ds  {\overline G}_{\cal S}(v) = \ds \sum_{r=0}^{\infty} v^r \Phi_{\cal S}(r+1) = \ds \frac{1}{v}\left(-1 + \sum_{r=0}^{\infty} v^r \Phi_{\cal S}(v)\right) = \frac{\overline{\Phi}_{\cal S}(v)-1}{v} , \hspace{1cm} 0 < |v| < 1 .
\eeq
This GF does not contain negative powers of $v$ and is hence useful to evaluate numerically matrix function 
(\ref{GF_upsilon_arb_tar_MRW}) (since $\widetilde{\pmb{W}}$ is not invertible for any non-empty target).

Let us analyze the THCP in a MRW. To that end, we consider the GF of the MRW propagator matrix, where
(\ref{thefollowing_important_MRW}) leads to
\beq
\label{GF_thefollowing_important_MRW}
\begin{array}{clr}
\ds \overline{\pmb{P}}^{(*)}(u,\xi;{\cal B}) & = \ds 
\left[\pmb{1} - \xi \overline{\Upsilon}(u)\right]^{-1} \cdot \overline{\pmb{\Pi}}(u) & \\[2ex]
& =\ds  \left[\pmb{1} - \xi \overline{\Upsilon}(u)\right]^{-1} \cdot \left[\pmb{1} - \overline{\Upsilon}(u)\right] \cdot   
\overline{\pmb{P}}^{(*)}(u; {\cal B}) &
\end{array}
, \hspace{1cm} |u|\leq 1 
\eeq
being the counterpart of (\ref{useful_identity}) and consistent with (\ref{hitting_rel_MRW}).
The survival probability (probability that the walker is alive at time $t$) is
\beq
\label{survival_MRW_walker}
{\cal P}_{\cal S}^{(*)}(t; i, {\cal B}) = [\pmb{P}^{(*)}(t;{\cal B})]_i .
\eeq
Evanescence of the MRW is reflected by the evanescence of the MRW propagator in the infinite time limit
\beq
\label{evan}
\pmb{P}^{(*)}(\infty; {\cal B}) = \lim_{u\to 1}
(1-u) \overline{\pmb{P}}^{(*)}(u; {\cal B}) =0 
\eeq
and so ${\cal P}_{\cal S}^{(*)}(\infty; i, {\cal B}) =0$. Consult Appendix \ref{invert} for related details.
(\ref{evan}) holds true, apart of some distinct situations  
discussed later on, in which the MRW is immortal. 

Now our goal is to derive the state probabilities in a MRW, i.e. the probabilities that the walker has hit ${\cal B}$ $n$ times up to and including time $t$.
One can introduce in full analogy to the immortal walker (\ref{moment_gen}), (\ref{state_proba_matrix} 
the mortal counterpart $\pmb{Q}^{(n,*)}(t) = [Q_{ij}^{(n,*)}(t)]$ 
defined by 
\beq
\label{defined_MRW_state_n_prob}
\pmb{P}^{(*)}(t,\xi;{\cal B}) =\sum_{n=0}^{\infty} \xi^n \pmb{Q}^{(n,*)}(t) ,
\eeq
where (see (\ref{MRW_H_matrix}) and (\ref{hitting_rel_MRW}) with (\ref{FHM_mortal_walker}))
\beq
\label{eval_Q_star_n}
\pmb{Q}^{(n,*)}(t) = \sum_{t=0}^t \pmb{\Upsilon}^{(n)}(r) \cdot \widetilde{\pmb{W}}^{t-r} \, \Phi_{\cal S}(t-r) , \hspace{1cm} n, t\in \mathbb{N}_0
\eeq
with $\pmb{Q}^{(0,*)}(t) =  \widetilde{\pmb{W}}^t \Phi_{\cal S}(t)$ and $\pmb{Q}^{(n,*)}(0) =\delta_{n,0} \pmb{1}$.
However, the normalization of
$[\pmb{Q}^{(n,*)}(t)]_i$ is improper yielding the evanescent survival probability $$\sum_{n=0}^{\infty}[\pmb{Q}^{(n,*)}(t)]_i = {\cal P}_{\cal S}^{(*)}(t; i, {\cal B}) \leq 1 $$ (equality for $t=0$) 
and so the $\pmb{Q}^{(n,*)}(t)$ are evanescent. We infer that the quantities $[\pmb{Q}^{(n,*)}(t)]_i$ are not the true state probabilities,
since they indicate the probabilities that the walker has hit ${\cal B}$ $n$ times {\it and} is alive at time instant $t$.
However, we seek the `true' non-evanescent state probabilities, which indicate the probabilities that the walker up to and including time $t$ has hit ${\cal B}$ 
$n$ times, {\it no matter} whether the walker is alive at time $t$ or not.
The true state probabilities should be non-evanescent as $t\to \infty$.
In the following, we denote the target hitting counting variable of the mortal walker with $\mathcal{N}^{(*)}_i(t;{\cal B})$ (departure node $i$).
To construct the true state probabilities, the following observation is helpful.
The effect of mortality is such that the non-decreasing MRW THCP counting variable saturates to a maximum value $\mathcal{N}^{(*)}_i(t;{\cal B}) \to \mathcal{N}^{(*)}_i(t_{*};{\cal B})$, which is maintained from the instant $t_{*}$ when the walker dies: 
$\mathcal{N}^{(*)}_i(t;{\cal B}) = \mathcal{N}^{(*)}_i(t_{*};{\cal B})$ for $t\geq t_{*}$. As a consequence all moments of $\mathcal{N}^{(*)}_i(t;{\cal B})$ are non-decreasing functions of $t$, saturating to maximum values as $t\to \infty$.
In order to capture the MRW THCP statistics, we
introduce the non-evanescent auxiliary matrix
\beq
\label{mean_hitting_mat}
\big\langle \pmb{\mathcal{N}}^{(*)}(t;{\cal B}) \big\rangle = \sum_{r=1}^t \pmb{\mathcal{H}}(r, 1;{\cal B}) \cdot \pmb{W}^{t-r} \neq \frac{d}{d\xi} 
\pmb{P}^{(*)}(t,\xi;{\cal B})\Big|_{\xi=1} = \sum_{r=1}^t \pmb{\mathcal{H}}(r, 1;{\cal B}) \cdot \pmb{P}^{(*)}(t-r;{\cal B}) .
\eeq
The matrix $\big\langle \pmb{\mathcal{N}}^{(*)}(t;{\cal B}) \big\rangle$ is the 
MRW counterpart of (\ref{mean_hits_eval}), which is recovered in the immortal limit. 
$\mathcal{N}_i^{(*)}(t;{\cal B}) \in \mathbb{N}_0$ (with $\mathcal{N}_i^{(*)}(0;{\cal B})=0$) counts the target hits 
in a walk starting from node $i$ up to and including time $t$ and remains constant from the moment $t_{*}$ of walker's death.
From the LHS we can construct in analogy to (\ref{useful_identity}) an auxiliary generating matrix, namely
\beq
\label{moment_GF}
\overline{\pmb{\mathcal{G}}}(u,\xi;{\cal B}) = \left[\pmb{1}-\xi \overline{\pmb{\Upsilon}}(u)\right]^{-1}\cdot 
\left[\pmb{1}- \overline{\pmb{\Upsilon}}(u)\right]\cdot [\pmb{1}-u\pmb{W}]^{-1} 
= \sum_{n=0}^{\infty} \xi^n \overline{\pmb{\mathcal{Q}}}^{(n,*)}(u) ,\hspace{1cm} |u| \leq 1
\eeq
in which $[\overline{\pmb{\mathcal{Q}}}^{(n,*)}(u)]_i$ are the GFs of the true state probabilities and
which is such that $\overline{\pmb{\mathcal{G}}}(u,\xi;{\cal B}) \to [\pmb{1}-u\pmb{\Omega}]^{-1}$ in the immortal limit.
$\overline{\pmb{\mathcal{G}}}(u,\xi;{\cal B})$ records the target hitting events of the MRW walker and one has
\beq
\label{inverse_moment_GF_MRW}
\big\langle \overline{\pmb{\mathcal{N}}}^{(*)}(u;{\cal B}) \big\rangle = 
\frac{\partial}{\partial \xi} \overline{\pmb{\mathcal{G}}}(u,\xi;{\cal B}) \Big|_{\xi=1} 
= \overline{\pmb{\mathcal{H}}}(u; {\cal B}) \cdot  \left(\pmb{1}-u\pmb{W}\right)^{-1} ,
\eeq
which contains the GF of (\ref{MRW_H_matrix}) 
\beq
\label{GF_of_H_mat}
\overline{\pmb{\mathcal{H}}}(u,\xi;{\cal B}) = 
\left(\pmb{1}- \xi \overline{\pmb{\Upsilon}}(u)\right)^{-1} \cdot \overline{\pmb{\Upsilon}}(u) \xi .
\eeq
It is straightforward to derive, from the generating function in Eq. (\ref{inverse_moment_GF_MRW}), the mortal counterpart of relation (\ref{further_mean}).
The proper normalization of the MRW state probabilities
is ensured by
\beq
\label{aux_walker}
\overline{\pmb{\mathcal{G}}}(u,1;{\cal B}) = 
\sum_{n=0}^{\infty} \overline{\pmb{\mathcal{Q}}}^{(n,*)}(u) = \left[\pmb{1}-u\pmb{W}\right]^{-1}, 
\eeq
thus $\pmb{\mathcal{G}}(t,1;{\cal B}) = \pmb{W}^t$, where the GFs of the true MRW state probability matrices read
\beq
\label{state_proba_mat}
\overline{\pmb{\mathcal{Q}}}^{(n,*)}(u) =  [\overline{\pmb{\Upsilon}}(u) ]^n \cdot
\left[\pmb{1}- \overline{\pmb{\Upsilon}}(u)\right] \cdot \left[\pmb{1}-u\pmb{W}\right]^{-1} 
\eeq
being the MRW counterpart of (\ref{extraction_state_proba_mats}).
The infinite time limit of $\pmb{\mathcal{G}}(t,\xi;{\cal B})$ can be retrieved from  
\beq
\label{dead_walkers_THCP}
 \pmb{\mathcal{G}}(\infty,\xi;{\cal B}) = \lim_{u\to 1} (1-u) \overline{\pmb{\mathcal{G}}}(u,\xi;{\cal B}) = 
\left[\pmb{1}-\xi \overline{\pmb{\Upsilon}}(1)\right]^{-1}\cdot 
\left[\pmb{1}- \overline{\pmb{\Upsilon}}(1)\right] \cdot \pmb{W}^{(\infty)}  
\eeq
containing the stationary transition matrix $\pmb{W}^{(\infty)} = |\phi_1\rangle \langle \overline{\phi}_1|$.
Correspondingly, the infinite time limits of the state probability matrices are obtained as
\beq
\label{non-evanscence_true_state}
\pmb{\mathcal{Q}}^{(n,*)}(\infty) = \lim_{u\to 1} (1-u) \overline{\pmb{\mathcal{Q}}}^{(n,*)}(u) =
 [\overline{\pmb{\Upsilon}}(1) ]^n \cdot
\left[\pmb{1}- \overline{\pmb{\Upsilon}}(1)\right] \cdot \pmb{W}^{(\infty)} , \hspace{1cm} n \in \mathbb{N}_0 .
\eeq
The true state probabilities take then the non-evanescent infinite time asymptotics
\beq
\label{saturated-stae_probas}
\mathbb{P}[\mathcal{N}_i^{(*)}(\infty;{\cal B}) = n] = [\pmb{\mathcal{Q}}^{(n,*)}(\infty)]_i = \left[ [\overline{\pmb{\Upsilon}}(1) ]^n \cdot
\left[\pmb{1}- \overline{\pmb{\Upsilon}}(1)\right] \right]_i , \hspace{1cm} n \in \mathbb{N}_0 
\eeq
indicating the probability that the walker hits ${\cal B}$ exactly $n$ times in the walker's life. For $n=0$ this is
$[\pmb{\mathcal{Q}}^{(0,*)}(\infty)]_i = 1 - [\overline{\pmb{\Upsilon}}(1)]_i $ the above-mentioned probability that the walker never hits ${\cal B}$ in its lifetime.
The non-decreasing expected number of t-node hits up to including time $t$, no matter whether the walker still is alive at time $t$ is with (\ref{inverse_moment_GF_MRW})
\beq
\label{expect_total_hit_MRW}
\big\langle \mathcal{N}_i^{(*)}(t;{\cal B}) \big\rangle = \sum_{r=1}^t [\pmb{\mathcal{H}}(t,1;{\cal B})]_i ,
\eeq
being consistent with (\ref{mean_hitting_mat}).
On the other hand, a consequence of the mortality is evanescence of the MHR
\beq
\label{number_ever_hits}
 [\pmb{\mathcal{H}}(\infty,1;{\cal B})]_i =
\lim_{u\to 1} (1-u) [\overline{\pmb{\mathcal{H}}}(u,1;{\cal B})]_i = 0 .
\eeq
The walker is expected to hit ${\cal B}$ only a finite number of times in a lifetime, expressed by the infinite time limit of (\ref{expect_total_hit_MRW}) yielding the saturated values
\beq
\label{finite_hitting_nums}
\big\langle \mathcal{N}_i^{(*)}(\infty;{\cal B}) \big\rangle = [\overline{\pmb{\mathcal{H}}}(u,1;{\cal B})]_i\Big|_{u=1} =
\left[\left(\pmb{1}- \overline{\pmb{\Upsilon}}(1)\right)^{-1} \cdot \overline{\pmb{\Upsilon}}(1)\right]_i 
< \infty .
\eeq
The finiteness of these numbers follows from invertibility of the matrix $\pmb{1} - \overline{\Upsilon}(1)$ (see Appendix \ref{invert}).

A further quantity of interest is the  `{\it mean residence time}' (MRT) a walker spends in a certain region \cite{fractionalbook-2019,Mi_Ria_frac_walk2017,Barkai2006}.
Let $\big\langle \mathcal{T}_{\cal S}(i,j;{\cal B}) \big\rangle$ be the MRT the walker spends on the node $j$ during its life for departure node $i$. The MRT is determined by the GF of the propagator matrix (\ref{GF_thefollowing_important_MRW}), namely 
\beq
\label{MRT_walker}
\big\langle \mathcal{T}_{\cal S}(i,j;{\cal B}) \big\rangle = \sum_{t=0}^{\infty} P^{(*)}_{ij}(t;{\cal B}) = 
\overline{P}^{(*)}_{ij}(1;{\cal B}) = \left[\left(\pmb{1} - \overline{\Upsilon}(1)\right)^{-1} \cdot \overline{\pmb{\Pi}}(1)\right]_{ij} .
\eeq
Consequently, the MRT the walker spends on a set of nodes ${\cal L}$ during its lifetime is 
$\langle \mathcal{T}_{\cal S}(i,{\cal L};{\cal B}\rangle =\sum_{j\in {\cal L}}\big\langle \mathcal{T}_{\cal S}(i,j;{\cal B}) \big\rangle$. When ${\cal L}$ comprises the whole network this quantity gives the expected lifetime 
of the walker, namely
\beq
\label{expect_lllife_time_MRW_walker_arb_target}
\begin{array}{clr}
\ds \big\langle \mathcal{T}_{\cal S}(i;{\cal B}) \big\rangle & = 
\ds \overline{\mathcal{P}}_{\cal S}(1;i,{\cal B}) =  
\left[\left[\pmb{1} - \overline{\Upsilon}(1)\right]^{-1} \cdot \overline{\pmb{\Pi}}(1)\right]_i  & \\[3ex]
& = \ds  \left[\overline{\pmb{\Pi}}(1)\right]_i +   \left[ \big\langle \pmb{\mathcal{N}}^{(*)}(\infty;{\cal B}) \big\rangle \cdot \overline{\pmb{\Pi}}(1)\right]_i .&
\end{array}
\eeq
We emphasize that the expected lifetime is finite since the matrix $\pmb{1} - \overline{\Upsilon}(1)$ is invertible. The expected lifetime becomes infinite in the limit of immortality. Inverting GF (\ref{state_proba_mat}) for $n=0$
yields for the probability that the walker has not hit ${\cal B}$ up to time $t$ no matter whether the walker still is alive at time $t$ as
$[\pmb{\mathcal{Q}}^{(0,*)}(t)]_i=  1- \sum_{r=1}^t  [\pmb{\Upsilon}(r)]_i$. 
As mentioned, a landmark of evanescence of the walker is that the large time asymptotics of this quantity is non-vanishing.
Consequently, the MFPT to hit ${\cal B}$ is infinite 
\beq
\label{MFPT_MRW}
\big\langle \mathcal{T}^{(*)}_{i \to {\cal B}} \big\rangle = \sum_{t=0}^{\infty} [\pmb{\mathcal{Q}}^{(0,*)}(t)]_i  = \lim_{u\to 1} \frac{1- [\overline{\pmb{\Upsilon}}(u)]_i}{1-u} \to \lim_{u\to 1}  
 \frac{1-[\overline{\pmb{\Upsilon}}(1)]_i}{1-u} = \infty .
\eeq
%
%
%
%
%
%
We point out that the MFPT defined in (\ref{MFPT_MRW}) consists of the average number of time steps until the walker reaches one of the t-nodes. 
This average contains the paths for which the walker hits a t-node alive, plus 
the paths for which the walker dies before hitting a t-node. The latter paths contribute to the average with infinite hitting times, leading to the divergence of (\ref{MFPT_MRW}). Instead, one can consider the 
{\it conditional mean first passage time (CMFPT)}. The CMFPT averages only over the fraction of paths for which the walker reaches the target (alive), 
and remains therefore finite. For thorough accounts, we refer to
\cite{Yuste_etal2013} and \cite{MeersonRedner2015}.
For the MRW the CMFPT reads
$$
\big\langle {\cal T}_{i\to {\cal B}}^{(*,\, c)} \big\rangle = 
\frac{\ds \sum_{t=1}^{\infty} [\pmb{\Upsilon}(t)]_i t}{\ds  \sum_{t=1}^{\infty} [\pmb{\Upsilon}(t)]_i} .
$$
In the immortal limit, the CMFPT as well as the MFPT (\ref{MFPT_MRW}) turn
into the finite expression (\ref{MFPT_to_B}) of the immortal Markov walker.
\\
The CMFPT is thoroughly analyzed in the reference \cite{MeersonRedner2015}. 
In that paper, the motion of a swarm of $M$ independent identical Brownian mortal searchers 
with constant mortality rate is considered in the one-dimensional infinite space. This model mimics
an idealized picture of the fertilization processes of an oocyte by sperm.
It is shown there, in the limit of vanishing mortality that the shortest CMFPT of $M$ walkers to reach the target, decreases as $1/\ln(M)$. 
On the other hand, in the case where mortality is sufficiently high, the shortest CMFPT becomes independent of $M$. 
The analysis of the hitting statistics of multiple MRW walkers is indeed calling for thorough investigation, meriting a dedicated study.
%
%
%
%
%
%
\paragraph{Special cases} The following cases stand out and merit our particular consideration.
\\[2ex]
{\it\large {\bf (a)} -- All nodes are t-nodes: ${\cal B} = {\hat {\cal B}}$ comprises the complete network} 
\\[1ex]
An instructive case emerges when the target comprises the complete network, thus
each step is a t-node hit.
Then one has $\widetilde{\pmb{W}} = \pmb{0}$ with $\widetilde{\pmb{W}}^t = \pmb{\Pi}(t) = \pmb{1} \delta_{t,0}$ and  $\pmb{W}^{({\hat {\cal B}})} = \pmb{W}$ and $\pmb{\Upsilon}(t) = \delta_{t,1} \Phi_{\cal S}(1) \pmb{W}$ (see 
(\ref{FHM_mortal_walker}), (\ref{Pi_dist})).
Eq. (\ref{hitting_rel_MRW}) then boils down to
\beq
\label{propa_com_tar}
\pmb{P}^{(*)}(t,\xi;{\hat {\cal B}}) = \delta_{t,0} \pmb{1} +\pmb{\mathcal{H}}(t,\xi;{\hat {\cal B}}) =  \delta_{t,0} \pmb{1} + \sum_{n=1}^{\infty} \xi^n (\Phi_{\cal S}(1))^n \pmb{W}^n\delta_{t,n} = \xi^t (\Phi_{\cal S}(1))^t \pmb{W}^t, \hspace{1cm} t \in \mathbb{N}_0
\eeq
with the propagator $\pmb{P}^{(*)}(t; {\hat {\cal B}}) = (\Phi_{\cal S}(1))^t \pmb{W}^t$ yielding the survival probability
\beq
\label{all_target}
{\cal P}_{\cal S}^{(*)}(t; i, {\hat {\cal B}}) =  (\Phi_{\cal S}(1))^t ,\hspace{1cm} t \in \mathbb{N}_0
\eeq
being independent of the departure node.
At each time step ${\hat B}$ is hit and the budget is renewed. Therefore, the survival of the walker 
is governed by a sequence of independent Bernoulli trials at each time step: The walker survives the subsequent step only if $T>1$ and 
dies at the subsequent step if $T=1$. Hence, the walker survives the subsequent step with probability 
$\mathbb{P}[T>1]=\Phi_{\cal S}(1)$ and dies with the complementary (`Bernoulli success') probability 
$\mathbb{P}[T=1] = 1-\Phi_{\cal S}(1) = \psi(1)$ (recall Sect. \ref{gen_mod}). The survival probability at step $t$ is therefore given by (\ref{all_target}), leading to the expected lifetime 
\beq
\label{lifetime_all_nodes}
\big\langle \mathcal{T}_{\cal S}(i;{\hat {\cal B}}) \big\rangle = 
\sum_{t=0}^{\infty} {\cal P}_{\cal S}^{(*)}(t; i, {\hat {\cal B}}) = \frac{1}{1-\Phi_{\cal S}(1)} .
\eeq
If the distribution is such that $\Phi_{\cal S}(1)=1$  (i.e. $\psi(1)= 0$) the walker is immortal and recovers the recurrent and ergodic Markov walk.
Interesting is here the connection of the expected lifetime and expected number of target visits in the walker's lifetime.
From (\ref{finite_hitting_nums}) we obtain
\beq
\label{infinite_time}
\big\langle \mathcal{N}^{(*)}_i(\infty; {\hat {\cal B}}) \big\rangle = 
\sum_{t=1}^{\infty} [\pmb{\mathcal{H}}(t;{\hat {\cal B}})]_i = \sum_{t=1}^{\infty} (\Phi_{\cal S}(1))^t = 
 \frac{\Phi_{\cal S}(1)}{1-\Phi_{\cal S}(1)} .
\eeq
As each node is a t-node, each step of the walker is a t-node hit, except for the last step when the walker dies.
Recall that, if the walker dies at time $t_*$ (i.e. ${\cal C}(t_{*}) =0$) the walker's lifetime is $t_*$, 
however the number of t-node hits is $t_{*}-1$ as step $t_*$ does not count as a target hitting event 
since per construction a t-node hit is counted at a time step $t$ only if the budget
${\cal C}(t) \geq 1$.
Consequently, one has here $\big\langle \mathcal{N}_i(\infty;{\hat {\cal B}}) \big\rangle = 
\big\langle \mathcal{T}(i;{\hat {\cal B}}) \big\rangle  -1$. 
Again, for the cases in which $\Phi_{\cal S}(1)=1$ the walker is immortal, thus 
$\big\langle \mathcal{N}^{(*)}_i(\infty; {\hat {\cal B}}) \big\rangle = \infty$ in (\ref{infinite_time}) as a landmark of recurrence and ergodicity of the immortal Markov walker.

Finally, instructive is to consider the infinite time limits of the state probabilities (\ref{saturated-stae_probas}) 
leading to a geometric distribution with respect to $n$, namely
\beq
\label{Q_zero_star_all_target}
\mathbb{P}[\mathcal{N}_i^{(*)}(\infty;{\hat {\cal B}}) = n] = [\pmb{Q}^{(n,*)}(\infty)]_i =  (\Phi_{\cal S}(1))^n (1 - \Phi_{\cal S}(1)) , \hspace{1cm} n \in \mathbb{N}_0.
\eeq
indicating the probabilities that ${\hat {\cal B}}$ is hit exactly $n$ times in the walker's life. In the distinct case in which $\Phi_{\cal S}(1)=1$
where the walker is immortal, the target is hit infinitely often, thus (\ref{Q_zero_star_all_target}) is null. For $n=0$ the probability that the walker has never hit the network ${\hat {\cal B}}$ in its life is 
$[\pmb{Q}^{(0,*)}(\infty)]_i = (1 - \Phi_{\cal S}(1)) =\psi(1) =\mathbb{P}[T=1]$ and equals to the probability that the walker dies at the first step. 
Also recall that $(\Phi_{\cal S}(1))^n$ is the probability that the walker survives $n$ successive steps, and $(1 - \Phi_{\cal S}(1))$ the probability that 
the walker dies at step $n+1$ (given survival up to step $n$). Hence (\ref{Q_zero_star_all_target}) coincides with the probability that the walker dies at time step $t_{*}=n+1$.
\\[3ex]
{\it\large {\bf (b)} -- ${\cal B} = \emptyset$: MRW dynamics without target}
\\[2ex]
Here, the walker acquires only once a budget, namely $T_1$ at $t=0$. The survival statistics of the walker should
be only governed by the distribution of $T$, independent of the walk on the network and the target.
To confirm this observation, we account for
$\widetilde{\pmb{W}} = \pmb{W}$ thus $\pmb{\Pi}(t)= \Phi_{\cal S}(t) \pmb{W}^t$ and $\pmb{W}^{(\emptyset)} = 
\pmb{\Upsilon}(t) = \pmb{0}$ thus $\pmb{\mathcal{H}}(t;\emptyset) = \pmb{0}$. 
The propagator then boils down to
\beq
\label{proagator_epsty_taret}
\pmb{P}^{(*)}(t;\emptyset) = \Phi_{\cal S}(t) \pmb{W}^t , \hspace{1cm} t \in \mathbb{N}_0 
\eeq
from which follows the survival probability 
\beq
\label{survival_empsty_set}
{\cal P}_{\cal S}(i;\emptyset) =  \Phi_{\cal S}(t)
\eeq
as expected. The expected lifetime of the walker then is equal to the mean of $T$, namely
\beq
\label{expeted_emptyset}
\big\langle \mathcal{T}_{\cal S}(i;\emptyset) \big\rangle = \sum_{t=0}^{\infty} \Phi_{\cal S}(t) =
\overline{\Phi}_{\cal S}(1) = \big\langle T \big\rangle
\eeq
and finally, we have the obvious feature (see (\ref{expect_total_hit_MRW}), (\ref{finite_hitting_nums}))
\beq
\label{confirm_zero_hits}
\big\langle \mathcal{N}_i(t;\emptyset) \big\rangle= 0 
\eeq
reflecting absence of a target.
\\[2ex]
\noindent {\it\large {\bf (c)} -- Neutral scenario} 
\\[2ex]
In the case in which $T$ is geometrically distributed (that is $T$ is the time of first success in a sequence of independent Bernoulli trials) with
$\mathbb{P}[t=T]= \psi(t)= P Q^{t-1}$ and $ \mathbb{P}[T>t]= \Phi_{\cal S}(t) = Q^t$ ($P=1-Q \in (0,1)$). We expect for this case that the survival of the walker should not depend on the network features and the number of target hits as a consequence of the memoryless feature of the Bernoulli process,
see remark (i) of Sect. \ref{gen_mod}. 
The GFs (\ref{Pi_GF_MRW}) and (\ref{GF_upsilon_arb_tar_MRW}) take the particularly simple forms, 
namely
$\overline{\pmb{\Pi}}^{Ber}(u) = [\pmb{1} -  uQ\widetilde{\pmb{W}}]^{-1}$  and $\overline{\pmb{\Upsilon}}^{Ber}(u) = \overline{\pmb{\chi}}(uQ)$.
Plugging these GFs into (\ref{GF_thefollowing_important_MRW}) and inverting the result to time
leads us to the simple result for the MRW propagator 
\beq
\label{Bernoulli_proba}
\pmb{P}^{(*)}_{Ber}(t) = Q^t \pmb{W}^t  
\eeq
with the survival probability 
\beq
\label{neutral_survival}
{\cal P}_{\cal S}(t;i,{\cal B}) =\Phi_{\cal S}(t) = Q^t
\eeq
independent of the network topology and the number of target hits as expected. 
The expected lifetime 
(\ref{expect_lllife_time_MRW_walker_arb_target}) then yields 
\beq
\label{exp_ber}
\big\langle {\cal T}^{Ber}_{\cal S}(i;{\cal B}) \big\rangle =[(\pmb{1}- Q \pmb{W})^{-1}]_i = \sum_{t=0}^{\infty} Q^t [\pmb{W}^t]_i =
\frac{1}{P} =\big\langle T \big\rangle .
\eeq
The number of budget renewals (t-node hits) does not have any effect on the lifetime statistics of the walker. We corroborate this result later on numerically (Fig. \ref{MRW_fig4}).
We observe that there is a connection to the survival statistics of above discussed case {\it{\bf (a)}}, since here as well
the survival probability of the walker is geometric and governed by a sequence of independent Bernoulli trials.
Both cases have the same survival statistics for $\Phi_{\cal S}(1)=Q$.
Consider for a moment geometrically distributed $T$ and that the budget renewal times $\tau_n$ are the arrival times of a proper renewal process. Let 
$\varphi(k)= \mathbb{P}[\Delta t = k]$ be the PDF from which are drawn the IID time intervals $\Delta t \in \mathbb{N}$ between the budget renewals
and $\Lambda(k) =\mathbb{P}[\Delta t > k] = 1- \sum_{r=1}^k \varphi(r)$ the probability that no budget renewal happens up to and including time $k$. Let $\overline{\varphi}(u)$, $\overline{\Lambda}(u)=\frac{1-\overline{\varphi}(u)}{1-u}$ their respective GFs. 
Then one can show in a similar way as in Appendix \ref{survival_derivation} that the GF of the 
survival probability (\ref{survival_proba_MRW}) reads
\beq
\label{GF}
\overline{{\cal P}}_{\cal S}(u) = \frac{\overline{\Lambda}(uQ)}{1-\overline{\varphi}(uQ)}= \frac{1}{1-uQ} .
\eeq
Inverting this relation yields the geometric distribution (\ref{neutral_survival}), a result which is universal for geometrically distributed $T$ and independent of the budget renewal process.
One can show that the geometric survival law (\ref{neutral_survival}) for geometrically distributed $T$ remains even true when the budget renewals do not follow a proper renewal process but an arbitrary renewal protocol.

We will explore at the end of the paper a prototypical example in which the cases {\bf (a)} -- {\bf (c)} are relevant. 

\subsection{MRW for a single target node}
\label{single_ressource}

The distinct case that target ${\cal B}$ consists of a single t-node (node $b$) merits some particular consideration.
We derive these results differently as in the previous section (see Appendix \ref{survival_derivation} for technical details of the derivations), and show consistency of both approaches.
The survival probability (\ref{survival_proba_MRW}) of the walker for departure node $i$ takes with (\ref{survival_indicator_function}) the representation 
(${\cal P}_S(t)=  {\cal P}_{\cal S}(t;i,b)$)
\beq
\label{starvation_survival}
 {\cal P}_{\cal S}(t;i,b)  =  \Big\langle\sum_{n=0}^{\infty}  {\cal M}_n(i,b) \,  \Theta\left[\tau_n(i,b),\, t, \, \tau_{n+1}(i,b)\right] 
\, \Theta\left[0,\, t- \tau_n(i,b), \, T_{n+1}\right]\Big\rangle , \hspace{1cm} t \in \mathbb{N}_0
\eeq
with ${\cal P}(0;i,b) =1$. 
In this relation, we identify the budget renewal times $\tau_n =\tau_n(i;b)$ with the hitting times of $b$ (\ref{indepndent_rand_interims}) and the time interval $\Delta t_1= \Delta t_{ib}$ with the first hitting time of node $b$ (for $i=b$ the recurrence time), and the $\Delta t_k =\Delta t_{bb}^{(k)}$ ($k=2, \ldots,n$) refer to
IID copies of the recurrence time to $b$. The survival indicator function (\ref{memory_budget_renewals}) then reads (${\cal M}_n = {\cal M}_n(i,b)$)
\beq
\label{memory_n}
{\cal M}_n(i,b) = \left\{\begin{array}{lc} \ds \Theta(T_1-1- \Delta t_{ib}) , & n=1 \\[1ex]
\ds \Theta(T_1-1- \Delta t_{ib}) \prod_{k=2}^n \Theta(T_k-1-\Delta t_{bb}^{(k)}) , & n \geq 2 
\end{array}\right. 
\eeq
and ${\cal M}_0(i,b) =1$.
In Appendix \ref{survival_derivation} we derive the GF of (\ref{starvation_survival}) in the form
\beq
\label{SP_GF_MRW}
\begin{array}{clr}
\overline{{\cal P}}_{\cal S}(u;i,b) & = {\overline \Pi}_i(u) + \overline{\mathcal{H}}_{ib}(u){\overline \Pi}_b(u) & \\[2ex]
     & =  {\overline \Pi}_i(u) + \overline{\Upsilon}_{ib}(u) \overline{{\cal P}}_{\cal S}(u;b,b)
     \end{array}
\eeq
being consistent with (\ref{GF_thefollowing_important_MRW}). 
$\overline{\mathcal{H}}_{ib}(u)$ is the GF of the MHR of $b$, namely
\beq
\label{hitting_rate_GF_MRW}
{\overline {\cal H}}_{ib}(u)  = \frac{{\overline \Upsilon}_{ib}(u)}{1- {\overline \Upsilon}_{bb}(u)} 
\eeq
containing the GFs $\overline{\Pi}_i(u), {\overline \Upsilon}_{ib}(u)$ of the
evanescent distributions $\Pi_i(t)= \Phi_{\cal S}(t)\Lambda_i(t)$ and 
$\Upsilon_{ib}(t)= \chi_{ib}(t) \Phi_{\cal S}(t)$.
These distributions can be defined by the following averaging procedures, namely
\beq
\label{not_hit_survival}
[\pmb{\Pi}(t)]_i = \Pi_i(t) = \Big\langle \Theta(\Delta t_{ib}-1-t) \Big\rangle \Big\langle \Theta(T_1-1-t) \Big\rangle =  \Lambda_i(t) \Phi_{\cal S}(t) , \hspace{1cm} t \in \mathbb{N}_0, 
\eeq
which is the (evanescent) probability that the walker is alive and has not yet reached $b$ at time instant $t$ being consistent with (\ref{Pi_dist}).
Then we have the evanescent first hitting PDF, 
i.e. the probability that the walker hits $b$ for the first time at time $t$ (alive)
\beq
\label{hitting_alive}
[\pmb{\Upsilon}(t)]_i  = \Upsilon_{ib}(t) = \Big\langle \Theta(T-1-\Delta t_{ib}) \delta_{t,\Delta t_{ib}} \Big\rangle  =  
\Big\langle \Theta(T-1-t)\Big\rangle \Big\langle \delta_{t,\Delta t_{ib}} \Big\rangle =
\Phi_{\cal S}(t) \chi_{ib}(t) 
\eeq
consistent with (\ref{FHM_mortal_walker}). In these averaging procedures, we used independence of $T$ and $\Delta t_{ib}$. 
Inverting (\ref{SP_GF_MRW}) yields 
\beq
\label{result_reads}
{\cal P}_{\cal S}(t;i,b) = \Lambda_i(t)\Phi_{\cal S}(t) + \sum_{r=0}^t {\cal H}_{ib}(t-r) \Lambda_b(r)\Phi_{\cal S}(r) ,
\eeq
solving
\beq
\label{second_result_reads}
{\cal P}_{\cal S}(t;i,b) = \Lambda_i(t)\Phi_{\cal S}(t) + \sum_{r=0}^t \Upsilon_{ib}(t-r) {\cal P}(r;b,b)
\eeq
with ${\cal P}(0;i,b) =1$. For $i=b$ (\ref{second_result_reads}) takes the form of a renewal equation, involving
the defective PDF $\Upsilon_{bb}(t)$ (the mean recurrence time to $b$ is infinite, see relation (\ref{MFPT_MRW})). 
Consequently, the hitting process for $i=b$ is not a proper recurrent renewal process, but a defective one:
For infinite time there are only a finite number of hitting events of $b$, which we confirm hereafter (Eq. (\ref{exp_num_hits})).
We refer to \cite{dono_etal2024} for a thorough analysis of such `defective' renewal processes.
The first hitting PDF of the mortal walker 
$\Upsilon_{ib}(t)$ is defective, lacking proper normalization
\beq
\label{FHT_mortality}
\mathbb{P}[T>\Delta t_{ib}]= {\overline \Upsilon}_{ib}(1) =  \sum_{r=1}^{\infty} \chi_{ib}(r) \Phi_{\cal S}(r) < \sum_{r=1}^{\infty} \chi_{ib}(r) = 1 
\eeq
reflecting lack of recurrence and ergodicity of the MRW.
A consequence of the evanescence of the MRW is
\beq
\label{large_time_hits_mortal_walker}
{\cal H}_{ib}(\infty) = \lim_{u\to 1} (1-u) \frac{{\overline \Upsilon}_{ib}(u)}{1- {\overline \Upsilon}_{bb}(u)} = 0 .
\eeq
It follows that the expected number of hitting events of $b$ in the lifetime of the walker indeed is finite
\beq
\label{exp_num_hits}
\big\langle {\cal N}^{(*)}_i(\infty;b) \big\rangle = \lim_{u\to 1} (1-u) \frac{{\overline {\cal H}}_{ib}(u)}{1-u} ={\overline {\cal H}}_{ib}(1) =\frac{{\overline \Upsilon}_{ib}(1)}{1- {\overline \Upsilon}_{bb}(1)} < \infty ,
\eeq
confirming that for $i=b$ the hitting process is not a proper (recurrent) renewal process.
We obtain the finite expected lifetime of the walker from the survival GF (\ref{Survival_GF_result}) as
\beq
\label{mean_survival_time}
\langle {\cal T}_{\cal S}(i;b) \rangle = {\overline {\cal P}}(1;i,b) = {\overline \Pi}_i(1) + 
\frac{{\overline \Upsilon}_{ib}(1)}{1- {\overline \Upsilon}_{bb}(1)}{\overline \Pi}_b(1) =  
{\overline \Pi}_i(1) + \big\langle {\cal N}^{(*)}_i(\infty;b) \big\rangle \,  {\overline \Pi}_b(1) .
\eeq
The relations (\ref{exp_num_hits}), (\ref{mean_survival_time}) are consistent with the general Eqs. (\ref{finite_hitting_nums}), 
(\ref{expect_lllife_time_MRW_walker_arb_target}).

Finally, we are interested to connect these results with the propagator $\pmb{P}^{(*)}(t;b) = [P^{(*)}_{ij}(t;b)]$
which takes with (\ref{hitting_rel_MRW}) the form
\beq
\label{propagator_mortal_walker}
P^{(*)}_{ij}(t;b) = [{\widetilde {\pmb W}}^t]_{ij} \Phi_{\cal S}(t) + 
\sum_{r=1}^t {\cal H}_{ib}(r) [{\widetilde {\pmb W}}^{t-r}]_{bj} \Phi_{\cal S}(t-r) 
\eeq
with initial condition $ \pmb{P}^{(*)}(0;b) =\pmb{1}$. Considering $j=b$ retrieves
$P^{(*)}_{ib}(t;b) = \delta_{ib}\delta_{t,0} + {\cal H}_{ib}(t)$ ($t\in \mathbb{N}_0$).
In addition, from (\ref{thefollowing_important_MRW}) one has
\beq
\label{memory_rep}
  P^{(*)}_{ij}(t;b) = [{\widetilde {\pmb W}}^t]_{ij} \Phi_{\cal S}(t) + \sum_{r=1}^t \chi_{ib}(r)\Phi_{\cal S}(r)  P^{(*)}_{bj}(t-r;b) .
\eeq
For the departure node $i=b$ this relation takes the form of a renewal equation 
with defective return PDF $\Upsilon_{bb}(t) = \chi_{bb}(r)\Phi_{\cal S}(r)$ and for $j=b$ it leads to the MRW counterpart
of (\ref{NohRieger_relation}). The connection of the relations (\ref{propagator_mortal_walker}), (\ref{memory_rep}) with (\ref{result_reads})) and (\ref{second_result_reads}) is straightforward.

Having established the general MRW model, we consider next a prototypical example, exploring some pertinent aspects of the complex MRW dynamics.

\subsection{Generic example}
\label{generic_example}
Here our goal is to explore numerically some pertinent MRW characteristics such as the expected lifetime and the expected number of target hits, where we perform a complete analysis of the asymptotic cases.
The walker travels again with power-law degree-biased steps (\ref{preferential_steps}) on the network.
We control the frequency of the t-node hits with the bias parameter $\alpha$. We consider a target ${\cal B}$ consisting of highly connected nodes. Consequently, large values of $\alpha$ trigger frequent t-node hits, whereas negative $\alpha$ rare visits of t-nodes. 
We specify distribution (\ref{walker_survival_T}) as follows
\beq
\label{adva}
\Phi_{\cal S}(t) =  \Phi_{\cal S}(t; Q_1,Q_2) =   Q_1^t\Theta(t_0-1-t) +Q_1^{t_0}Q_2^{t-t_0} \Theta(t-t_0)  = 
\left\{\begin{array}{clr} Q_1^t \, , & t \leq t_0 & \\[2ex]
        Q_1^{t_0} Q_2^{t-t_0} \, ,& t > t_0 \end{array} \right. , \hspace{0.25cm} Q_i \in [0,1]
\eeq
with some positive integer $t_0 >1$. We will need its GF
\beq
\label{adva_GF}
{\overline \Phi}_{\cal S}(u) =  {\overline \Phi}_{\cal S}(u; Q_1,Q_2) =\frac{1-(Q_1u)^{t_0}}{1-Q_1u} +  \frac{(Q_1u)^{t_0}}{1-Q_2u} .
\eeq
The $\alpha_t$ defined in (\ref{represPhiS})
are then given by
\beq
\label{verify_III}
\alpha_t = 1-Q_1 + \Theta(t-t_0-1)(Q_1-Q_2) = \left\{\begin{array}{clr}  1-Q_1 = P_1\, , & t \leq t_0 & \\[2ex]
1-Q_2 =P_2 \, ,& t > t_0 \end{array} \right. .
\eeq
The distribution (\ref{adva}) covers all previously mentioned scenarios:
For $Q_2 < Q_1$ (forager's scenario) conditions (\ref{condition_ineq}), (\ref{advant}) are holding true. 
Frequent budget renewals (t-node hits) should extend the walker's lifespan. The forager's scenario contains the deterministic case $T=t_0+1$ for $Q_2=0$, $Q_1=1$ with $\Phi_{\cal S}(t;1,0)= \Theta(t_0-t)$ and
$\psi(t)=\delta_{t,t_0+1}$.

On the other hand, for $Q_2 > Q_1$ frequent budget renewals should be detrimental, reducing 
the walker's expectancy of life.
For $Q_2=Q_1$ (\ref{adva}) is a geometric distribution
$\Phi_{\cal S}(t) =Q_1^t$ where budget renewals occurring for $t>0$ should not affect the expected lifetime of the walker. 

\begin{figure}[t!]
\centerline{
\includegraphics[width=0.7\textwidth]{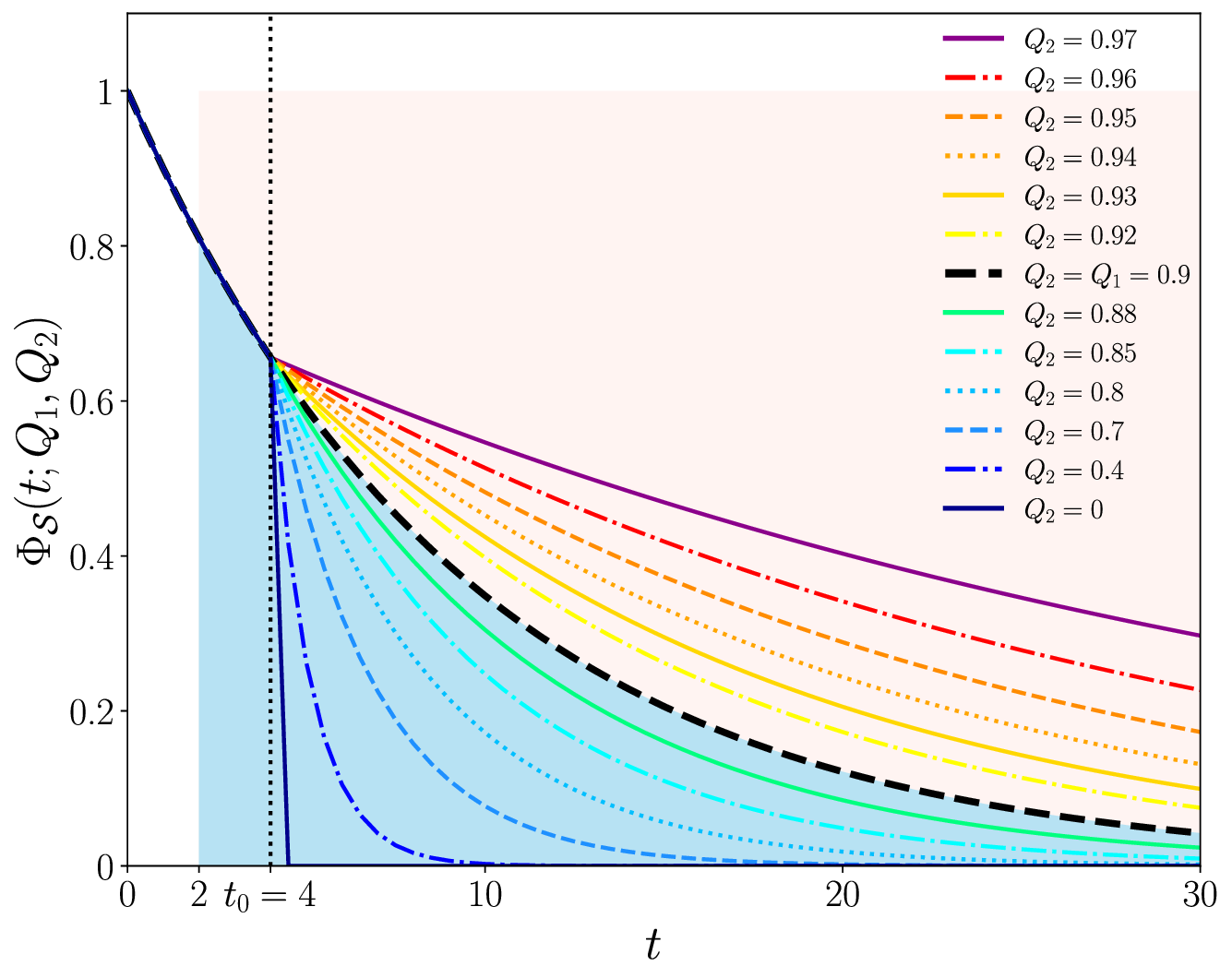}
}
\caption{Time dependence of $\Phi_{\cal S}(t; Q_1,Q_2)$ (Eq. (\ref{adva})) for some values of $Q_2$ with fixed $Q_1=0.9$ and $t_0=4$. The blue shaded region corresponds to forager's scenarios ($\Phi_{\cal S}(t) < 
(\Phi_{\cal S}(1))^t$), the red shaded region corresponds to detrimental scenarios. Both scenarios are separated by the neutral scenario (black dashed line).}
\label{FigPhiS}
\end{figure}

We depict the time dependence of distribution $\Phi_{\cal S}(t; Q_1,Q_2)$ in Fig. \ref{FigPhiS}. 
The curves in the blue shaded region below the black dashed line refer to $Q_2 < Q_1$ (forager's scenario) and the curves in the red shaded region correspond to situations where frequent budget renewals are detrimental. The black dashed curve ($Q_1=Q_2$) indicates the geometric distribution of $T$ of the neutral scenario. 
%
%
the scenarios (shaded regions) of the distributions $\Phi_{\cal S}(t; Q_1,Q_2)$ in Fig. \ref{FigPhiS} starts from $t=2$, since 
the comparison of the decay of $\Phi_{\cal S}(t)$ with the geometric fall-off $[\Phi_{\cal S}(1)]^t$ requires at least two time increments.
%
%

%
%
\begin{figure}[t!]
\centerline{
\includegraphics[width=0.85\textwidth]{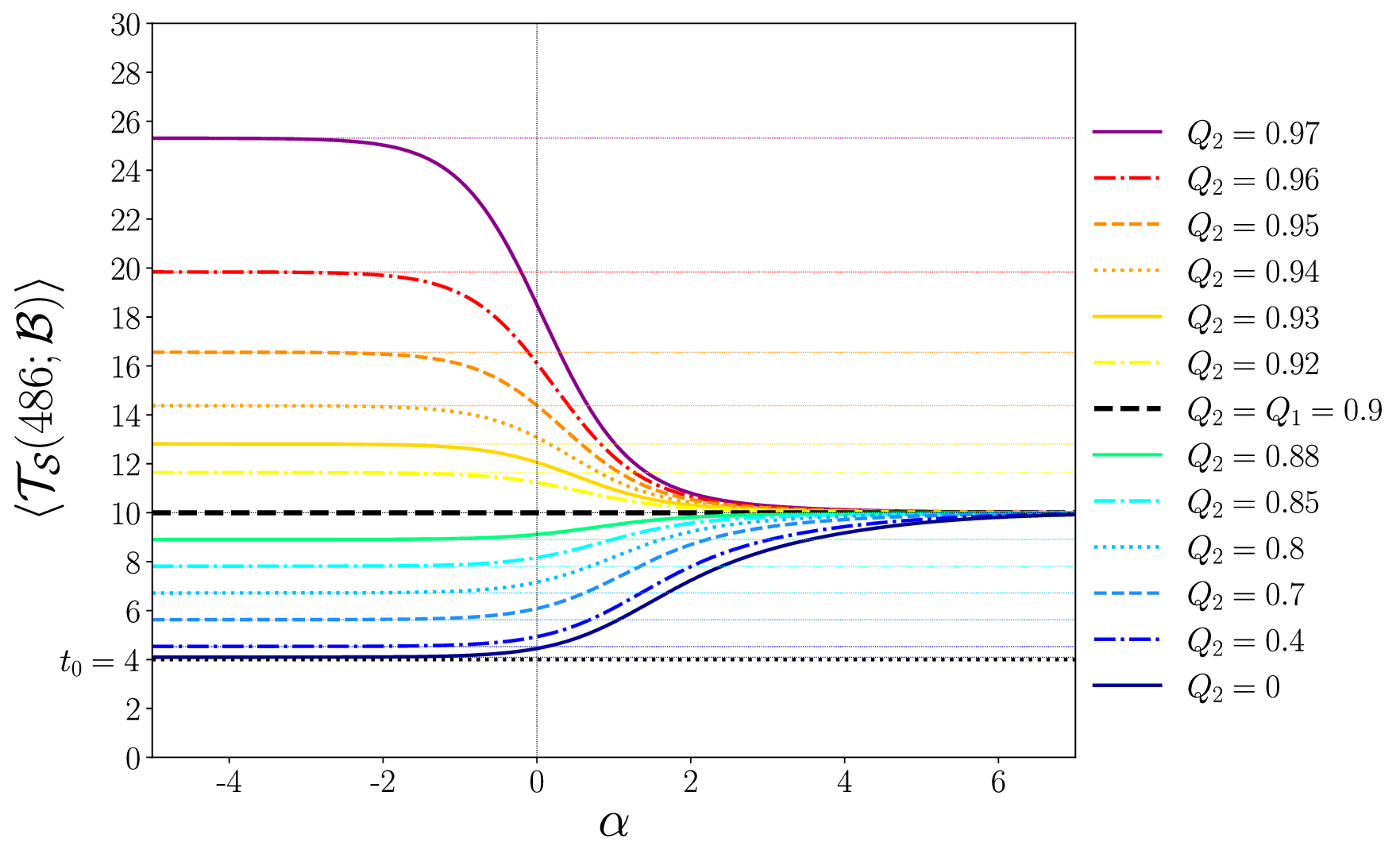}
}
\vspace{-3mm}
\caption{Expected lifetime $\big\langle \mathcal{T}_{\cal S}(486;{\cal B}) \big\rangle$ (Eq.   
(\ref{expect_lllife_time_MRW_walker_arb_target})) versus $\alpha$ for 
same parameters and color codes as in Fig. \ref{FigPhiS} with fixed values $Q_1=0.9$ and $t_0=4$ (see (\ref{adva})). We consider the same BA graph and departure node $486$ as in Figs. \ref{fig3}, \ref{fig2}.
The target ${\cal B}$ is highly connected with average degree $95.9$ and contains the distinguished highly connected nodes ${\hat j}$ and ${\hat n}$ (see main text).
For $Q_2 < Q_1$ the expected lifetime increases monotonously with $\alpha$
corresponding to the forager's scenario (blue curves). 
For $Q_2> Q_1$ the lifetime decreases monotonously with $\alpha$ (detrimental scenario). The neutral scenario refers to $Q_1=Q_2$ (black dashed line).}
\label{MRW_fig4}
\end{figure}

\begin{figure}[t]
\centerline{
\includegraphics[width=0.7\textwidth]{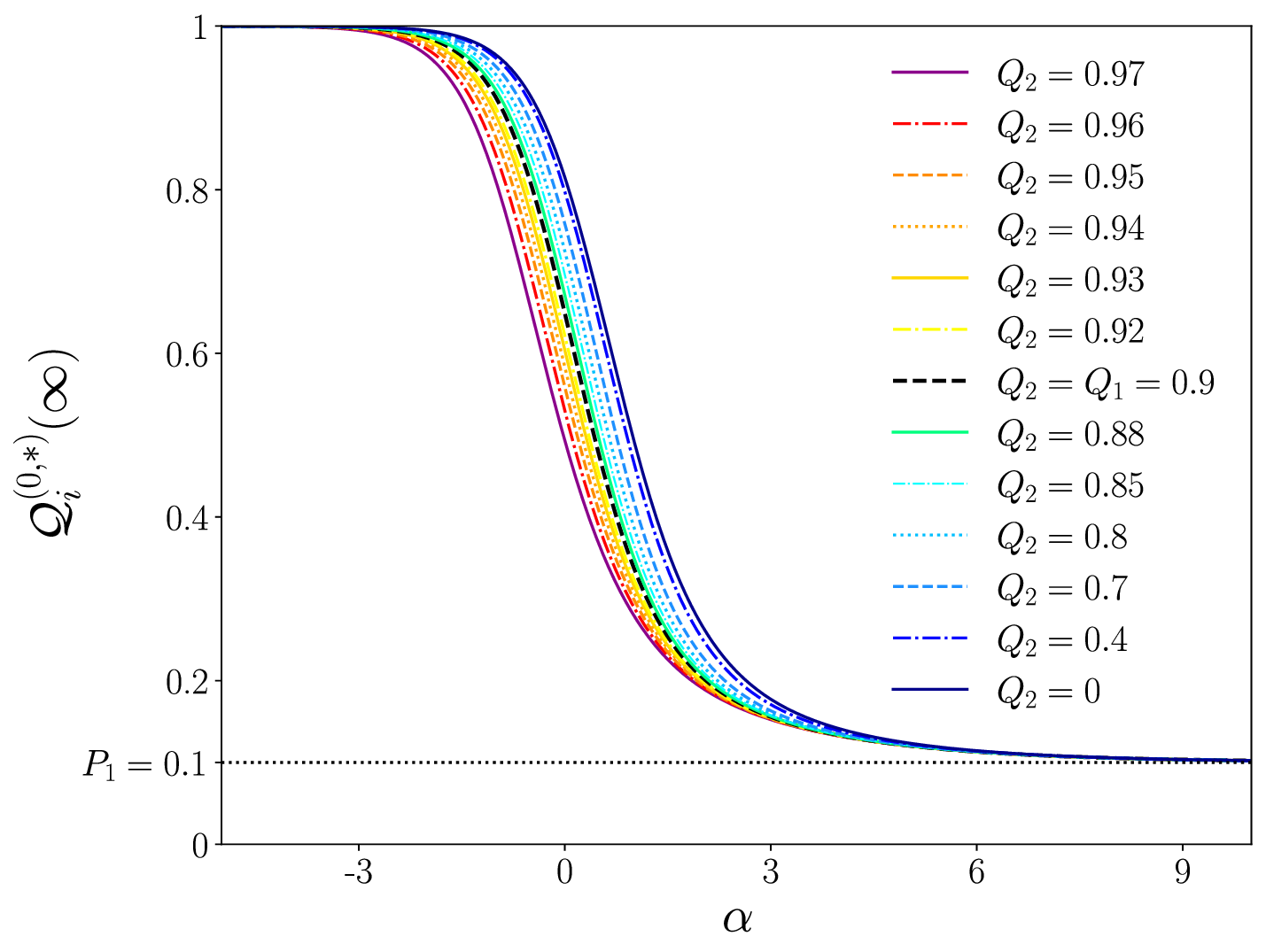}
}
\vspace{-3mm}
\caption{Probability $[\pmb{\mathcal{Q}}^{(0,*)}(\infty)]_{486}$ that the walker never hits the target (Eq.   
(\ref{saturated-stae_probas})) versus bias parameter $\alpha$. We use for all curves the same setting, color code, and target as in Fig. \ref{MRW_fig4} (keeping $Q_1=0.9, t_0=4$ fixed). For $\alpha$ fixed, the probability of never hitting ${\cal B}$ decreases monotonously with $Q_2$ and the expected lifetime, see Fig. \ref{MRW_fig4}: The longer the walker lives, the smaller is the probability $[\pmb{\mathcal{Q}}^{(0,*)}(\infty)]_{486}$ of never visiting the target.
For $\alpha \to \infty$ the asymptotic value $P_1=0.1$ is taken (see main text). 
}
\label{MRW_fig6b}
\end{figure}

\begin{figure}[t!]
\centerline{
\includegraphics[width=0.75\textwidth]{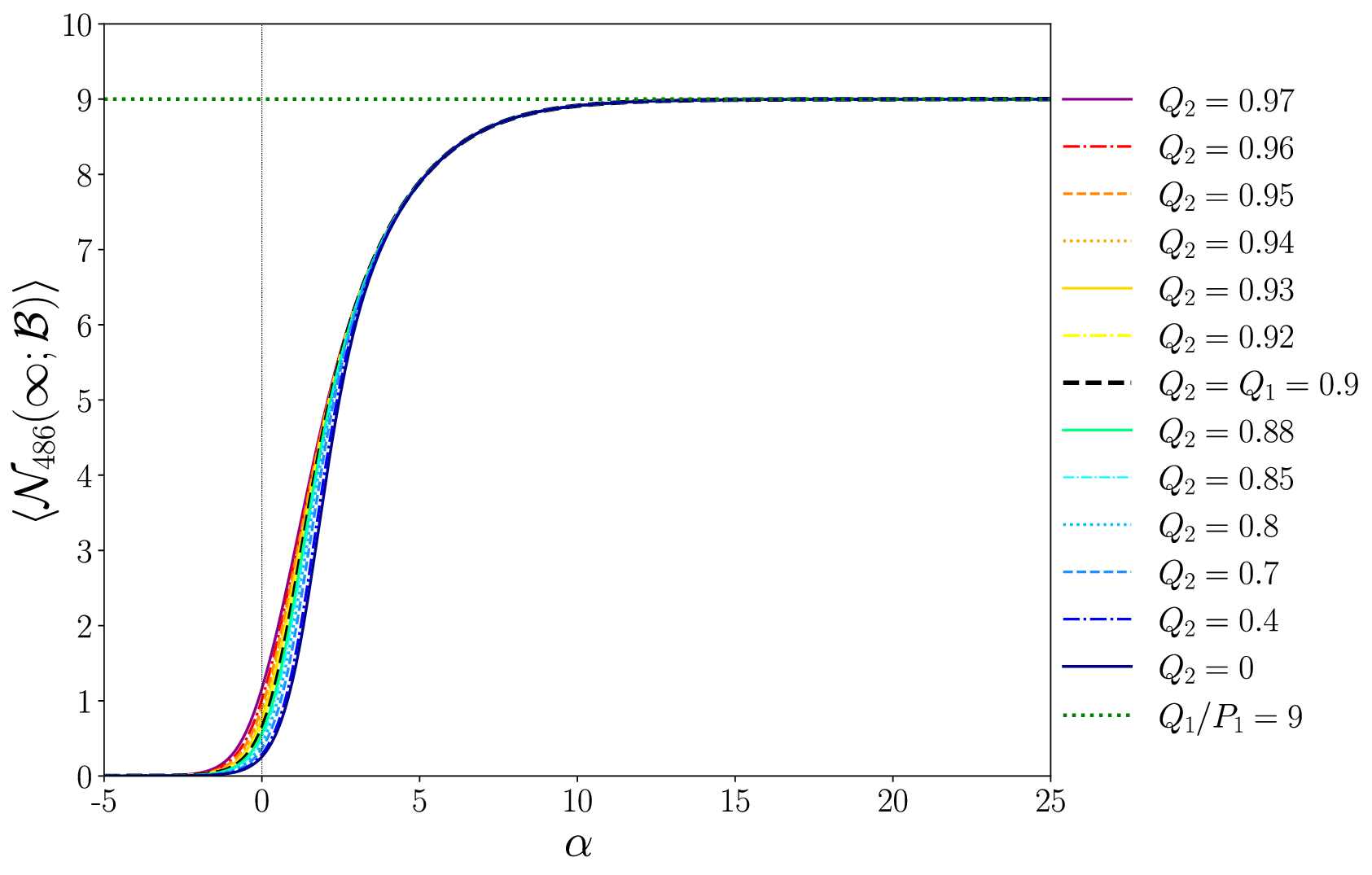}
}
\caption{$\big\langle \mathcal{N}_{486}(\infty;{\cal B}) \big\rangle$ of Eq.   
(\ref{finite_hitting_nums}) versus $\alpha$ for the same setting, target and color code as in Fig. \ref{MRW_fig4}. 
}
\label{MRW_fig6}
\end{figure}

\begin{figure}[H]
\centerline{
\includegraphics[width=0.75\textwidth]{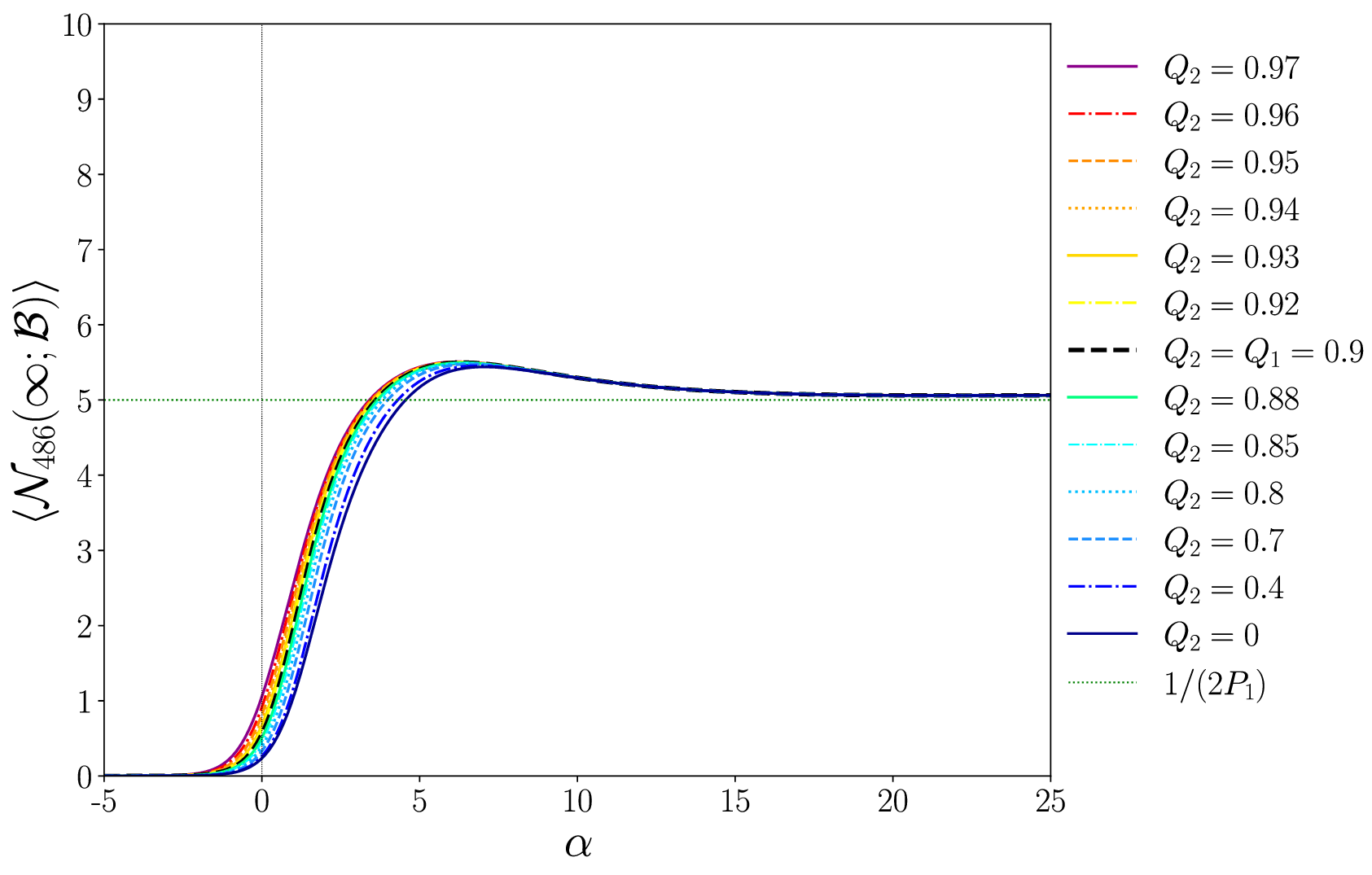}
}
\caption{$\big\langle \mathcal{N}_{486}(\infty;{\cal B}) \big\rangle$ of Eq.   
(\ref{finite_hitting_nums}) versus $\alpha$ for some values of $Q_2$ 
with $Q_1=0.9$ and $t_0=4$ for the same setting and target as in Figs \ref{fig3}, \ref{fig2} (${\hat j} \in {\cal B}$, ${\hat n} \notin {\cal B}$) and color code of Fig. \ref{MRW_fig6}.
}
\label{MRW_fig5}
\end{figure}

We investigate numerically the survival statistics resulting from the complex random walk dynamics in Figs. \ref{MRW_fig4} - \ref{MRW_fig5} for the same BA graph 
and the same departure node $i=486$ as in Figs. \ref{fig3}, \ref{fig2} for a highly connected target specified subsequently.
The expected lifetime (\ref{expect_lllife_time_MRW_walker_arb_target}) and the number of target visits (\ref{finite_hitting_nums}) 
are completely determined by the matrix GFs $\overline{\pmb{\Pi}}(u) = {\overline \Phi}_{\cal S}(u \widetilde{\pmb{W}})$
and $\overline{\pmb{\Upsilon}}(u) = u\overline{G}_{\cal S}(u \widetilde{\pmb{W}}) \cdot \pmb{W}^{(\cal B)}$ (Eqs. (\ref{Pi_GF_MRW}), (\ref{GF_upsilon_arb_tar_MRW})) where here
\beq
\label{GS_function}
u\overline{G}_{\cal S}(u v) = \sum_{t=1}^{\infty} u^t v^{t-1} \Phi_{\cal S}(t) = \frac{u Q_1(1-(uvQ_1)^{t_0})}{1- uvQ_1} +\frac{uQ_2(Q_1uv)^{t_0}}{1-u v Q_2} .
\eeq
\paragraph{Discussion}
The plot of Fig. \ref{MRW_fig4} shows a numerical evaluation of the expected lifetime 
$\big\langle \mathcal{T}_{\cal S}(486;{\cal B}) \big\rangle$ as a function of bias parameter $\alpha$. We consider the same cases
for $\Phi_{\cal S}(t)$ and color codes as in Fig. \ref{FigPhiS}.
For a fixed $\alpha$, the expected lifetime increases
monotonously with $Q_2$. 
For any $Q_2< Q_1$ (forager's scenario) the expected lifetime increases monotonously with $\alpha$ (blue curves, below the black dashed line). The smaller $Q_2$, the more pronounced is the increase of the expected lifetime with $\alpha$, i.e. the more beneficial frequent t-nodes hits are.
These numerical results corroborate impressively our prediction that frequent t-node hits extend the walker's lifetime in forager's scenarios.

On the other hand, for $Q_2 > Q_1$ (reddish curves above the black dashed line) the expected lifetime is decreasing monotonously 
with $\alpha$, confirming that frequent budget renewals are here indeed detrimental. The decrease is the more pronounced the larger
$Q_2$. 
Moreover, the numerical evaluation also shows
that the case $Q_1=Q_2$ is neutral, where the expected lifetime is independent of $\alpha$ (horizontal dashed line) for which one has
\beq
\label{Q1_equal_Q2}
\langle \mathcal{T}_{\cal S}(i;{\cal B}\rangle = \overline{\Phi}_{\cal S}(1) = \langle T \rangle  = 10 = \frac{1}{P_1} ,
\hspace{0.5cm} \forall \alpha  , \hspace{0.5cm} P_1+Q_1=1 ,
\eeq
being in perfect agreement with Eq. (\ref{exp_ber}). These results again are an impressive confirmation of our predictions.
The value $1/P_1$ is also the common asymptotics for $\alpha \to \infty$ for all $Q_2$,
where the survival dynamics converges to a Bernoulli trial process,
which corresponds perfectly to case {\bf (a)} of Sect. \ref{arbitrary_target_MRW} with Eq. (\ref{lifetime_all_nodes}).
Indeed we will prove analytically a little later that for the choice of the target of Fig. \ref{MRW_fig4} 
the walker navigates uniquely on t-nodes when $\alpha \to \infty$.

In Fig. \ref{MRW_fig6b} we explore numerically the state probability $[\pmb{\mathcal{Q}}^{(0,*)}(\infty)]_{486}$ (the probability that the walker never hits ${\cal B}$ in a lifetime, Eq. (\ref{saturated-stae_probas})).
For $\alpha \to -\infty$ the highly connected target is not accessible with $[\pmb{\mathcal{Q}}^{(0,*)}(\infty)]_{486} \to 1$.
In the limit $\alpha \to \infty$ every step is a t-node hit, the numerical evaluation yields
$[\pmb{\mathcal{Q}}^{(0,*)}(\infty)]_{486} \to P_1 >0$ agreeing perfectly with the prediction of Eq. (\ref{Q_zero_star_all_target}) referring to
case {\bf (a)} of Sect. \ref{arbitrary_target_MRW}. Let us analyze these limiting cases more closely.
\\[3ex]
\noindent {\it\large Limit $\alpha \to \infty$: Each step is a target hit if two specific nodes are in ${\cal B}$}
\\ [1ex]
In this limit, the walker navigates uniquely on a set of highly connected nodes specified hereafter. We will see a little later analytically and numerically that if these nodes are included into the target,
each step is a target hit, thus the case {\bf (a)} in Sect. \ref{arbitrary_target_MRW} applies.
In the BA graph considered in our numerical evaluations, two specific highly connected nodes stand out, which we denote as ${\hat j}$ and ${\hat n}$.
${\hat j}=8$ is the only node with the highest degree $K_8=141$ of that BA network.
Among the neighbor nodes of ${\hat j}$ the node ${\hat n}$ has the largest degree of all neighbor nodes of ${\hat j}$.
We identified a single node with this feature, namely ${\hat n}= 0$ with degree $K_0=135$. We point out that in the Python NetworkX library, the $N$ nodes of a network are labeled from $0$ to $N-1$.
Let us retain for the following discussion that in Figs. \ref{MRW_fig4}, \ref{MRW_fig6b} and \ref{MRW_fig6} both of these nodes are included into the target
(${\hat j}, {\hat n} \in {\cal B}$), whereas in Fig. \ref{MRW_fig5} ${\hat j} \in {\cal B}$
and ${\hat n} \notin {\cal B}$. This target is identical as considered in Figs. \ref{fig3}, \ref{fig2} and visualized there (cyan colored nodes).
Since we are interested in the large time asymptotics, we assume that walker behaves stationary
$W_{ij}(\alpha) = W_j(\alpha) = \frac{\mathcal{A}_j(\alpha)}{\mathcal{B}(\alpha)}$, see (\ref{stationary_dist}).
For large $\alpha$ the normalization factor ${\cal B}(\alpha)$ behaves as (see (\ref{definiteion_A}))\footnote{We denote with 
"$\sim$" asymptotic equality.}
\beq
\label{B_alpha_large}
\mathcal{B}(\alpha) \sim  2 K_{\hat j}^{\alpha} \sum_{n \in {\cal Z}_{\hat j}} K_n^{\alpha} = 2 \mathcal{A}_{\hat j} , \hspace{1cm} (\alpha \to \infty)
\eeq
scaling with $K_{\hat j}^{\alpha}$
where ${\cal Z}_{\hat j}$ denotes the set of neighbor nodes of ${\hat j}$. 
Then only two categories of nodes exist for which $W_j(\alpha)$ remain non-null, namely
$$W_{\hat j} \sim \frac{\mathcal{A}_{\hat j}}{2\mathcal{A}_{\hat j}} = \frac{1}{2}  ,  \hspace{1cm} (\alpha \to \infty),  $$
and 
$$W_{\hat n} \sim \frac{K_n^{\alpha} K_{\hat j}^{\alpha}}{2 K_{\hat j}^{\alpha} \sum_{n \in {\cal Z}_{\hat j}} K_n^{\alpha}} =
\frac{1}{2} \frac{K_{\hat n}^{\alpha}}{K_{\hat n}^{\alpha} + \sum_{m \in {\cal Z}_{\hat j} (m\neq {\hat n})} K_m^{\alpha} } \to \frac{1}{2}  , \hspace{1cm} (\alpha \to \infty). $$ 
All other nodes $W_j(\alpha) \to 0$ ($j \neq {\hat j}, {\hat n}$) become inaccessible. Hence, the stationary transition matrix reads 
\beq
\label{alpha_infty_transition_matrix} W_j(\infty) \sim  \frac{1}{2}\left( \delta_{j,{\hat j}} + \delta_{j,{\hat n}} \right).
\eeq
The walker is located with equal probability on nodes ${\hat j}$ and ${\hat n}$, which we checked also numerically.
Consider the situation in which
${\hat j}, {\hat n} \in {\cal B}$, where the walker navigates for $\alpha\to \infty$ exclusively on these two t-nodes. This situation is hence captured by case {\bf (a)} of Sect. \ref{arbitrary_target_MRW}, leading to (see (\ref{lifetime_all_nodes}))
\beq
\label{heuristic_der}
\big\langle \mathcal{T}_{\cal S}(i;{\cal B}) \big\rangle \sim \sum_{r=0}^{\infty} \left[ [\overline{\pmb{\Upsilon}}(1)]^r \cdot \overline{\pmb{\Pi}}(1) \right]_i = \frac{1}{1-Q_1} [\pmb{1}]_i = \frac{1}{P_1} ,\hspace{1cm} (\alpha \to \infty) 
\eeq
with $P_1=0.1$, being in perfect accordance with the asymptotic value obtained numerically for all cases of $Q_2$
in Fig. \ref{MRW_fig4}.
\\[2mm]
In the same way, we obtain for
the expected number of t-node visits (budget renewals) in the lifetime of the walker (see Eq. (\ref{infinite_time})) 
\beq
\label{heuristic_N_star}
\big\langle \mathcal{N}_i^{(*)}(\infty;{\cal B}) \big\rangle \sim
 \sum_{r=0}^{\infty} \left[ [\overline{\pmb{\Upsilon}}(1)]^r \cdot [\overline{\pmb{\Upsilon}}(1) \right]_i = \frac{Q_1}{P_1} ,\hspace{1cm} (\alpha \to \infty) ,
\eeq
which is in perfect agreement with the asymptotics observed numerically for all cases in Fig. \ref{MRW_fig6}. 
\\[2mm]
The asymptotic behavior of $\big\langle \mathcal{N}_{486}^{(*)}(\infty;{\cal B}) \big\rangle$ changes drastically, if one of the nodes ${\hat j}, {\hat n}$ is not in the target.
This situation is depicted in Fig. \ref{MRW_fig5}, where ${\hat j} =8 \in {\cal B}$, ${\hat n} =0 \notin {\cal B}$. The infinite time limit $\big\langle \mathcal{N}_{486}^{(*)}(\infty;{\cal B}) \big\rangle$ decreases slightly for large $\alpha$ after having reached a maximum at some value 
$\alpha_0 \approx 5$. 
We explain this non-monotonic behavior by the fact that the walker for $\alpha \to \infty$
is located on ${\hat j}$ and ${\hat n}$ with equal probability. However, only ${\hat j}$ is a t-node thus approximately only every second step is a t-node
hit in the walker's life. So we expect roughly $\big\langle \mathcal{N}_{486}^{(*)}(\infty;{\cal B}) \big\rangle \sim 
\frac{1}{2P_1}= 5$, i.e. half of the expected lifetime when ${\hat j}, {\hat n} \in {\cal B}$ as in Fig. \ref{MRW_fig4}. This prediction agrees impressively with the asymptotics observed in Fig. \ref{MRW_fig5} (green dashed line).
\\[2ex]
\noindent {\it\large Limit $\alpha \to -\infty$: Dynamics without target hits}
\\[1ex]
In this limit, highly connected nodes including the t-nodes are not accessible.
Consequently, the settings of case {\bf (b)} in Sect. \ref{arbitrary_target_MRW} apply. The expected lifetime of the walker is then given by (Eq. (\ref{expeted_emptyset})) 
$$ \big\langle \mathcal{T}_{\cal S}(i;{\cal B}) \big\rangle \sim {\overline \Phi}_{\cal S}(1; Q_1,Q_2)  = \big\langle T \big\rangle , $$
which is in perfect agreement with the asymptotics obtained numerically for $\alpha \to -\infty$, see Fig. \ref{MRW_fig4} (colored dotted lines). 
We also observe numerically in Figs. \ref{MRW_fig6}, \ref{MRW_fig5} that
$\big\langle  \mathcal{N}_i^{(*)}(\infty;{\cal B}) \big\rangle \to 0$ whenever $\alpha$ is sufficiently small.
\section{Conclusions and future directions}
\label{conclusions}
The first part of this work was devoted to the analysis of the {\it target hitting counting process} (THCP) for an ergodic immortal Markov walker and its connection with first-passage statistics.
Our formulation generalizes well-known results for a single target node to an arbitrary set of target nodes, providing explicit expressions for the first-hitting quantities in terms of defective matrices that encode the target information [see Eq. (\ref{transition_mat_mod})].
For the special case of a single target node, we showed that the THCP reduces to a renewal counting process when the departure node coincides with the target. In the large time limit, the THCP converges to a Markovian Bernoulli counting process [see Eqs. (\ref{MHR_large_time})--(\ref{large_time_THCP_asymptotics})]. These theoretical results were corroborated numerically by considering degree-biased random walks,
allowing for both preferential and anti-preferential visits to targets with specific connectivity.
The proposed framework is general and can be extended to non-Markovian random walks on networks. In particular, the discrete-time Markov dynamics can be subordinated to a continuous-time renewal process, naturally embedding the model into the continuous-time random walk (CTRW) framework \cite{MontrollWeiss1965,Scher_Montroll1975,Shlesinger2017}.
\\[2mm]
In the second part of this work, we investigated the dynamics of an evanescent random walker, referred to as the \emph{mortal random walk} (MRW).  The survival of the walker depends on the positivity of a budget variable, which evolves under the competing effects of budget renewals at target-hitting events and budget costs incurred during inter-hitting intervals. For the stochastic process governing the budget evolution [Eq. (\ref{budget})], we highlighted connections with stochastic resetting \cite{Mujamda2011}. By construction, the MRW is non-recurrent and lacks ergodicity, except for the distinguished situation described in case (a) of Sec. \ref{arbitrary_target_MRW}. We derived the probability distributions that characterize the dynamics and survival statistics of the MRW, including the \emph{evanescent propagator} matrix, which encodes both the survival probability and the expected lifetime of the walker. In addition, we introduced a non-evanescent \emph{auxiliary propagator} [see Eq. (\ref{moment_GF})] that captures the target-hitting statistics, such as state occupation probabilities and the expected number of target visits during the walker's lifetime. We obtained the infinite-time limits of these quantities [see Eq. (\ref{dead_walkers_THCP}) and subsequent relations]. A hallmark of walker evanescence is that the mean first-passage time (MFPT) to any target is infinite [Eq. (\ref{MFPT_MRW})]. We also examined specific scenarios exhibiting distinct survival behaviors. In one extreme case, every node of the network is a target node, reducing the survival process to a Bernoulli trial sequence. More generally, we identified three representative MRW scenarios in which the frequency of target hits exerts qualitatively different influences on the lifetime of the walker  (Sec. \ref{generic_example}).
\\[2mm]
A broad class of stochastic processes can be defined by the survival statistics introduced in Eqs. (\ref{survival_indicator_function})--(\ref{survival_proba_MRW}), in which budget renewals follow a proper renewal process. Interesting extensions of the MRW framework arise when modifying the budget dynamics [Eq. (\ref{budget})]. For instance, in random walks with long-range steps such as L\'evy flights \cite{MetzlerKlafter2000}, each step could incur a cost that depends on its length. Another variant is obtained when the budget decreases at a constant rate and is increased by independent and identically distributed (IID) increments occurring at the arrival times of a renewal process \cite{Burunev_Mujamdar2025}.
\\[2mm]
A further challenging problem concerns the survival statistics of a MRW subjected to stochastic resetting \cite{Radice2023}. Resetting events may incur larger costs than ordinary steps, yet they can also shorten inter-hitting times, thereby altering the survival statistics of the walker. In this context, it would be of interest to determine whether an optimal resetting rate exists that maximizes the life expectancy of the walker. Relatedly, the problem of a mortal random searcher under resetting presents a promising direction  \cite{Pal-Sandev2023}. Other extensions of the MRW arise in foraging scenarios with depletion and regeneration of resources \cite{Chupeau_Beni2016,Chupeau_eta2016}.
\\[2mm]
Potential applications and extensions of the MRW framework are diverse, spanning domains such as finance, gambling, population dynamics, and epidemic spreading \cite{Pastor-SatorrasVespignani2001,Granger-et-al2023,SISI_Entropy_2024}; the propagation of wildfires, where the ``lifetime'' of the fire depends on encountering advantageous conditions; chemical reactions in which reactants may be absorbed; and biological processes involving agents with limited lifetimes, such as protein degradation in gene networks \cite{Zhang-etal2025}, among many others.

\section{Acknowledgement}
T.M.M. gratefully acknowledges fruitful discussions with Trifce Sandev. We thank two anonymous reviewers for their valuable comments, helping us to substantially improve the presentation.
\hfill
\appendix
\setcounter{equation}{0}
\renewcommand{\theequation}{A\arabic{equation}}
\section{Appendix and supplementary materials}

\subsection{Spectral features of degree biased random walks}
\label{degree_biased}
We recall some basic features of degree biased random walks. 
To that end, rewrite (\ref{preferential_steps}) in matrix form
\beq
\label{mat_rep_W}
\pmb{W} = \pmb{{\cal D}}_{\cal A}^{-1} \cdot \pmb{{\cal A}} = (\pmb{{\cal D}}_{\cal A})^{-\frac{1}{2}}\cdot 
\left[(\pmb{{\cal D}}_{\cal A})^{-\frac{1}{2}} \cdot \pmb{{\cal A}} \cdot (\pmb{{\cal D}}_{\cal A})^{-\frac{1}{2}}\right]\cdot (\pmb{{\cal D}}_{\cal A})^{\frac{1}{2}} .
\eeq
We skip in all notations the dependence of $\alpha$.
Here we have introduced $\pmb{{\cal A}} = [{\cal A}_{ij}]$ and
the diagonal matrix $\pmb{{\cal D}}_{\cal A} = [\delta_{ij}{\cal A}_i]$ with
\beq
\label{definiteion_A}
\begin{array}{clr}
\ds {\cal A}_{ij} & =\ds  A_{ij}f_if_j ,\hspace{1cm} f_j = K_j^{\alpha} & \\ \\
 \ds {\cal A}_i  & = \ds \sum_{j=1}^N {\cal A}_{ij} & \\ \\
 \ds {\cal B} & = \ds  \sum_{j=1}^N {\cal A}_j .&
 \end{array}
\eeq
Important is that the matrix in the brackets in (\ref{mat_rep_W})
$\pmb{{\cal S}} =[{\cal S}_{ij}]$ is symmetric (Hermitian) and has uniquely real eigenvalues with the canonical representation
\beq
\label{spectral_W_bias}
{\cal S}_{ij} = \frac{{\cal A}_{ij}}{\sqrt{{\cal A}_i{\cal A}_j}}  =\sum_{k=1}^N \lambda_k \langle i|\varphi_k\rangle\langle \varphi_k|j\rangle ,
\eeq
in which $\langle \varphi_k|$ stand for the left eigenvectors and $|\varphi_k\rangle$ for the adjoint (transposed, complex conjugated) right eigenvectors fulfilling
$\delta_{ij} =\sum_{k=1}^N \langle i|\varphi_k\rangle\langle \varphi_k|j\rangle$ and $\langle\varphi_k |\varphi_m\rangle =\delta_{km}$.
The matrix $\pmb{{\cal S}}$ has the same eigenvalues $\lambda_k$
as the transition matrix (\ref{mat_rep_W}) with $\lambda_1=1$ and $|\lambda_m|<1$ ($m=2,\ldots,N$). 
This takes us to the spectral representation of the transition matrix (\ref{mat_rep_W}), namely \cite{Riascos_Mateos2021}
\beq
\label{canonic_W}
W_{ij} = {\cal S}_{ij}\sqrt{\frac{{\cal A}_j}{{\cal A}_i}} =\sum_{k=1}^N \lambda_k \frac{\langle i|
\varphi_k\rangle}{\sqrt{{\cal A}_i}} \langle \varphi_k|j\rangle \sqrt{{\cal A}_j} = 
\sum_{k=1}^N\lambda_k |\phi_k\rangle\langle{\overline \phi}_k| ,
\eeq
which maintains the same type of spectral structure as in the unbiased case (Eq. (\ref{purely_markovian})).
The unbiased standard walk is contained by construction for $\alpha=0$ (where ${\cal A}_i = K_i$ as $f_j=1$). We also can readily retrieve the stationary distribution. Noting that $\pmb{W}|\phi_1\rangle = |\phi_1\rangle$ and choosing the normalization such that the constant right eigenvector $\langle i|\phi_1\rangle =1$ to the
Perron-Frobenius eigenvalue $\lambda_1=1$, leads to the components of the left eigenvector $\langle {\overline \phi}_1|j\rangle = {\cal A}_j/{\cal B}$ taking us to the stationary distribution for the degree-biased walk
\beq
\label{stationary_dist}
P_{ij}(\infty)= W_j^{(\infty)} = \lim_{t\to \infty} [\pmb{W}^t]_{ij} =
\langle i|\phi_1\rangle \langle {\overline \phi}_1|j\rangle  =  {\cal A}_j/{\cal B} ,
\eeq
which is independent of the departure node $i$ of the walker and with $ W_j^{(\infty)} >0$ (per construction no disconnected nodes exist)
reflecting ergodicity of the degree biased Markov walk. This result can be easily reconfirmed by the observation 
$${\cal A}_i[\pmb{W}^t]_{ij} ={\cal A}_j[\pmb{W}^t]_{ji} $$ and using $[\pmb{W}^t]_{ij} \to  W_j^{(\infty)}$ 
(loss of memory of the departure node for $t \to \infty$).

\begin{figure}[t!]
\centerline{
\includegraphics[width=1.0\textwidth]{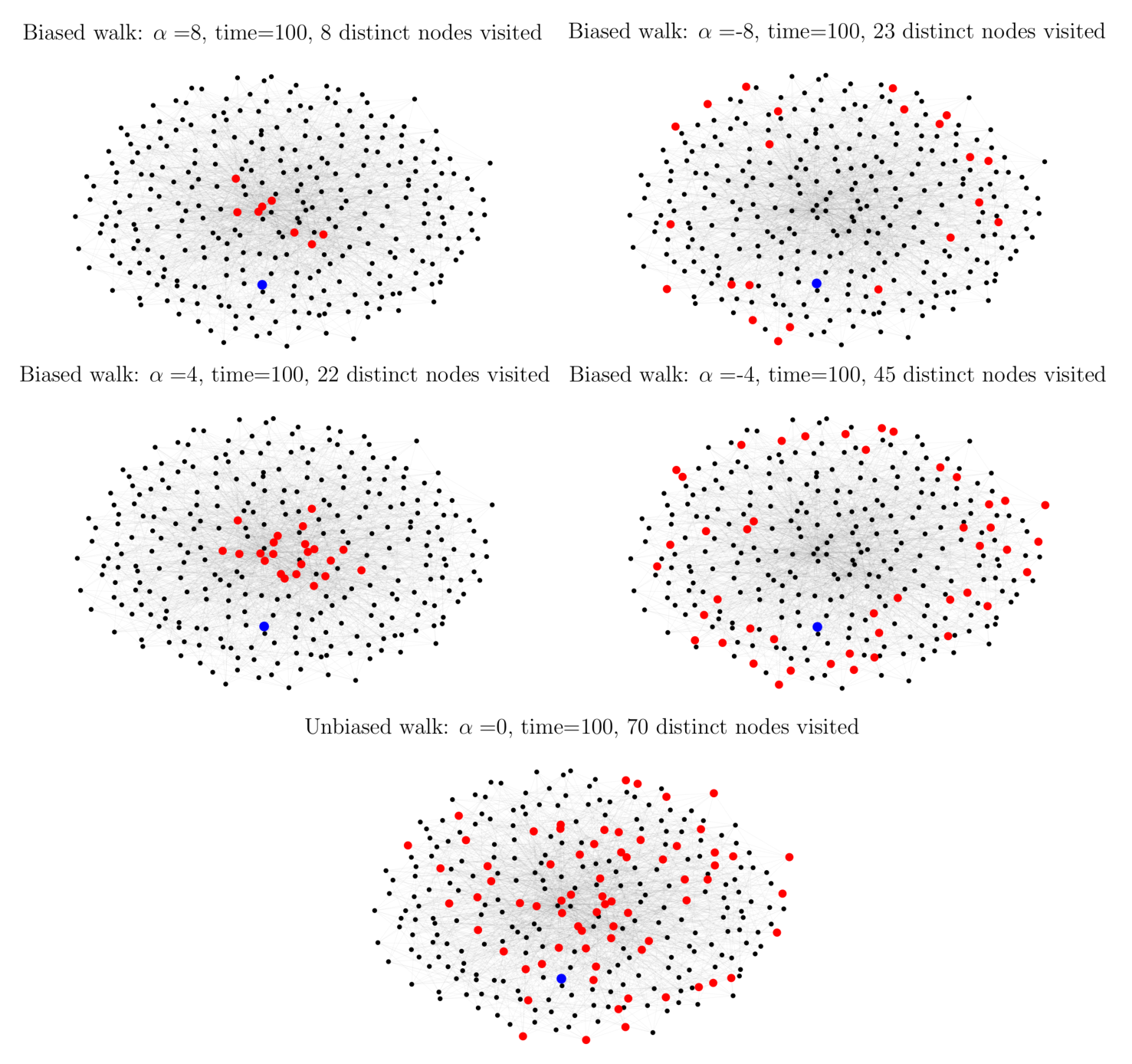}
}
\vspace{-2mm}
\caption{Distinct nodes visited (red) up to the runtime $100$ (departure node blue) in a power-law degree-biased Markov walk for some values of $\alpha$. We consider a Barabási-Albert network with $300$ nodes generated using the Python NetworkX library:  $nx.barabasi\_albert\_graph(300, m=7, seed=0)$.
For $\alpha < 0$ the visited nodes have small degrees  and are peripherally located. For $\alpha >0$ highly connected nodes located around the center are preferentially visited. $\alpha=0$ corresponds to the unbiased walk in which approximately the largest number of distinct nodes are visited.}
\label{fig1}
\end{figure}

We depict in red the set of distinct nodes visited during runtime $100$ in a power-law degree-biased walk (\ref{preferential_steps}) in Fig. \ref{fig1}. One can clearly see that for $\alpha =8$ nodes with high degree are preferentially visited, which are represented in the center of the network. On the other hand, for $\alpha =-8$ nodes with small degrees 
are preferentially visited, represented in the
periphery of the graph. 
The plots also show that the smaller the absolute value of $\alpha$ the larger is the number of distinct nodes which are visited for a certain runtime, where this number is here maximal for $\alpha \approx 0$, close to the unbiased walk.
For very large absolute values of $\alpha$ the walker gets trapped to nodes with highest (lowest) 
degrees, an effect which was highlighted in the literature \cite{Calva-Riascos2022}.
\subsection{Recurrence and ergodicity of Markov chain (\ref{purely_markovian})}
\label{rec_mar_cha}
Here we show recurrence and ergodicity of the (immortal) Markov walk (\ref{purely_markovian}), which is equivalent to the proper normalization of first hitting PDF $\chi_{i \to {\cal B}}(t)$: any non-empty target is eventually (recurrently) hit with probability one.
The FHM of a t-node set ${\cal B}$ is given by
\beq
\label{FHM_gen_target}
\pmb{\chi}(t) = \widetilde{\pmb{W}}^{t-1}\cdot \pmb{W}^{({\cal B})} , \hspace{0.5cm} t = \{1,2,\ldots \} \in \mathbb{N} ,
\eeq
see (\ref{first_hitting_mat}) with (\ref{transition_mat_mod}). 
The first hitting PDF of target ${\cal B}$
for departure node is $i$ $$\chi_{i\to {\cal B}}(t) = [\widetilde{\pmb{W}}^{t-1}\cdot \pmb{W}^{({\cal B})}]_i =
\sum_{b \in {\cal B}} [\widetilde{\pmb{W}}^{t-1}\cdot \pmb{W}]_{ib} ,\hspace{1cm} \chi_{i\to {\cal B}}(0) =0 .$$
Summing up the first hitting PDF over $t$ gives
\beq
\label{FHM_gen_tar_GF}
\sum_{t=1}^{\infty} \pmb{\chi}(t)\cdot |\phi_1 \rangle  = \left( \pmb{1} - \widetilde{\pmb{W}}\right)^{-1} \cdot \pmb{W}^{({\cal B})} \cdot |\phi_1 \rangle  = |C \rangle 
\eeq
where $\langle i| \pmb{\chi}(t) |\phi_1 \rangle  = [\pmb{\chi}(t)]_i = \chi_{i\to{\cal B}}(t)$ and
$|\phi_1 \rangle$ denotes the column vector with all components
$\langle i | \phi_1 \rangle = 1$. Let us determine the (unknown) constants $C_i = \langle i | C\rangle $.
Multiplying (\ref{FHM_gen_tar_GF}) with the matrix $[\pmb{1}- \widetilde{\pmb{W}}]$ from the left yields
\beq
\label{vect_eq}
\pmb{W}^{({\cal B})} \cdot |\phi_1 \rangle = \left( \pmb{1} - \widetilde{\pmb{W}}\right) \cdot |C \rangle .
\eeq
The LHS is a well-defined column vector with the components $ [\pmb{W}^{({\cal B})}]_i= \sum_{j \in {\cal B}} W_{ij}$
and on the RHS one has $det\left( \pmb{1} - \widetilde{\pmb{W}}\right) \neq 0$ (this matrix is invertible for any non-empty t-node set). 
Consequently, the system (\ref{vect_eq}) has the unique solution
\beq
\label{unique_sol}
C_i = 1 ,\hspace{1cm} i=1,\ldots, N 
\eeq
proving proper normalization of
the first hitting PDF $ \chi_{i\to{\cal B}}(t)$ with recurrence and ergodicity of the Markov chain.

\subsection{Invertibility of $\pmb{1}-\overline{\pmb{\Upsilon}}(1)$}
\label{invert}
Here we show that the matrix $\pmb{1}-\overline{\pmb{\Upsilon}}(1)$ is invertible, or equivalently that $\overline{\pmb{\Upsilon}}(1)$ defined in (\ref{GF_upsilon_arb_tar_MRW}) has 
spectral radius $\rho[\overline{\pmb{\Upsilon}}(1)] < 1$. 
A consequence of this feature is evanescence of the MRW propagator (see (\ref{evan} with (\ref{GF_thefollowing_important_MRW})).
To this end, it is sufficient to show that $(\overline{\pmb{\Upsilon}}(1))^n \to \pmb{0}$ as $n\to \infty$.
From inequality (\ref{not_properly}) we have that $[\overline{\pmb{\Upsilon}}(1)]_i < 1$ for all $i=1,\ldots,N$. 
Let ${\hat \Upsilon} = max([\overline{\pmb{\Upsilon}}(1)]_i) < 1$. Then we have that
\beq
\label{matrix_power_1}
[\overline{\pmb{\Upsilon}}(1)]_{ij} \leq [\overline{\pmb{\Upsilon}}(1)]_i \leq  {\hat \Upsilon} , \hspace{1cm} i, j = 1, \ldots, N ,
\eeq
thus $[((\overline{\pmb{\Upsilon}}(1))^2]_{ij} = \sum_{k=1}^N \overline{\pmb{\Upsilon}}(1)]_{ik} \overline{\pmb{\Upsilon}}(1)]_{kj} \leq {\hat \Upsilon} \, [\overline{\pmb{\Upsilon}}(1)]_i  
\leq  {\hat \Upsilon}^2 $ and so
\beq
\label{matrix_power_n}
[(\overline{\pmb{\Upsilon}}(1))^n]_{ij} \leq  {\hat \Upsilon}^n \to 0 ,\hspace{1cm} (n \to \infty).
\eeq

\subsection{The THCP for stationary Markov chains}
\label{stat_Markov_chains}
Here we consider the THCP for a stationary Markov chain. The transition matrix for the steps then is
$\pmb{W} =[W_{ij}]=[w_j]$ with identical rows and positivity $w_j>0$ ($j=1,\ldots,N$) as there are no disconnected nodes. 
It is easy to see that matrix powers of $\pmb{W}$ remain stationary,
$\pmb{W}^t = \pmb{W}$ ($t \in \mathbb{N}$) with canonical representation $\pmb{W} = |\phi_1\rangle \langle w|$, where $\langle i|\phi_1\rangle =1$ and 
$ \langle w|j\rangle =w_j$ with the eigenvalues $\lambda_1=1$ and $N-1$ eigenvalues $\lambda_m=0$ ($m=2,\ldots, N$).

We consider an arbitrary t-node set ${\cal B}$. The involved defective transition matrices 
are here $\widetilde{\pmb{W}} = [ w_j(1-\Theta(j;{\cal B})) ]$ and $\pmb{W}^{({\cal B})} = [w_j\Theta(j;{\cal B})]$ (see (\ref{transition_mat_mod})).
The generating matrix (\ref{Omega}) has the entries $\pmb{\Omega}(\xi;{\cal B}) = [ w_j\xi^{\Theta(j;{\cal B})} ]$.
Then we observe the commuting feature
\beq
\label{observation}
[\widetilde{\pmb{W}} \cdot \pmb{W}^{({\cal B})}]_i = [\pmb{W}^{({\cal B})} \cdot \widetilde{\pmb{W}}]_i  = 
\sum_{j=1}^N\sum_{k=1}^N  w_j(1-\Theta(j;{\cal B})) w_k\Theta(k;{\cal B}) =
[\widetilde{\pmb{W}}]_i [\pmb{W}^{({\cal B})}]_i =  p_{\cal B} q_{\cal B} ,
\eeq
leading to the first hitting PDF (\ref{first_hitting_PDF}) $\chi_{i\to {\cal B}}(t) =  [\widetilde{\pmb{W}}^{t-1} \cdot \pmb{W}^{({\cal B})}]_i= q_{\cal B}^{t-1}p_{\cal B}$ ($t \in \mathbb{N}$), 
in which $p_{\cal B} =  [\pmb{W}^{({\cal B})}]_i = \sum_{b \in {\cal B}} w_b$ is the probability to hit ${\cal B}$ in one step
and $q_{\cal B} = [\widetilde{\pmb{W}}]_i =1-p_{\cal B}$ the complementary probability, all independent of $i$. 
Now we evaluate the moment GF 
\beq
\label{hitting_power_GF}
\Big\langle \xi^{{\cal N}_{ij}(t; path, {\cal B})} \Big\rangle_{path(i;t)} 
=[(\pmb{\Omega}(\xi;{\cal B}))^t]_i = [\left(\widetilde{\pmb{W}}+ \xi\pmb{W}^{({\cal B})}\right)^t]_i = 
\left([\widetilde{\pmb{W}}]_i+ \xi[\pmb{W}^{({\cal B})}]_i \right)^t =
\left(q_{\cal B} +\xi p_{\cal B}\right)^t ,
\eeq
which we identify with the moment GF of the counting variable of a Bernoulli counting process, see \cite{Pachon_etal2025}.
From this moment GF it is straightforward to obtain for the state probabilities the {\it Binomial distribution}, namely
$$Q_i^{(n)}(t) = \frac{1}{n!} \frac{d^n}{d\xi^n} [(\pmb{\Omega}(\xi;{\cal B}))^t]_i\big|_{\xi=0} = \frac{t!}{(t-n)!n!} p_{\cal B}^n q_{\cal B}^{t-n} , \hspace{1cm} n, t \in \mathbb{N}_0 , $$
yielding for $n=0$
\beq
\label{proba_not_hit}
\Lambda_i(t) = [\widetilde{\pmb{W}}^t]_i = \left[\left(\sum_{j=1}^N w_j(1-\Theta(j;{\cal B}))\right)^t\right]_i  = q_{\cal B}^t.
\eeq
We infer that the THCP in a stationary Markov chain is a (Markovian) Bernoulli counting process.
The expected number of t-node hits up and including time $t$ is 
$\langle {\cal N}_i(t;{\cal B}) \rangle =
\frac{d}{d\xi}[(\pmb{\Omega}(\xi;{\cal B}))^t]_i\bigg|_{\xi=1} = p_{\cal B} t$ being the characteristic linear relation of the Bernoulli counting process.
All these quantities are independent of the departure node, reflecting the memoryless feature of the Bernoulli THCP. Finally, we evaluate the MFPT from (\ref{MFPT_to_B})
which we can expand using (\ref{proba_not_hit}) to arrive at
\beq
\label{MFPT_to_B_stat_MC}
\big\langle {\cal T}_{i \to {\cal B}} \big\rangle = [(\pmb {1}-{\widetilde {\pmb W}})^{-1}]_i = 
\sum_{r=0}^{\infty} [{\widetilde {\pmb W}}^r]_i = \sum_{r=0}^{\infty} q_{\cal B}^r = \frac{1}{p_{\cal B}} ,
\eeq
which is the expected time between successive Bernoulli events (hitting events of ${\cal B}$).
Since (\ref{MFPT_to_B_stat_MC}) is independent of the departure node, it is also the mean recurrence time
to ${\cal B}$. This relation is therefore a generalization of Kac's theorem to arbitrary t-node sets ${\cal B}$ holding for stationary Markov chains.

\subsection{MRW survival probability for a single target node}
\label{survival_derivation}
Point of departure is expression (\ref{starvation_survival}), which can be rewritten as
\beq
\label{starvation_survival_MRW}
\begin{array}{clr}
\ds {\cal P}_{\cal S}(t;i,b)  = & \ds \Big\langle \Theta(\Delta t_{ib}-1-t) \Theta(T_1-1-t])  \Big\rangle & \\  & & \hspace{-1cm}\ds  t \in \mathbb{N}_0 
\\ \\ & \ds + \Big\langle\sum_{n=1}^{\infty}  {\cal M}_n(i,b)  \Theta[\tau_n(i,b),t,\tau_{n+1}(i,b)]  \Theta(T_{n+1}-1-[t-\tau_n(i,b)]) \Big\rangle  & 
\end{array}
\eeq
with the indicator function
${\cal M}_n(i,b)$ of (\ref{memory_n}). We denote the t-node with $b$, and consider 
departure node $i$. Performing first the average with respect to the IID $T_k$ yields
\beq
\label{simpler_survival}
 {\cal P}_{\cal S}(t;i,b)  = \Lambda_i(t) \Phi_{\cal S}(t) + \sum_{n=1}^{\infty} \Big\langle \Phi_{\cal S}(\Delta t_{ib}) \prod_{r=2}^n 
 \Phi_{\cal S}(\Delta t_{bb}^{(r)}) \Theta[\tau_n(i,b),t,\tau_{n+1}(i,b)] \Phi_{\cal S}(t-\tau_n(i,b)) \Big\rangle 
\eeq
with $\big\langle \Theta(\Delta t_{ib}-1-t) \big\rangle =\Lambda_i(t)$.
Now take the GF and use the feature $ \Theta[\tau_n(i,b),t,\tau_{n+1}(i,b)] =1 $ for $\tau_n(i,b) \leq t \leq \tau_{n+1}(i,b) -1$ and null else to arrive at
\beq
\label{GF_survival_prob}
\begin{array}{clr}
\ds {\overline {\cal P}}_{\cal S}(u;i,b) & =\ds  \sum_{t=0}^{\infty} {\cal P}_{\cal S}(t;i,b) u^t = {\overline \Pi}_i(u) + \sum_{n=1}^{\infty}
 \Big\langle \Phi_{\cal S}(\Delta t_{ib}) u^{\Delta t_{ib}} \Big\rangle 
  \prod_{r=2}^n \Big\langle \Phi_{\cal S}(\Delta t^{(r)}_{bb}) u^{\Delta t^{(r)}_{bb}}  \Big\rangle \times & \\ \\
 & \ds \times  \Big\langle \sum_{k=0}^{\infty}  \Phi_{\cal S}(k)  \Lambda_b(k) u^k   \Big\rangle ,   &
 \end{array}
\eeq
where in the last line we have put $k=t-\tau_n(i,b)$. This relation contains the GFs
\beq
\label{Pi_i_GF}
{\overline \Pi}_i(u) =\big\langle \sum_{k=0}^{\Delta t_{ib}^{(n+1)}-1} u^k  \Phi_{\cal S}(k)  \big\rangle  
= \sum_{k=0}^{\infty} u^k  \Phi_{\cal S}(k)  \big\langle  \Theta(\Delta t_{ib}-1-k)  \big\rangle =  \sum_{k=0}^{\infty}  \Phi_{\cal S}(k)  \Lambda_i(k) u^k 
\eeq
and
\beq
\label{upsilons}
{\overline \Upsilon}_{ib}(u) =  \Big\langle \Phi_{\cal S}(\Delta t_{ib}) u^{\Delta t_{ib}} \Big\rangle = 
\sum_{k=0}^{\infty} u^{k} \chi_{ib}(k) \Phi_{\cal S}(k) .
\eeq
We can further evaluate these GFs, namely

\beq
\label{GF_walker_survival_not_reach_b}
{\overline \Pi}_i(u) = \sum_{r=0}^{\infty} u^r \Lambda_i(r) \Phi_{\cal S}(r) = \sum_{r=0}^{\infty} [\left(u \widetilde{\pmb{W}}\right)^t]_i \Phi_{\cal S}(r) = 
\left[ {\overline \Phi}_{\cal S}(u \widetilde{\pmb{W}}) \right]_i 
\eeq
and
\beq
\label{survival_excursion}
{\overline \Upsilon}_{ib}(u) = \sum_{r=1}^{\infty}  [\widetilde{\pmb{W}}^{r-1}\cdot \pmb{W}]_{ib}  \Phi_{\cal S}(r) u^r  =
u \left[{\overline G}_{\cal S}(u\widetilde{\pmb{W}})\cdot\pmb{W}\right]_{ib}  , \hspace{1cm} |u|\leq 1 
\eeq
being consistent with (\ref{Pi_GF_MRW}) and (\ref{GF_upsilon_arb_tar_MRW}), respectively.
Plugging (\ref{GF_walker_survival_not_reach_b}) and (\ref{survival_excursion}) into (\ref{GF_survival_prob}) yields
\beq
\label{Survival_GF_result}
\begin{array}{clr}
\ds {\overline {\cal P}}_{\cal S}(u;i,b) & =\ds  {\overline \Pi}_i(u) + {\overline \Upsilon}_{ib}(u) {\overline \Pi}_b(u)
\sum_{n=1}^{\infty} ({\overline \Upsilon}_{bb}(u))^{n-1} = {\overline \Pi}_i(u) + 
\frac{{\overline \Upsilon}_{ib}(u)}{1- {\overline \Upsilon}_{bb}(u)}{\overline \Pi}_b(u) &  \\[2ex]
 & =\ds  {\overline \Pi}_i(u) + {\overline {\cal H}}_{ib}(u) {\overline \Pi}_b(u) & \\[2ex]
 & = \ds {\overline \Pi}_i(u) + {\overline \Upsilon}_{ib}(u) {\overline {\cal P}}_{\cal S}(u;b,b) . &
 \end{array}
\eeq
This relation contains the GF of the MHR of node $b$ 
\beq
\label{hitting_rate_GF}
{\overline {\cal H}}_{ib}(u) = \frac{{\overline \Upsilon}_{ib}(u)}{1- {\overline \Upsilon}_{bb}(u)} .
\eeq

\hfill

\end{document}